\documentclass[a4paper,12pt]{article}

\usepackage[utf8]{inputenc}
\usepackage[numbers]{natbib}
\usepackage[T1]{fontenc}
\usepackage{amsmath, amssymb, amsfonts}
\usepackage{graphicx}
\usepackage{hyperref}
\usepackage{subcaption}
\usepackage{geometry}
\usepackage{cleveref}
\usepackage{color}
\usepackage{bm}
\usepackage{xcolor}
\usepackage{url}

\title{Introduction to Stability and Turbulent Transport in Magnetic Confinement Fusion Plasmas}
\author{J. F. Parisi$^{1,2}$  \\ \small{$^{1}$ Marathon Fusion, $^{2}$ Princeton Plasma Physics Laboratory} \\ \small{\texttt{jason@marathonfusion.com}}}
\date{\today}

\begin{document}

\maketitle

\begin{abstract}
This tutorial provides an accessible introduction to the principles of stability and turbulent transport in magnetic confinement fusion plasmas. Key concepts, models, and practical implications are discussed to guide researchers new to the field. Some challenges and opportunities are discussed.
\end{abstract}

\tableofcontents

\section{Short Problem Description} \label{sec:problem_description}

We begin this tutorial with a short motivational problem. A thorough introduction is provided in \Cref{sec:introduction}. The concepts introduced in this section are described in more detail in later sections.

Building a fusion power plant that produces substantial net energy requires effective confinement of hot plasma, so that fusion reactions occur frequently enough to generate significant power. However, plasma confinement is challenging, as energy continuously escapes through transport processes. For a given heating source, improved confinement  -- i.e., reduced transport and higher temperature gradients -- allows the plasma core to reach higher temperatures. Historically, as fusion devices achieved hotter plasmas, transport losses also increased, limiting further gains in core temperature. The main drivers of these losses are plasma instabilities and turbulence. This tutorial introduces key concepts and tools used to study plasma stability and turbulent transport.

To motivate the importance of these processes in magnetic confinement fusion (MCF) -- which includes machines such as tokamaks, stellarators, and mirrors -- we begin with a simplified example. This example illustrates how the total fusion power output can depend sensitively on three core transport properties: the critical temperature gradient, the heat diffusivity, and the transport stiffness.

\begin{figure}[t]
\centering
    \includegraphics[width=0.88\textwidth]{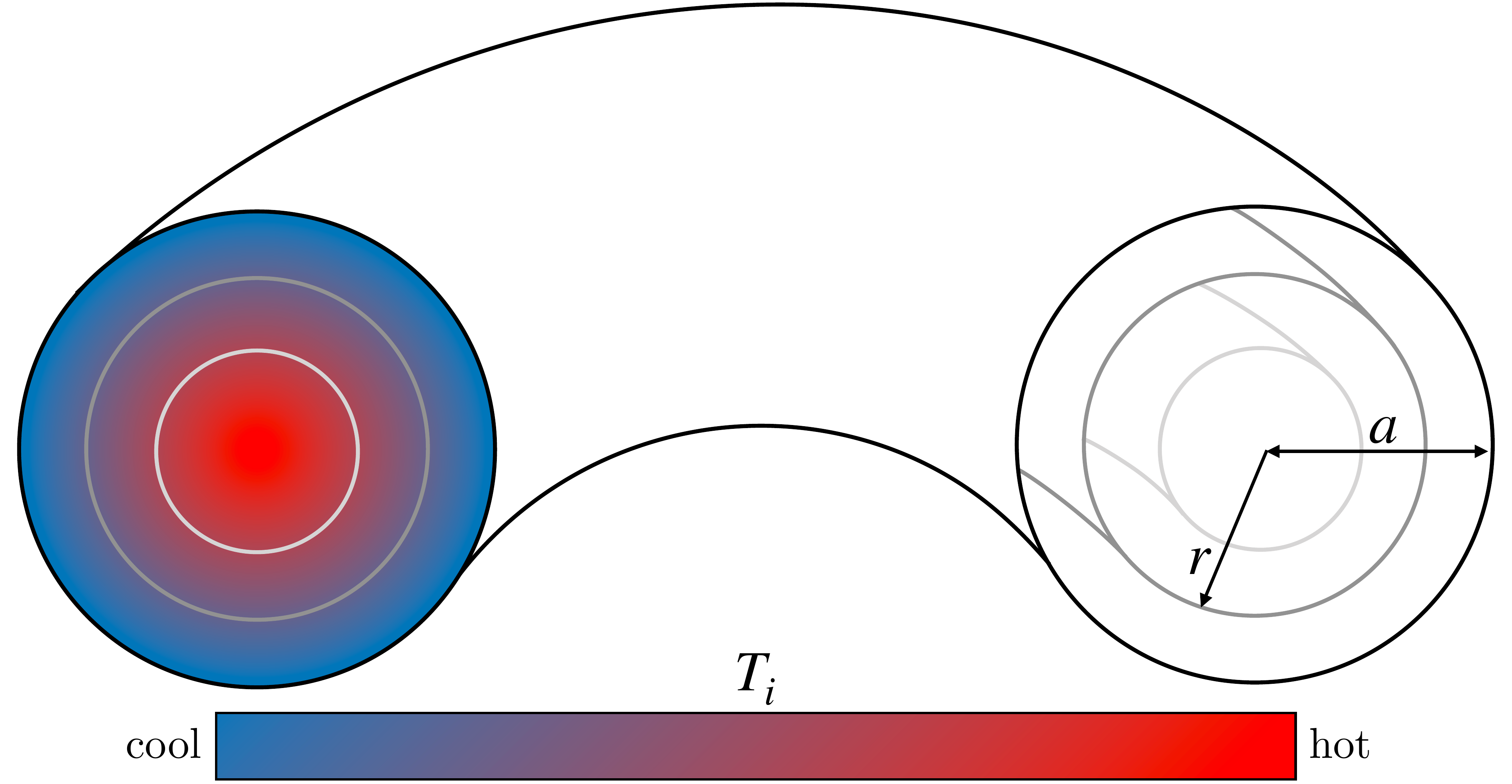}
 \caption{Cartoon of ion temperature profile, radial coordinate $r$, and minor radius $a$ for \Cref{eq:simple_steady_energy,eq:qi_simple_intro}.}
 \label{fig:simple_torus}
\end{figure}

Consider a torus where energy is injected at a volumetric rate $p_\mathrm{ext,i}$ (in Watts per cubic meter) around the plasma center. We are interested in the ion temperature $T_i$ as a function of the radial coordinate $r$. A cartoon of $T_i$ and the radial coordinate $r$ is shown in \Cref{fig:simple_torus}. In steady-state the plasma ion pressure is constant in time and the conservation equation for radial energy transport is
\begin{equation}
\frac{1}{r} \frac{\partial}{\partial r} \left( r q_i \right)  = p_\mathrm{ext,i},
\label{eq:simple_steady_energy}
\end{equation}
where $q_i$ is the ion heat flux (in Watts per square meter) and $r$ is the minor radial coordinate (in meters). The left-hand side of \Cref{eq:simple_steady_energy} describes energy flux in the radial direction and is equal to a divergence, $\nabla \cdot \mathbf{q}_i$. The right-hand side describes an energy source. Therefore, \Cref{eq:simple_steady_energy} is similar to Gauss's law $\nabla \cdot \mathbf{E} = 4 \pi \tilde{\rho}$ for charge density $\tilde{\rho}$ that sources the electric field $\mathbf{E}$. 

We assume that the normalized, dimensionless heat transport is described by
\begin{equation}
\hat{q}_i = \hat{\chi}_i \left( \frac{a}{L_{T,i}} - \frac{a}{L_{T,i}^{\mathrm{crit}}} \right)^{\alpha_\mathrm{stiff} } \mathrm{H} \left(\frac{a}{L_{T,i}} - \frac{a}{L_{T,i}^{\mathrm{crit}}} \right),
\label{eq:qi_simple_intro}
\end{equation}
where the heat flux is normalized by a `gyroBohm' value $q_{i,\mathrm{gB} }$,
\begin{equation}
\hat{q}_i \equiv \frac{q_i}{q_{i,\mathrm{gB} }}, \;\;\;\;\;\; q_{i,\mathrm{gB} } \equiv \left( \frac{\rho_i}{a} \right)^2 n_i v_{ti} T_i,
\end{equation}
and the heat diffusivity $\chi_i$ is also accordingly normalized
\begin{equation}
\hat{\chi}_i \equiv \frac{\chi_i}{\chi_{i,\mathrm{gB} }}, \;\;\;\;\;\; \chi_{i,\mathrm{gB} } \equiv \frac{\rho_i^2 v_{ti}}{a}. 
\end{equation}
The quantity $a$ is the torus minor radius, $L_{T,i}$ is the logarithmic ion temperature gradient,
\begin{equation}
L_{T,i} \equiv - \left( \partial T_i / \partial r \right)^{-1},
\end{equation}
$T_i$ is the ion temperature, $L_{T,i}^\mathrm{crit}$ is the critical ion temperature gradient below which there is zero transport, $\alpha_\mathrm{stiff}$ is a `stiffness' parameter, $\mathrm{H}$ is a Heaviside function, $n_i$ is the ion density, $v_{ti} = \sqrt{2 T_i / m_i} $ is the ion thermal speed, $m_i$ is the ion mass, $\rho_i = v_{ti} / \Omega_{c,i}$ is the ion gyroradius, and $\Omega_{c,i} = q_i B / c m_i$ is the ion cyclotron frequency where $q_i$ is the total ion charge, $B$ is the magnetic field amplitude, and $c$ is the speed of light. The form of $\hat{q}_i$ in \Cref{eq:qi_simple_intro} assumes that turbulence drives all of heat transport; in following sections, we show that other mechanisms also contribute to the heat transport, although they are typically much smaller than turbulence in current machines.

The form of $\hat{q}_i$ in \Cref{eq:qi_simple_intro} is motivated by experiments and theory showing that transport (1) has varying `stiffness' -- \Cref{eq:qi_simple_intro} $\alpha_\mathrm{stiff}$ is typically larger than one but can vary significantly -- (2) has a critical gradient $a/L_{T,i}^\mathrm{crit} \neq 0$, below which there is zero transport -- and (3) can be written as proportional to a diffusivity $\hat{\chi}_i$ \cite{Horton1999,Jenko2001,Hoang_2001,Baker_2001,Ryter_2005,Camenen_2005,Mantica_2009,Hillesheim_2013,Horton1999,Stober_2000,Mantica_2011,Sauter2014,Luce_2018,Garbet_2004b,Barnes2011,Holland_2013,Citrin_2014,Kim2024}. We emphasize that while the transport equation in \Cref{eq:simple_steady_energy} and the form of $\hat{q}_i$ in \Cref{eq:qi_simple_intro} are highly simplified, they illustrate the importance of stability and transport.

\begin{figure*}[bt!]
    \centering
    \begin{subfigure}[t]{0.49\textwidth}
    \includegraphics[width=0.99\textwidth]{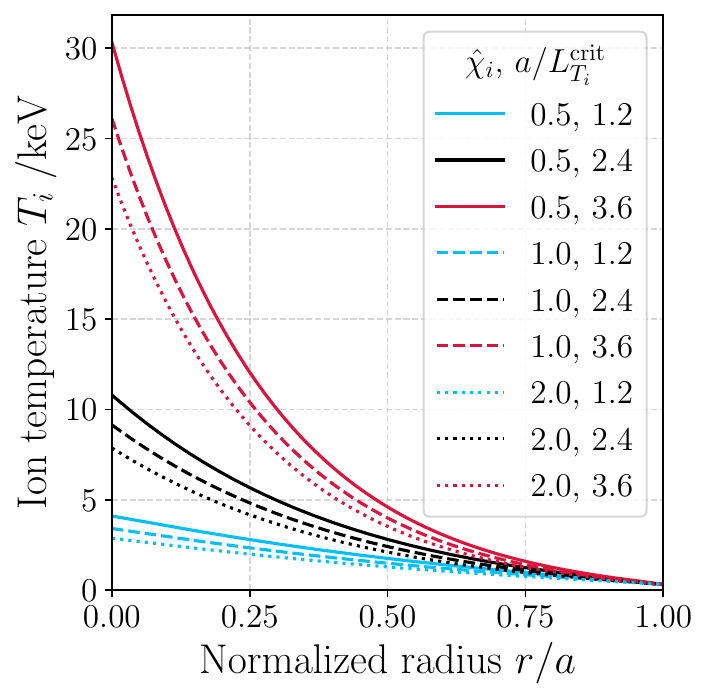}
    \caption{$\alpha_\mathrm{stiff} = 3.0$.}
    \end{subfigure}
    \centering
    \begin{subfigure}[t]{0.49\textwidth}
    \includegraphics[width=0.99\textwidth]{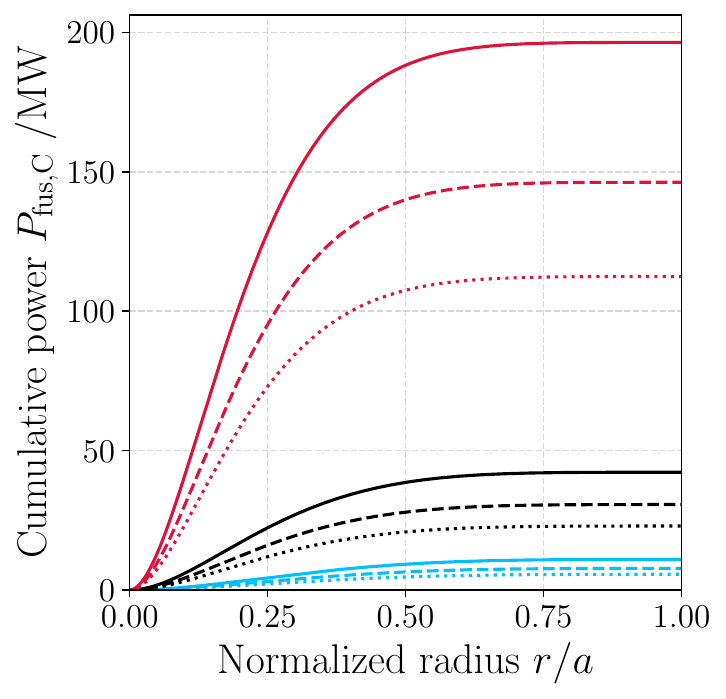}
    \caption{$\alpha_\mathrm{stiff} = 3.0$.}
    \end{subfigure}
    \centering
    \begin{subfigure}[t]{0.49\textwidth}
    \includegraphics[width=0.99\textwidth]{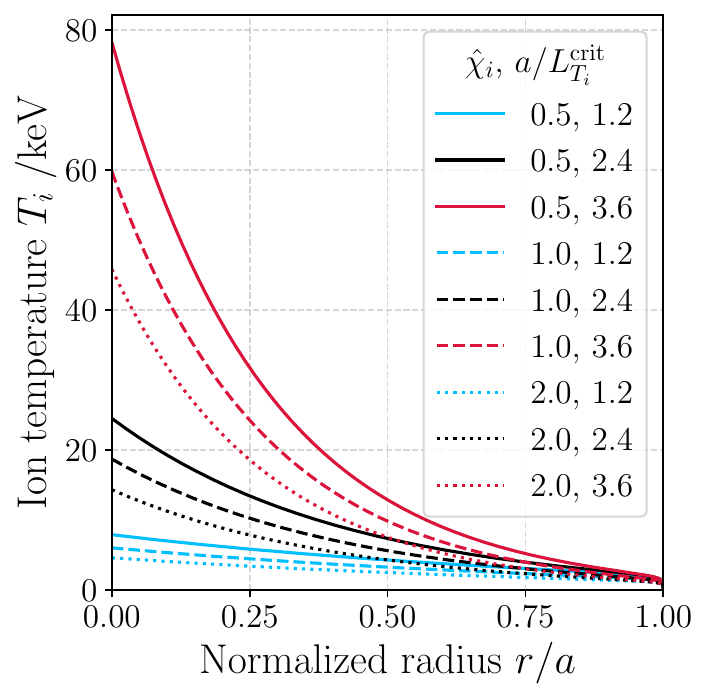}
    \caption{$\alpha_\mathrm{stiff} = 1.0$.}
    \end{subfigure}
    \centering
    \begin{subfigure}[t]{0.49\textwidth}
    \includegraphics[width=0.99\textwidth]{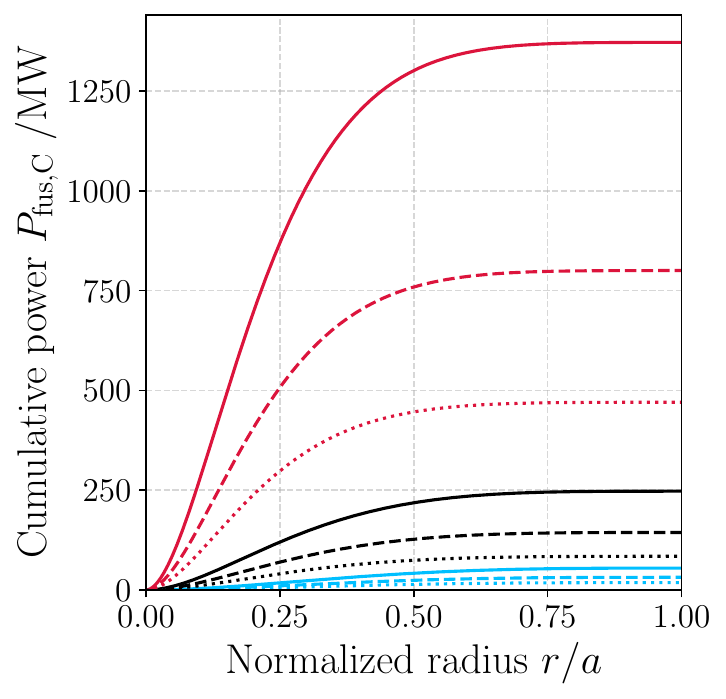}
    \caption{$\alpha_\mathrm{stiff} = 1.0$.}
    \end{subfigure}
    \caption{(a), (c) ion temperature profiles for different $\hat{\chi}_i$ and $a/L_{T,i}^\mathrm{crit}$ values for $\alpha_\mathrm{stiff} = 3.0$ and $1.0$ respectively; (b), (d) corresponding cumulative fusion power profiles (see \Cref{eq:PfusC}).}
    \label{fig:simple_example}
\end{figure*}

We solve \Cref{eq:simple_steady_energy} using three values of the critical gradient and three values of the diffusivity,
\begin{equation}
\frac{a}{L_{T,i}^{\mathrm{crit}}} \in [1.2, 2.4, 3.6], \;\;\;\;\; \hat{\chi}_i = [0.5, 1.0, 2.0].
\end{equation}
The resulting ion temperature profiles are shown in \Cref{fig:simple_example}(a) for $\alpha_\mathrm{stiff} = 3.0$: each of the three linestyles corresponds to a different $\hat{\chi}_i$ value and each of the three colors corresponds to a different $a/L_{T,i}^\mathrm{crit}$ value. In \Cref{fig:simple_example}(b), we plot the cumulative fusion power up to a given radial location,
\begin{equation}
P_\mathrm{fus,C} (r) = \int_0^{V(r)} p_\mathrm{fus} dV,
\label{eq:PfusC}
\end{equation}
where $p_\mathrm{fus}$ is the fusion power density (see \Cref{eq:powerdensity}) and $V(r)$ is the plasma volume enclosed by a surface at radius $r$. We assume the volume element $dV$ satisfies circular flux surfaces shown in \Cref{fig:simple_torus}. The results are striking: a 50\% change in $a/L_{T,i}^\mathrm{crit}$ changes the total fusion power by roughly a factor of five, whereas a 50\% change in $\hat{\chi}_i$ changes the total fusion power by less than 50\%. We repeat this exercise for a less `stiff' transport mechanism, $\alpha_\mathrm{stiff} = 1.0$, in \Cref{fig:simple_example}(c) and (d). For transport that is less stiff, $\hat{\chi}_i$ becomes increasingly important for the variation in the cumulative fusion power but in this simple model, is still less important than $a/L_{T,i}^\mathrm{crit}$.

This exercise illustrates some important features of turbulent transport. The energy fluxes are determined by at least three important parameters: the critical gradient, the diffusivity, and the stiffness. Plasma `stability' typically determines $a/L_{T,i}^\mathrm{crit}$ and `transport' typically determines $\hat{\chi}_i$ and $\alpha_\mathrm{stiff}$ although there are turbulent phenomena that can modify $a/L_{T,i}^\mathrm{crit}$ \cite{Dimits2000,Mikkelsen2008,Zhu2020,Pueschel2021,Ivanov2022,Hoffmann2023,Cao2023}. Therefore, if one is looking to increase the core temperature of fusion power plants, there are multiple approaches: increase the critical gradient, and decrease the turbulent diffusivity, and decrease the stiffness.  We are ultimately interested interested in the total enclosed fusion power; the distribution of the integrated volume in \Cref{eq:PfusC} has a big effect on the fusion power \cite{parisi2025a,Di2025burn}. There are many more subtleties, some of which we will cover in later sections.

A second important feature is that predicting profiles can be much less challenging in systems with high transport stiffness. The reason is that finding $\hat{\chi}_i$ is usually orders of magnitude harder than finding $a/L_{T,i}^\mathrm{crit}$. This is because finding $\hat{\chi}_i$ involves nonlinear numerical simulations, whereas $a/L_{T,i}^\mathrm{crit}$ only requires linear simulations, or sometimes even pen and paper theory. Nonlinear simulations are typically much more demanding in numerical resources, implementation, and convergence checks, therefore they require more numerical and human time to complete. If transport is highly stiff ($\alpha_\mathrm{stiff} > 1$) then the steady-state temperature gradients might only be marginally above the critical linear gradient $a/L_{T,i}^\mathrm{crit}$. In this case, using the critical profile gradient would give a close estimate for the profiles. If the transport is not stiff \cite{Sauter2014}, high-fidelity transport simulations are often needed to calculate the achievable temperature gradients. In reality, profiles in devices such as tokamaks are set by many more effects than those laid out in \Cref{eq:simple_steady_energy,eq:qi_simple_intro} -- high-fidelity plasma turbulence modeling is \textit{sometimes} a necessary, but never sufficient, condition for accurately finding profiles.

\section{Introduction} \label{sec:introduction}

\subsection{Who is This for?}

The primary objective of this tutorial is to serve as an accessible entry point for understanding stability and transport in MCF plasmas. While the focus is on plasma stability and turbulence, the intended audience spans the full range of researchers and engineers, and it should be accessible to physics and engineering students at the Sophomore/Junior level of college. In fusion systems, many components are highly interdependent, and plasma confinement is no exception. This tutorial emphasizes practical applications of plasma stability and turbulent transport to MCF devices rather than the theoretical foundations of turbulence, for which there are many excellent resources \cite{Burgers1948,Orszag1970,Frisch1999,Horton1999,Krommes2002,Scott2007,Horton2012}. Some sections conclude with references for those interested in exploring the material further.

\subsection{What is Instability and Turbulence?}

First, let us define precisely what we mean by instability and turbulence in the context of fusion plasmas. Instability is when an initially small perturbation such as density or temperature grows over time. If this instability grows to become sufficiently large, the resulting plasma state can become turbulent. Turbulence, \textit{multiscale disorder} (a description borrowed from \cite{Schekochihin2008}), is a state where energy can be transferred across a range of spatial and temporal scales. One particularly interesting, and often challenging, aspect of turbulence in plasmas is that it is \textit{kinetic} -- that is, in additional to the problem having spatial and temporal dependence, there is also a velocity-space dependence.

\subsection{Nuclear Fusion Reactions in Plasmas}

The practical motivation to study stability and transport in MCF plasmas is that it is a powerful tool for improving the viability of fusion energy. Nuclear fusion energy promises to produce energy through nuclear reactions with the highest energy released per unit mass of any known physical mechanism with the exception of matter anti-matter annihilation. Nuclear fusion reactions release energy by fusing two nuclei into new configurations with a lower total nuclear binding energy. A particularly attractive fusion reaction is the deuterium-tritium (D-T) fusion reaction \cite{Strachan1994short,Keilhacker1999,Wurzel2022}, which fusions a deuterium and tritium nucleus into a fast helium-4 (alpha particle) and a fast neutron, 
\begin{align}
& \mathrm{D} + \mathrm{T} \longrightarrow \alpha + \mathrm{n} \;\;\;\;\; \left( E_{\mathrm{DT}} = 17.6 \; \mathrm{MeV} \right)
\end{align}
where $E_{\mathrm{DT}}$ is the energy released from a D-T fusion reaction. The neutron carries roughly four fifths of the energy released. Because helium-4 has an exceptionally high binding energy per nucleon, the energy released per D-T reaction is very high. The difference in binding energy between the incoming and outgoing particles is released as kinetic energy, which can be converted to heat and/or electricity.

\begin{figure}[t]
\centering
    \includegraphics[width=0.85\textwidth]{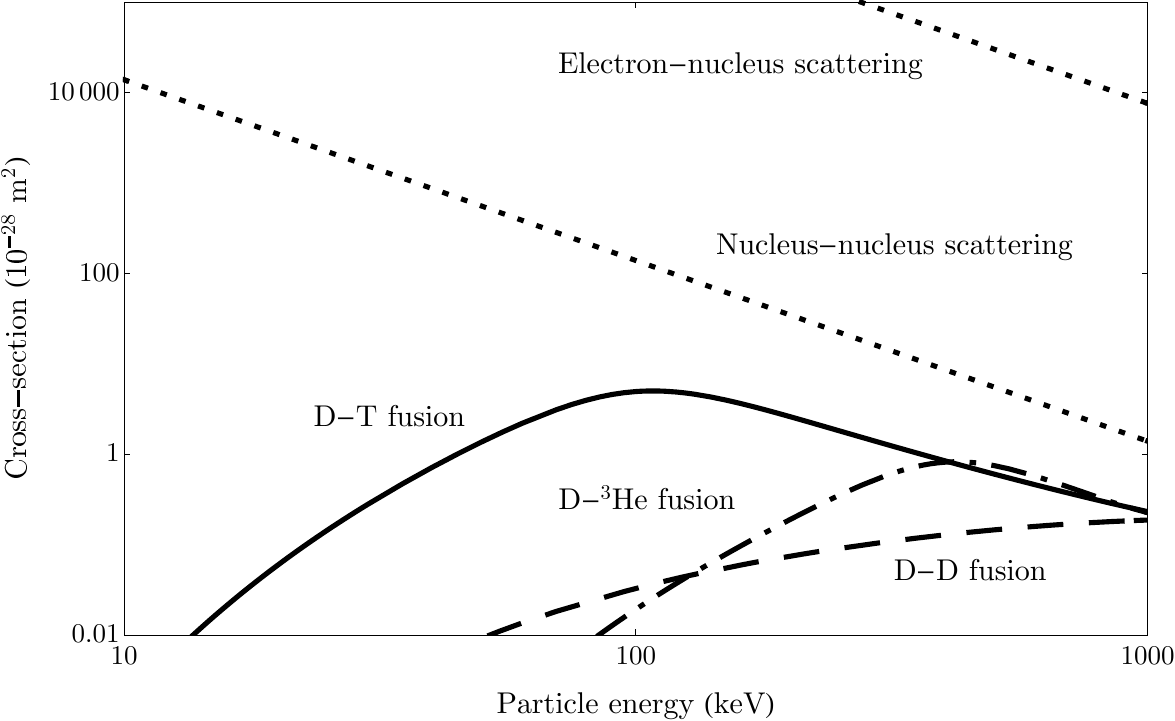}
 \caption{Cross sections for Coulomb scattering and fusion reactions versus particle energy. Adapted from \cite{Parisi2019}.}
 \label{fig:crosssections}
\end{figure}

Because the cross section for electrostatic repulsion between ions - shown by the nucleus-nucleus scattering cross section in \Cref{fig:crosssections} - is so much higher than fusion reactions, a viable fusion energy generation scheme must allow particles to scatter many times before fusion occurs. The probability of a deuterium and tritium tunneling through the electrostatic Coulomb potential barrier increases exponentially with energy \cite{Mott1965}. However, above a certain energy -- for D-T fusion approximately 100 keV -- the probability of fusion decreases \cite{Li2008}. Therefore according to \Cref{fig:crosssections} ideally our D-T fuel will have energies in the tens to hundreds of keV. At such high energies, the electrons and nuclei are disassociated, leaving a collection of positively charged nuclei (ions) and electrons called a plasma. In most magnetically confined plasmas, the ion distribution function is approximately a Maxwellian, 
\begin{equation}
F_{Ms} (\mathcal{E}) = n_s \Big{(} \frac{m_s}{2 \pi T_s } \Big{)}^{3/2} \exp \Big{(} - \frac{\mathcal{E}}{T_{s} } \Big{)},
\label{eq:particledistribution}
\end{equation}
where $s$ is the species label, $\mathcal{E} = m_s v/2$ is the particle energy with speed $v$ and mass $m_s$, and $n_{s}$ and $T_{s}$ are the density and temperature.

In plasmas with temperatures of tens of keV, despite the higher fusion cross section, Coulomb collisions between ions are still orders of magnitude more frequent than D-T fusion reactions -- see nucleus-nucleus scattering in \Cref{fig:crosssections}. This necessitates the plasma be \textit{confined} so that particles can be scattered many times before eventually undergoing fusion. In addition to the particles constituting the fusion fuel being well-confined, we also require that energy be well-confined. A fairly good analogy to confinement in a fusion plasma is a wood fire. The wood fire needs particles to burn. And heat must be sufficiently well-confined, otherwise the fire will eventually stop. On the other hand, the waste products from the fire must also be ventilated, otherwise there will be insufficient oxygen for ignition. Analogously, alpha particles from D-T fusion reactions must deposit their energy and then be removed from the plasma to prevent them from diluting the fusion fuel. Just as there are techniques for improving a wood fire's heat output, wood burn-up, and longevity, there are many techniques in development for improving the properties of fusion fire. These requirements -- and others we introduce later -- necessitate a \textit{confinement scheme}.

Because magnetic fields permeate plasma, one mainstream fusion energy confinement approach is to use strong magnetic fields to confine plasma in a toroidal cage. The plasma needs to be hot in the center of the torus -- so that the fusion rate is high -- and cold at the edge --so that the machine walls do not melt. These are two of the most stringent boundary conditions for MCF, but alone, are insufficient information to build a power plant: some of the next most important questions are how to configure the plasma temperature and pressure profiles given that we wish to generate significant fusion power and not melt the walls.

\begin{figure}[t]
\centering
    \includegraphics[width=0.8\textwidth]{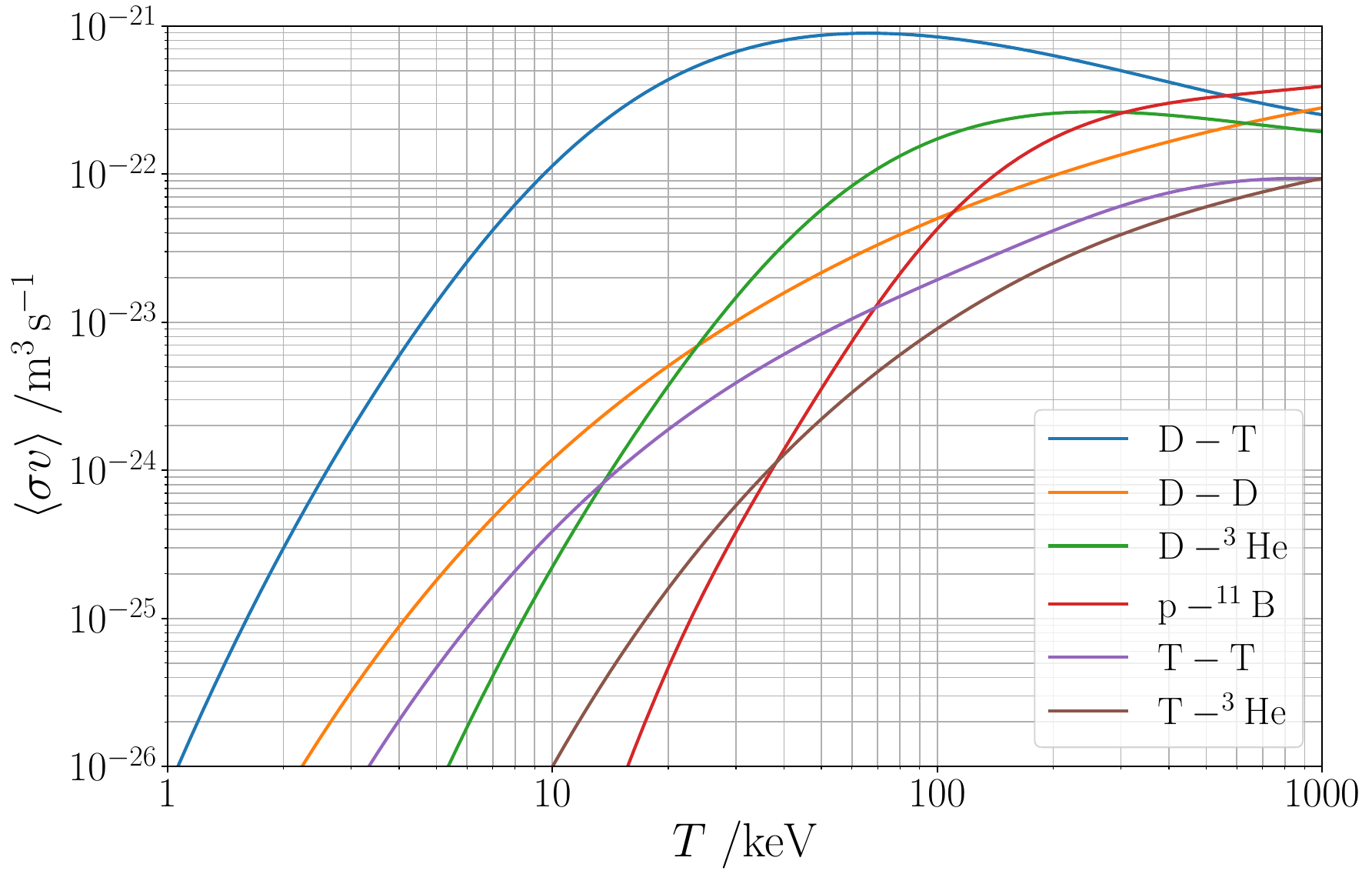}
 \caption{Reactivity of some fusion reactions with Maxwellian energy distributions.}
 \label{fig:reactivity}
\end{figure}

According to \Cref{fig:crosssections}, only particles with energies
\begin{equation}
20 \; \mathrm{keV} \lesssim \mathcal{E} \lesssim 500 \; \mathrm{keV},
\label{eq:E_requirement_simple}
\end{equation}
have a significant likelihood of undergoing fusion reactions. In fusion plasmas, this usually means that the overwhelming majority of fusion reactions come from ions in the high energy tails of $F_{Ms}$ in \Cref{eq:particledistribution} -- the remainder have insufficient energy to have a high likelihood of tunneling through the Coulomb barrier. Intuitively, one strategy to increase fusion power might be to increase the plasma temperature so that most particles satisfy \Cref{eq:E_requirement_simple}. However, this is usually a fruitless task for at least three reasons. 

First, for D-T fusion reactions, the fusion power peaks around $T \simeq 60$ keV, meaning that if many particles can significantly more energy than $\sim$500 keV, the cross section and therefore the fusion power drops. The D-T fusion power per unit volume is
\begin{equation}
p_\mathrm{fus} =  \frac{1}{4} n^2 \langle \sigma v \rangle E_\mathrm{DT},
\label{eq:powerdensity}
\end{equation}
where $n$ is the fuel number density, $\sigma$ is the cross section, and $\langle \sigma v \rangle$ is the fusion reactivity -- a velocity integral over the Maxwellian distribution function $F_{Ms}$ in \Cref{eq:particledistribution}. We plot the reactivity of some common fusion reactions in \Cref{fig:reactivity} -- while plasmas that burn too hot will have lower fusion power with increasing temperature, achieving such high temperatures in a MCF device is very challenging due to energy loss mechanisms such as radiation.

Second, because the ion and electron temperatures are usually quite similar, a plasma with higher ion temperature will also have higher electron temperature and therefore a higher Brehmstrahlung power per unit volume
\begin{equation}
    p_\mathrm{B} = C_\mathrm{B} n n_e T_e^{1/2}.
    \label{eq:pBreh_density}
\end{equation}
Here $n$ is total hydrogen density, $n_e$ is the electron density, and $C_\mathrm{B} = 3.34 \times 10^{-21}$ keV $\mathrm{m}^3$/s is a constant. Therefore, while \Cref{eq:powerdensity} and \Cref{fig:reactivity} show that higher power density might be achievable with higher temperature, the radiative power losses can become prohibitive.

Third, plasma confinement typically degrades with higher temperature. This is due to two main reasons: (1) Because of the requirement that a plasma is cool at the edge and hot in the center, an increasingly hot center results in steeper temperature gradients. These steep temperature gradients are a source of free energy that drive plasma instabilities, which cause further plasma losses -- sustaining a plasma with very high temperature gradients usually requires prohibitively high power; (2) Charged particles in a magnetic field undergo gyromotion in the plane perpendicular to the field; as a particle becomes more energetic, its gyroradius increases and the particle confinement time decreases due to various transport mechanisms. We will discuss this in detail in coming sections.

It is important to distinguish the transport discussed in this tutorial from other transport mechanisms. Here, we focus on `diffusive'-like transport as opposed to larger disruptive events. For example, magnetohydrodynamic (MHD) instability can cause very large displacements of energy and particles through events such as edge-localized modes (ELMs) \cite{Federici2019b,Creely2020,Muldrew2024,Maingi_2014,Hughes2020,Kuang2020,Viezzer2023} and plasma disruptions \cite{Wesson1989,Hender2007,Zakharov2012,Boozer2012,Strait2019,Kates2019}.

This tutorial is organized as follows: we introduce the Lawson criterion and other confinement considerations in \Cref{sec:Lawson}. In \Cref{sec:confimentschemes}, we introduce magnetic confinement schemes. In \Cref{sec:transport_overview}, we give an overview of the various transport mechanisms. In \Cref{sec:KT_GK_Transp}, we describe the theoretical framework for self-consistent transport calculations. In \Cref{sec:linearstab}, we review the gyrokinetic instabilities that give rise to turbulent transport. In \Cref{sec:codesworkflows}, we introduce some commonly used codes for gyrokinetic, neoclassical, transport, and magnetic equilibrium calculation. In \Cref{sec:toka_confinement}, we cover the main tokamak confinement regimes. In \Cref{sec:challenges}, we discuss some challenges and questions. Due to time constraints, I have omitted many important topics -- some of these are briefly covered in \Cref{sec:otherimportant}. We conclude in \Cref{sec:summary}.




\begin{figure}[bt]
    \centering
    \begin{subfigure}[t]{0.49\textwidth}
    \includegraphics[width=0.99\textwidth]{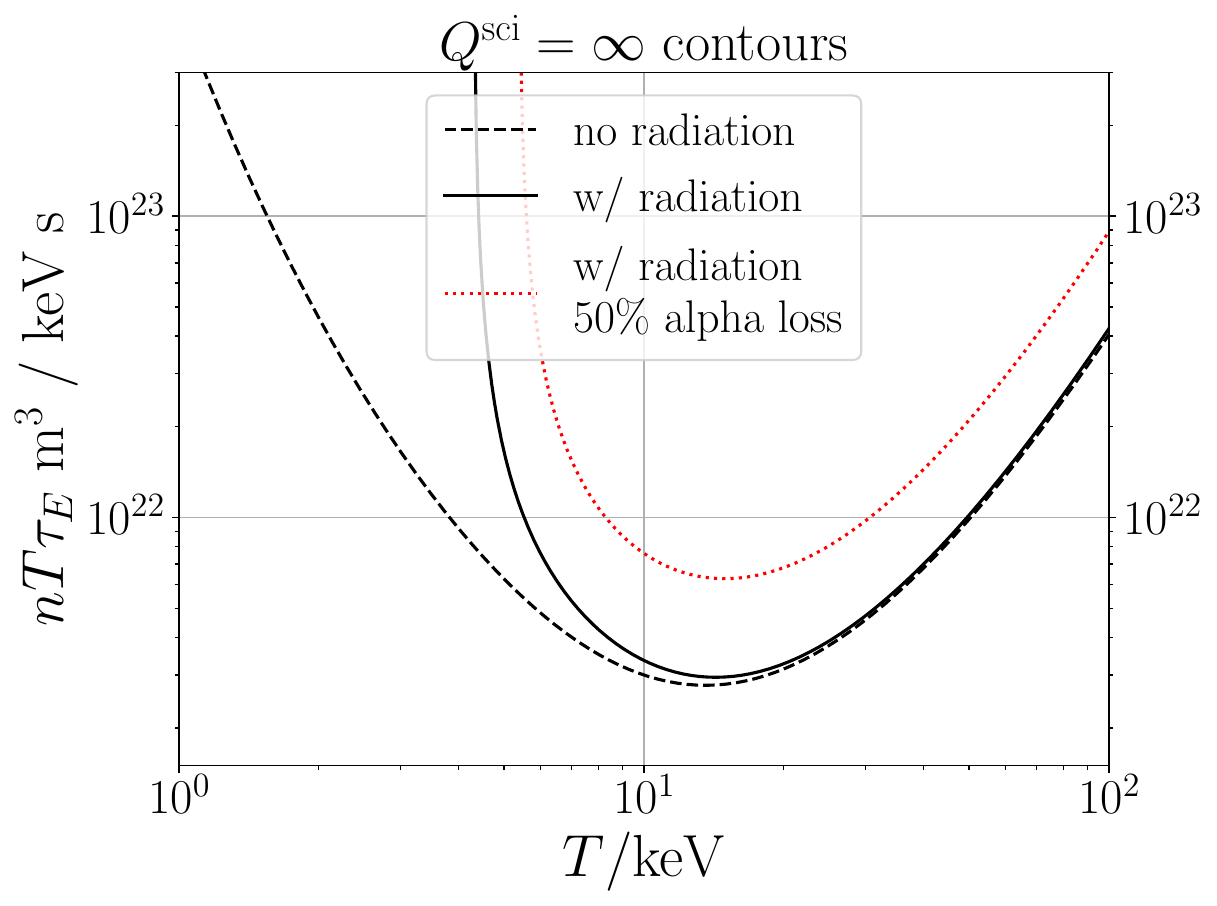}
    \caption{}
    \end{subfigure}
    \centering
    \begin{subfigure}[t]{0.49\textwidth}
    \includegraphics[width=0.99\textwidth]{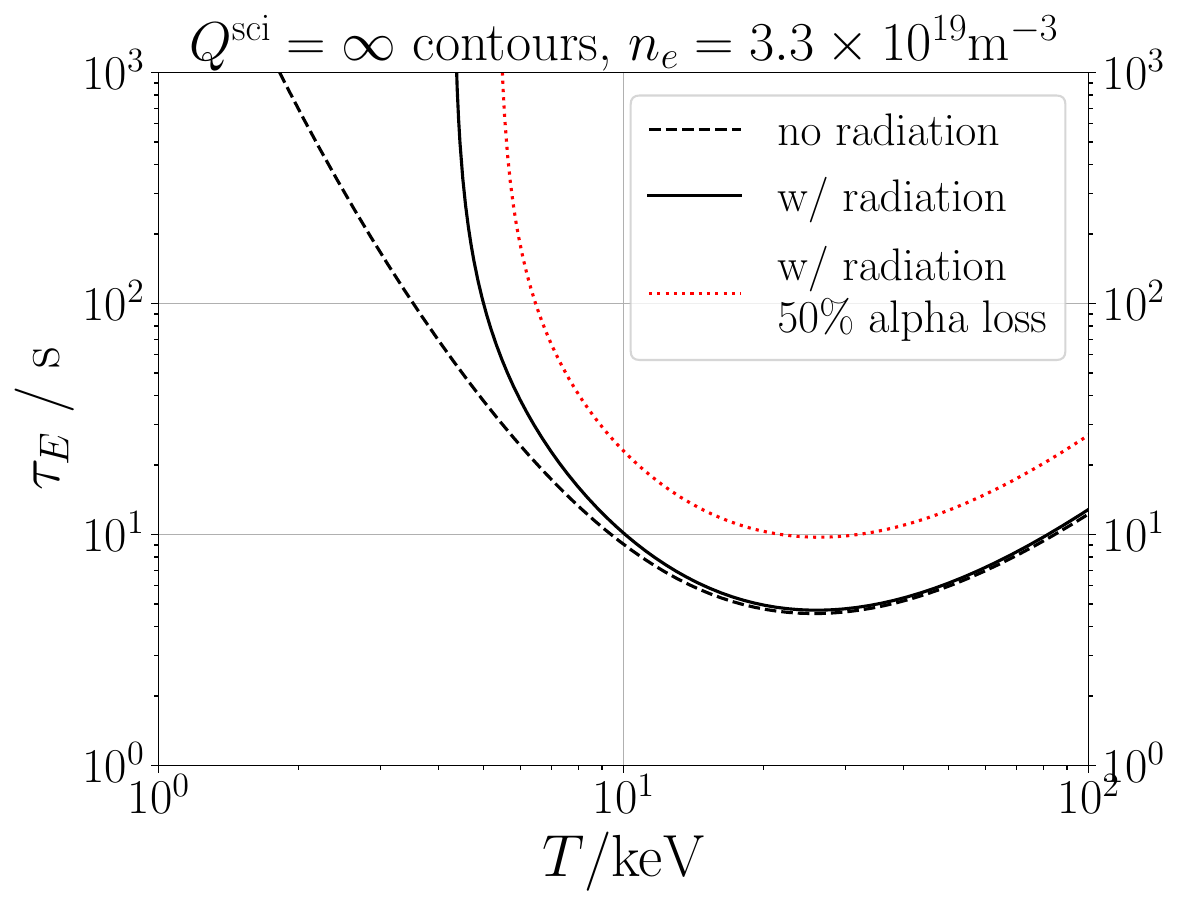}
    \caption{}
    \end{subfigure}
    \caption{(a) Lawson triple product and (b) energy confinement time, both for an ignited plasma. In (b), we assume a constant plasma density of $n = 3.3 \times 10^{19} \mathrm{m}^{-3}$.}
    \label{fig:lawsoncriterion}
\end{figure}

\section{Energy and Particle Confinement} \label{sec:Lawson}

In order to estimate the required quality of energy confinement in a fusion device, we defer to a derivation called the Lawson criterion \cite{Lawson1957,Wurzel2022}. The result will be a quantitative measure of the conditions required for fusion ignition, analogous to what we intuitively know for a wood fire: a roaring fire needs enough wood fuel (fusion fuel), temperature, and the heat must be confined for sufficiently long. We will also present an analogous argument for particle confinement.

\subsection{Lawson Criterion: Energy Confinement}

Here, we determine the required quality of energy confinement in a fusion plasma. The argument \cite{Lawson1957,Wurzel2022} begins with the requirement that power losses $p_\mathrm{L}$ should balance the power sources from alpha heating $p_\alpha$ and external heating $p_\mathrm{ext}$,
\begin{equation}
p_\mathrm{L} = p_\alpha + p_\mathrm{ext}.
\label{eq:powerbalance}
\end{equation}
We define the energy confinement time as
\begin{equation}
\tau_E \equiv \frac{w}{p_\mathrm{L} },
\label{eq:tauE_def}
\end{equation}
where $w = 3 n T/2$ is the stored thermal energy per unit volume for a plasma composed of ion density $n$ with ion temperature $T$. $\tau_E$ measures the time for a unit of energy deposited in the plasma center to leave the plasma. As formulated in \Cref{eq:tauE_def}, $\tau_E$ includes the effects of any physical mechanism that causes energy to leave the plasma.

A self-sustaining fusion plasma uses the alpha particles produced from D-T reactions to keep it hot. Because an alpha particle is approximately four times the neutron mass, roughly one fifth of the total fusion power is carried as kinetic energy of alpha particles. Therefore, the heating power provided by alpha particles $p_\alpha$ assuming no alpha particle losses is
\begin{equation}
p_\alpha = \frac{ p_\mathrm{fus}}{5}.
\end{equation}
Using \Cref{eq:powerbalance}, the external heating power required to sustain a plasma is
\begin{equation}
p_\mathrm{ext} = \frac{3 n T}{\tau_E}  - \frac{1}{4}  n^2 \langle \sigma v \rangle  \frac{E_\mathrm{DT}}{5}.
\label{eq:Pext_equation}
\end{equation}
When the plasma can supply all of its heating requirements from alpha particles the external power requirement vanishes, $p_\mathrm{ext} = 0$. Such a plasma is said to have \textit{ignited}. For an ignited plasma, the value of $n \tau_E$ must satisfy
\begin{equation}
n \tau_E > \frac{60}{\langle \sigma v \rangle } \frac{T}{E_\mathrm{DT} }.
\label{eq:ne}
\end{equation}
It is customary to multiply \Cref{eq:ne} by $T$ in order to obtain a triple product,
\begin{equation}
n T \tau_E > \frac{60}{\langle \sigma v \rangle } \frac{T^2}{E_\mathrm{DT} }.
\label{eq:nTe}
\end{equation}
It is often used that for temperature in the range 10-20 keV, the D-T fusion reactivity satisfies
\begin{equation}
\langle \sigma v \rangle  = 1.1 \times 10^{-24} \frac{T^2}{\left(\mathrm{keV}\right)^2} \mathrm{m}^3 \mathrm{s}^{-1},
\end{equation}
with a maximum of 10\% error. In this range, $n T \tau_E$ in \Cref{eq:nTe} is temperature independent. We plot solutions to \Cref{eq:nTe} on the `no radiation' curve in \Cref{fig:lawsoncriterion}(a) using realistic cross section data. However, there are other important loss mechanisms such as Bremsstrahlung radiation (\Cref{eq:pBreh_density}). We show the modified triple product to include Bremsstrahlung power in the `w/ radiation' curve in \Cref{fig:lawsoncriterion}(a), where we have added the Bremsstrahlung loss term in power balance (\Cref{eq:powerbalance}). In \Cref{eq:nTe} we also assumed that 100\% of the alpha particles deposit all of their power. In the 'w/ radiation 50\% alpha loss' curve in \Cref{fig:lawsoncriterion}(a), we show the effect of 50\% of the alpha particle energy being lost. This significantly increases the $n T \tau_E$ required for ignition \cite{Darrow1996,Kiptily2024}.

We can finally make a very rough quantitative estimate of how large $\tau_E$ must be in order for the plasma to ignite. We plot $\tau_E$ versus $T$ in \Cref{fig:lawsoncriterion}(b) for a plasma with $n = 3 \times 10^{19} \mathrm{m}^{-3}$, which is the density of a record JET D-T experiment 42976 with \cite{Keilhacker1999}. For the 'w/ radiation' curve at a minimum with $T \simeq 20-30$ keV, the confinement time for ignition satisfies $\tau_E \gtrsim 4$ seconds. The JET D-T discharge had a predicted ion temperature of $T=28$ keV and a predicted confinement time of $\tau_E = 0.90$ seconds.

A common measure of plasma performance is the scientific gain parameter $Q^\mathrm{sci} = P_\mathrm{fus} / P_\mathrm{ext}$ where $P_\mathrm{fus}$ is the total fusion power. At t = 13.13 seconds, JET discharge 42976 achieved $Q^\mathrm{sci} = 0.60$. Yet, according to the simple arguments that gave the curve in \Cref{fig:lawsoncriterion}(b), only a factor of four to five increase in $n \tau_E$ is required to bring the plasma to ignition (where $Q^\mathrm{sci} = \infty$). It is important to keep in mind that the Lawson criterion makes several approximations -- higher fidelity modeling is required to more accurately determine the plasma conditions required for very high fusion gain \cite{Ikeda2007,Betti2010,Ongena2016,Rodriguez-Fernandez2020}.

We need to understand what determines $\tau_E$ so that we might increase its value, which in tokamaks is set largely by turbulent transport. The first step is to write power balance in \Cref{eq:powerbalance} as an energy continuity equation,
\begin{equation}
\frac{3}{2} \frac{\partial n T}{\partial t} + \nabla \cdot \mathbf{q}  = p_\alpha + p_\mathrm{ext},
\label{eq:energy_continuity}
\end{equation}
where the quantity $\mathbf{q}$ is the heat flux. The transport term $\nabla \cdot \mathbf{q}$ measures the rate of heat being transported into or out of a volume. In steady state where $\partial T / \partial_t  =0$, we can write
\begin{equation}
\tau_E = \frac{w}{\nabla \cdot \mathbf{q}}. 
\end{equation}
Therefore, the energy confinement time increases as the rate of heat leaking out of a volume $\nabla \cdot \mathbf{q}$ decreases in size. In the limit of perfect heat insulation, $\nabla \cdot \mathbf{q} = 0$, the energy confinement time tends to infinity. The study of energy confinement in fusion plasmas centers on making $\nabla \cdot \mathbf{q}$ as small as possible.

\subsection{Particle Confinement} \label{subsec:particleconf}

Controlling particle confinement is also crucial for a workable confinement scheme \cite{Reiter1990,Sakasai1999,Maisonnier_2007,Angioni2012,Whyte2023,Parisi2025b}. For D-T MCF power plants, the conditions for particle confinement are more complicated than for energy confinement. For energy confinement, it is almost always more desirable to achieve a higher particle confinement time. For particle confinement, there is a more complicated set of tradeoffs \cite{Yoshida2025}. On the one hand, if the particle fueling rate is constant, it is desirable for deuterium and tritium to remain in the hot core for as long as possible to increase the probability of them undergoing fusion reactions. Furthermore, tritium is very expensive and challenging to handle, so it is undesirable to have large tritium flows through the plasma and ancillary power plant systems \cite{Jung1984,Abdou1986,Roth2008,El-Guebaly2009,Jackson2013,Kovari2018,Boozer2021,Abdou2021,Meschini2023,Whyte2023,Parisi2025b}. Additionally, it is desirable for alpha particles born from D-T reactions to stay confined to the plasma core while they slow down, depositing their energy into the hot core. On the other hand, once alpha particles are slow and thermalized, they become helium ash, which dilute the core fuel and decrease the total fusion power -- it is desirable for the helium ash to be transported from the fusion core as quickly as possible \cite{Uckan1988,Synakowski1990,Furth1990,Reiter1991,Synakowski1995,Heidbrink2002,Sharapov2008,Rodriguez-Fernandez2022,Kiptily2023,Meyer2024}. There are also further requirements that non-fuel impurities be transported from the core \cite{Hawryluk1979,Hirshman1981,Dux2000,Burhenn2009,Helander2017,Angioni2021}, but sometimes confined in the outer-core to edge region to radiate power \cite{Jensen1977,Kallenbach2012,Fable_2022,Eldon2024,Rutherford_2024,Wilson2024}.

In this section, we focus on D-T fuel confinement, neglecting the subtleties discussed above for the other particle species. We basic require D-T fuel to stay in the core for as long as possible to maximize the probability of fusion occurring. We can write a similar equation to power balance in \Cref{eq:powerbalance} for D-T particle balance. To simplify the arguments, the following equations will be for the deuterium species in a D-T plasma, although they will also be applicable to tritium. The particle balance equation is
\begin{equation}
S_\mathrm{L} = S_\mathrm{fus} + S_\mathrm{ext},
\label{eq:particlebalance}
\end{equation}
where $S_\mathrm{L}$ is the loss of deuterium particles by transport (units of particles per second), $S_\mathrm{fus}$ is a sink of deuterium particles into D-T fusion reactions, and $S_\mathrm{ext}$ is external fueling of deuterium provided by neutral beams \cite{Duesing1987,Sonato2017,Hopf2021} and pellet injectors \cite{Combs1993,Baylor2007,Hemsworth2017,Morris2022}. We define the particle confinement time as
\begin{equation}
\tau_p \equiv \frac{n_\mathrm{D}}{S_\mathrm{L} } ,
\end{equation}
where $n_\mathrm{D}$ is the deuterium particle density. As we did before for the Lawson criterion, we assume that the plasma density and temperature profiles are constant. The sink of deuterium particles is equal to minus the D-T fusion reaction rate
\begin{equation}
S_\mathrm{fus} = - n^2_\mathrm{D} \langle \sigma v \rangle,
\end{equation}
where we have assumed equal densities of deuterium and tritium. Rearranging particle balance in \Cref{eq:particlebalance} gives a required deuterium particle confinement time
\begin{equation}
\tau_p = \frac{n_\mathrm{D}}{S_\mathrm{ext} -  n^2_\mathrm{D} \langle \sigma v \rangle} = \frac{n_\mathrm{D}}{S_\mathrm{ext} -  P_\mathrm{fus}/ E_\mathrm{DT} }. 
\label{eq:taup}
\end{equation}
There is a fundamental difference between particle confinement and energy confinement: for energy confinement, a plasma can `ignite' since fusion reactions release energy that can heat the plasma. However, a D-T plasma cannot `ignite' for the particle supply, because D-T fusion reactions are a sink of D-T fuel. Therefore, for a fusion plasma with perfect particle confinement, $\tau_p \to \infty$, the particle fueling source $S_\mathrm{ext}$ must balance the D-T particle consumption rate $P_\mathrm{fus}/ E_\mathrm{DT}$. In reality, there are always mechanisms that cause particle transport, which means we require the denominator of \Cref{eq:taup} to be positive definite over sufficiently long time periods,
\begin{equation}
S_\mathrm{ext} >  P_\mathrm{fus}/ E_\mathrm{DT}.
\end{equation}
Therefore, \Cref{eq:taup} shows that in a burning plasma, particle and energy transport are coupled by fusion reactions. There are many more subtle ways that particle and energy transport are coupled, some of which we will review in this article.

Finally, in addition to particle and energy transport, we are also often interested in momentum transport. Tokamak plasmas can rotate, often very quickly \cite{Brau1983,Rozhansky1996,Solomon2010,Ida2014rotation}. Both the rotation and the spatial gradient of the rotation (`rotation shear') can have a significant impact on the stability properties of toroidal plasmas \cite{Waelbroeck1991,Hahm1995,Taylor1997physics,Wahlberg2000,Guazzotto2004,Chapman2006,Barnes2011b,Yan2014}. Angular momentum can even be redistributed internally in tokamaks \cite{Rice2007inter,Diamond2009,Parra2015}. In this article, we will focus on particle and energy transport, generally omitting momentum transport.

\section{Magnetic Confinement Schemes} \label{sec:confimentschemes}

In \Cref{sec:introduction} we claimed that Coulomb scattering being much faster than any fusion reaction rate motivates a confinement scheme. Mainstream fusion research efforts are centered on two distinct confinement schemes: inertial confinement fusion and magnetic confinement fusion. In inertial confinement fusion, the fusion fuel is rapidly compressed and heated, relying on its own inertia to hold it together briefly until fusion reactions occur. The goal is for the fuel to have undergone many fusion reactions at extremely high densities before the fuel flies apart. In contrast, magnetic confinement fusion occurs at roughly ten orders of magnitude lower density, using strong magnetic fields to contain a hot plasma for extended periods.

\begin{figure}[t]
\centering
    \includegraphics[width=\textwidth]{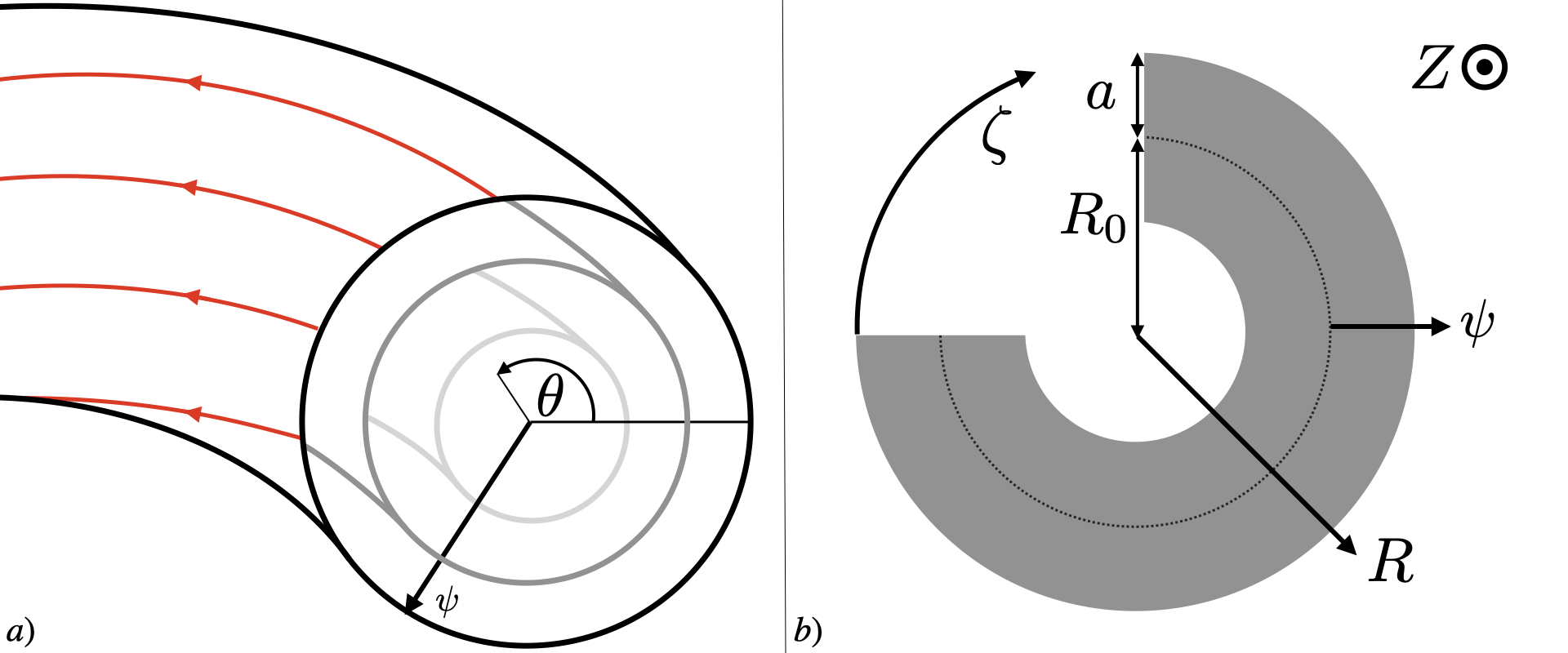}
 \caption{(a): Three nested magnetic flux surfaces. Red lines on the outermost surface represent a field line sampling the surface; $\theta$ is a poloidal angle and $\psi$ is the poloidal flux divided by $2\pi$. (b): Bird's eye view of a tokamak; $R$ is the major radial coordinate, $Z$ is the axial coordinate, $\zeta$ is the toroidal angle, $R_0$ is the $R$ location of the LCFS, and $a$ is the minor radius.}
\label{fig:coordinates}
\end{figure}

This tutorial focuses on magnetic confinement fusion. We will generally use the coordinate system in tokamaks, although many concepts will also carry over to other magnetic confinement schemes such as stellarators. A tokamak is a donut-shaped (toroidal) machine designed to confine plasma. Inside a tokamak, magnetic fields force the charged particles to move along circular paths, tracing out layers called magnetic flux surfaces. Think of these flux surfaces as nested donut-shaped layers inside the tokamak -- see \Cref{fig:coordinates}(a) for an illustration. Charged particles move rapidly along these surfaces but move much more slowly from one surface to another.

The relatively slow particle motion across flux surfaces occurs because charged particles spiral around magnetic field lines, a motion called Larmor gyration. An ion with thermal speed
\begin{equation}
v_{ti} = \sqrt{ \frac{2 T}{m}  } 
\end{equation}
has a gyroradius
\begin{equation}
\rho_i = \frac{v_{ti}}{\Omega_i} 
\end{equation}
with gyrofrequency
\begin{equation}
\Omega_i = \frac{Z_i e B}{m_i},
\label{eq:Omega_c}
\end{equation}
for a charge number $Z_i$, proton charge $e$, magnetic field strength $B$, and mass $m_i$. Although particles can move freely and quickly along magnetic field lines, their perpendicular motion across lines is much more limited. This is the basic premise of magnetic confinement.

Because particles mostly remain confined to their flux surfaces, certain properties of the plasma, such as temperature, density, and electric potential, remain nearly uniform on each surface. These properties are called `flux functions' because they are constant a single flux surface. Flux functions $g$ therefore satisfy
\begin{equation}
\mathbf{B} \cdot \nabla g = 0, 
\end{equation}
where $\mathbf{B}$ is the magnetic field vector. The equilibrium magnetic field for tokamaks can be written as
\begin{equation}
\mathbf{B} = I\nabla \zeta + \nabla \zeta \times \nabla \psi.
\label{eq:B_eq_tok}
\end{equation}
Here, $I(\psi(R, Z)) = RB_T$ is a flux function where $B_T$ is the toroidal component of the magnetic field, R is the major radial coordinate, $Z$ is the axial coordinate, $\psi$ is the poloidal flux divided by $2\pi$, and $\zeta$ is the toroidal angle. The toroidal field in \Cref{eq:B_eq_tok} is $\mathbf{B}_T = I \nabla \zeta$ and the poloidal field is $\mathbf{B}_p = \nabla \zeta \times \nabla \psi$. The variables $(R,Z,\zeta)$ form a cylindrical coordinate system, shown in \Cref{fig:coordinates}(a).

The outer boundary of the nested magnetic surfaces is called the Last Closed Flux Surface (LCFS). In the region beyond the LCFS called the scrape-off layer (SOL), particles stream along open field lines until they intersect with a divertor. Inside the LCFS, however, particles remain well-contained, significantly reducing heat loss and making efficient confinement possible. Other useful quantities are $a$, the minor radius, and $R_0$, the $R$ location of the LCFS, illustrated in \Cref{fig:coordinates}(b).

\begin{figure}[t]
\centering
    \includegraphics[width=\textwidth]{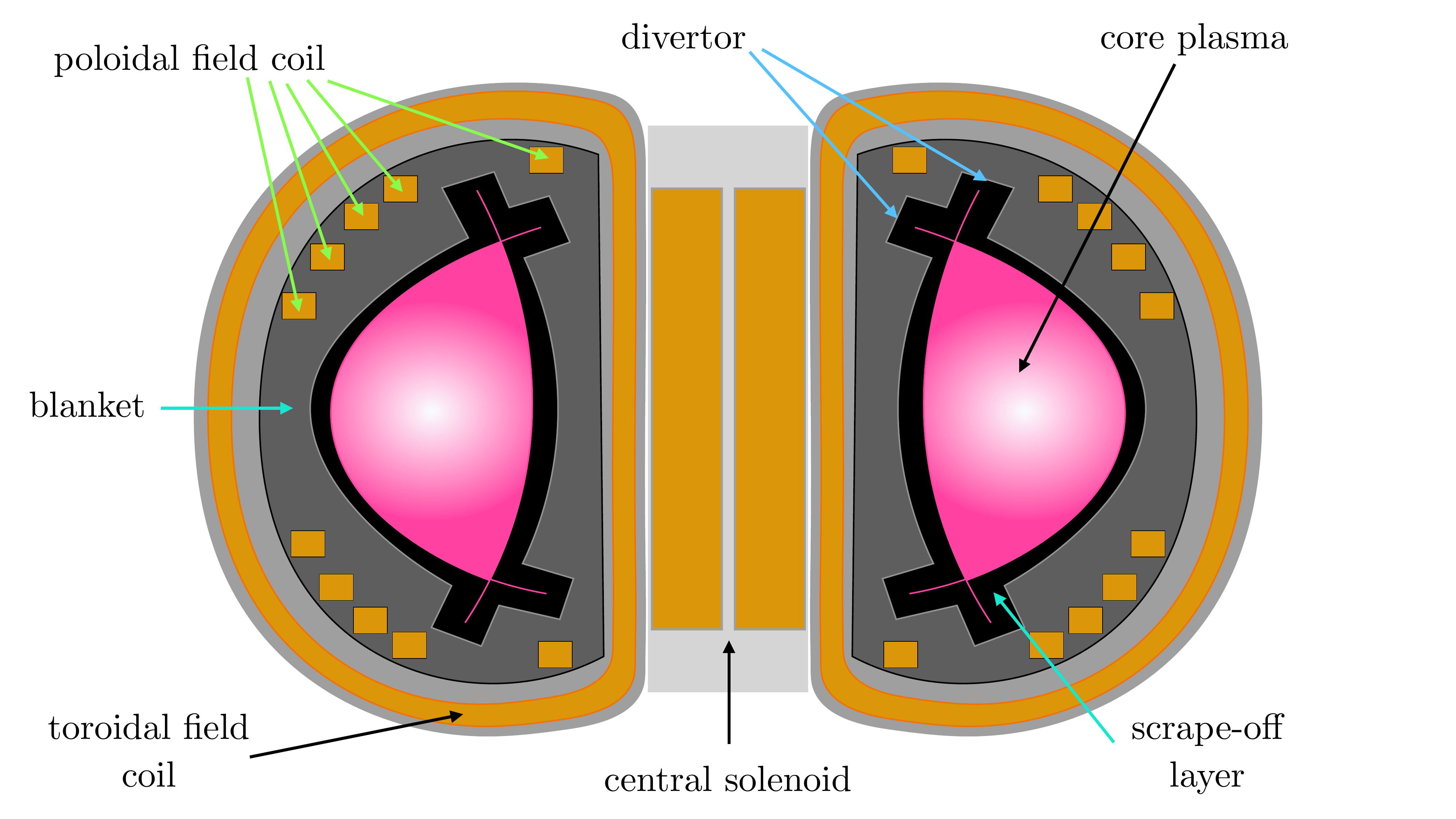}
 \caption{Some key features of a tokamak.}
\label{fig:tokamak_general}
\end{figure}

Before moving ahead, it is worth placing the problem we are studying in context. As described in the Lawson criterion in \Cref{sec:Lawson}, the fusion power, confinement, plasma heating, and other losses are strongly coupled. However, there are more strong couplings: in \Cref{fig:tokamak_general} we show a schematic of a tokamak, showing the core plasma and scrape-off layer. This review will primarily be focused on turbulence within the LCFS, although SOL turbulence is also an important research area. Surrounding the vacuum vessel containing the plasma is a blanket where tritium fuel is bred by lithium neutron capture reactions. These neutrons are produced by D-T reactions in the plasma core, and so the neutron product rate and their spatial trajectories in the plasma core impacts tritium breeding.  There are poloidal field coils used mainly for plasma shaping, the central solenoid for current drive and heating, and toroidal field coils for generating the toroidal field. There are many other important systems we have omitted. We emphasize that stability and transport in the plasma core should not be studied in isolation -- many of the power plant systems are coupled in interesting and challenging ways. These couplings will become stronger as we move toward routine operation of burning plasmas.

\section{Transport Mechanisms} \label{sec:transport_overview}

In this section, we describe the physical mechanisms causing transport in MCF plasmas: classical, neoclassical, and turbulent. Since the main subject of this tutorial is turbulent transport, we will spend only a short amount of time covering classical and neoclassical transport.

The task of transport calculations is to self-consistently find the evolution of the magnetic equilibrium and plasma profiles. This is an involved task, accomplished by solving a system of equations across a range of time and spatial scales: (1) the gyrokinetic equation, which describes the evolution of fast plasma perturbations on small gyroradius scales, (2) the neoclassical equation, which describes the evolution of slow plasma perturbations on large equilibrium scales, (3) the transport equation, which describes the slow evolution of the magnetic equilibrium and plasma profiles and (4) Maxwell's equations. We will cover these equations in detail in \Cref{sec:KT_GK_Transp}.

In this section, we focus on order of magnitude estimates of the transport coefficients in the transport equation. This involves calculating the particle and heat fluxes $\mathbf{\Gamma}_s$ and $\mathbf{q}_s$ that enter the particle and heat transport equations,
\begin{equation}
\begin{aligned}
& \frac{\partial n_s}{\partial t} + \nabla \cdot \mathbf{\Gamma}_s = S_s, \;\;\; \frac{3}{2} \frac{\partial n_s T_s}{\partial t} + \nabla \cdot \mathbf{q}_s = P_s,
\end{aligned}
\end{equation}
where $n_s$ is the number of particles per unit volume for a species $s$ (an ion or electron species), $T_s$ is the temperature, and $S_s$ and $P_s$ are the particle and energy sources. Typically, the fluxes can be written in `diffusive' form,
\begin{equation}
\mathbf{\Gamma}_s = - D_s \nabla n_s, \;\;\;\; \mathbf{q}_s = - \chi_s n_s \nabla T_s,
\label{eq:Gamma_q_diff}
\end{equation}
$\mathbf{\Gamma}_s$ and $\mathbf{q}_s$ in \Cref{eq:Gamma_q_diff} describe the radial transport of particles and heat that tends to flatten density and temperature gradients $\nabla n_s$ and $\nabla T_s$.

In the following subsections, we will estimate the relative size of $\chi_s$ for classical, neoclassical, and turbulent heat transport. For each of these three transport mechanisms, our mental model for transport will be as follows: transport occurs as a spatially two-dimensional random walk with step size $l$, and each step occurs with a frequency $f$. The random walk is two dimensional because the plasma is magnetized, and within the LCFS motion along the field lines does not lead to radial transport. We estimate the diffusion coefficient as a random walk,
\begin{equation}
\chi_s \sim l^2 f.
\label{eq:randomwalk}
\end{equation}
We will see that for tokamaks, only turbulent transport gives sufficiently fast radial energy diffusion to explain the observed energy confinement time in high-performance experiments. Recall that 

\begin{figure}[bt]
    \centering
    \begin{subfigure}[t]{0.49\textwidth}
    \includegraphics[width=0.99\textwidth]{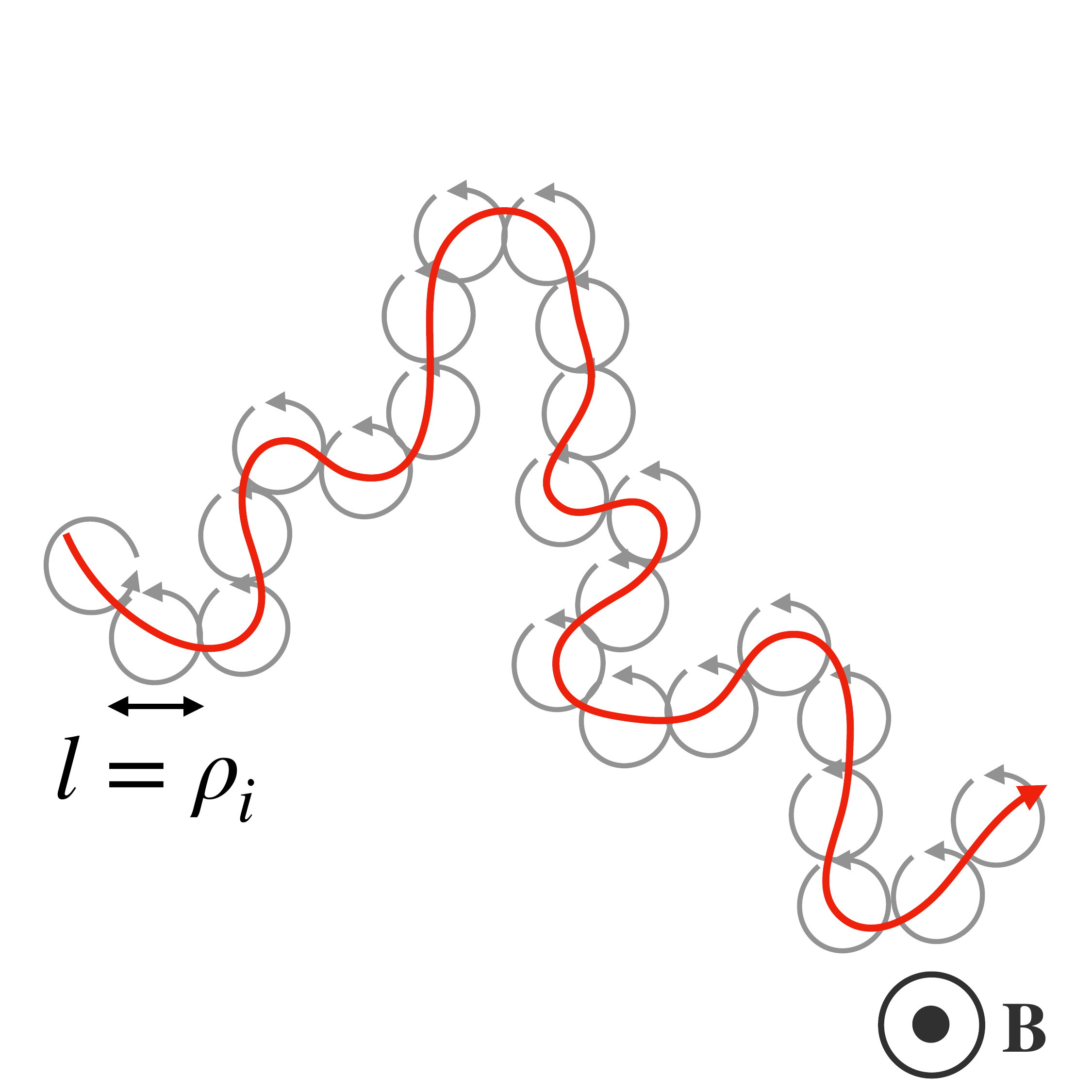}
    \caption{Classical.}
    \end{subfigure}
    \centering
    \begin{subfigure}[t]{0.49\textwidth}
    \includegraphics[width=0.99\textwidth]{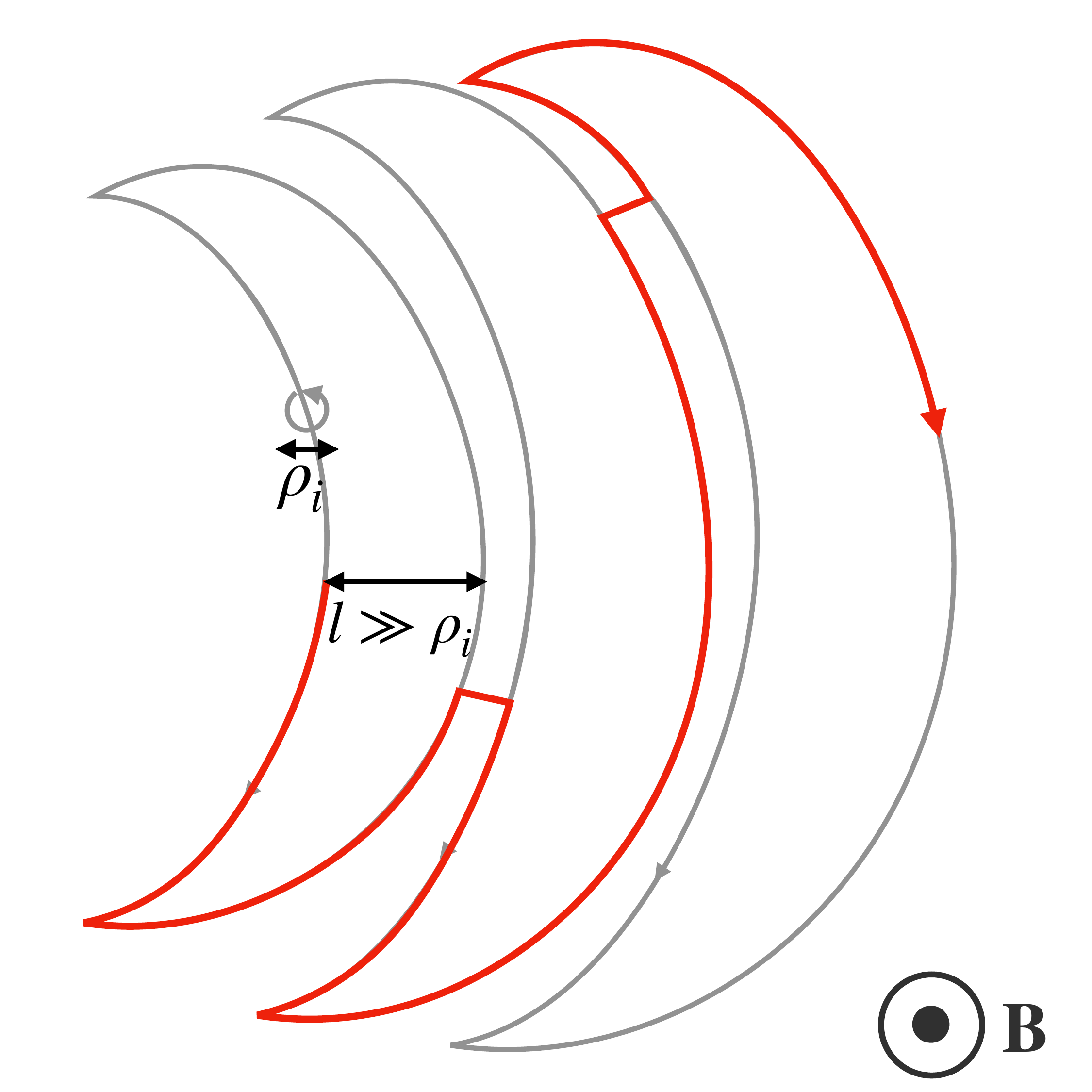}
    \caption{Neoclassical.}
    \end{subfigure}
    \caption{Heuristic random walk models for (a) classical and (b) neoclassical transport.}
    \label{fig:classical_neoc}
\end{figure}

\subsection{Classical Transport}

The first regime of transport that causes particles and energy to diffuse from the plasma core to the edge is called \textit{classical} transport \cite{Braginskii1958,Balescu1960,Lenard1960,Helander2002}. In the classical transport regime, particles undergo a random walk with a step size of the Larmor radius $\rho_i$ and a frequency of the collision frequency -- a cartoon of this process is shown in \Cref{fig:classical_neoc}(a). Collisional transport gives relatively low transport levels and so is tolerable within MCF schemes such as tokamaks. If this were the only transport mechanism, energy confinement would be excellent, and as of 2025, we would almost certainly have fusion power plants delivering electricity to the grid.

The transport coefficient estimate in \Cref{eq:randomwalk} using a step size equal to the ion gyroradius $\rho_i$ and a step frequency equal to the ion-ion collision frequency $\nu_{ii}$ gives
\begin{equation}
\chi_i^{\mathrm{classical}} \sim \rho_i^2 \nu_{ii}.
\end{equation}
Revisiting the JET discharge 42976 \cite{Keilhacker1999} we introduced earlier, we now estimate the energy confinement time in a plasma where energy transport is dominated by collisional transport. Heuristically, we can estimate the collisional confinement time as
\begin{equation}
\tau_{E}^{\mathrm{classical}} \approx \frac{a^2}{\chi_i^{\mathrm{classical}}} = \left( \frac{a}{\rho_i} \right)^2 \frac{1}{\nu_{ii}}. 
\end{equation}
For this JET discharge, $\rho_i \approx 0.0015$ meters, $\nu_{ii} \approx 600 $ Hertz, the minor radius is $a = 0.95$ meters, and toroidal field $B_\mathrm{T}$ = 3.6 Tesla. This gives
\begin{equation}
\tau_{E}^{\mathrm{classical}} \approx 670 \; \mathrm{seconds}. 
\label{eq:tauE_classical_estimate}
\end{equation}
Given that the experimental value is $\tau_E = 0.90$ seconds, the classical rough estimate gives a confinement time over 700 times larger than the experimental value. Additional transport mechanisms are required to explain the experimental $\tau_E$ value.

\subsection{Neoclassical Transport}

Unfortunately, there are additional transport mechanisms with much larger diffusivity than classical transport. One is \textit{neoclassical} transport \cite{Hirshman1981,Balescu1998,Helander2002}. The toroidicity of tokamaks and stellarators causes the diffusive step size to increase, which increases the diffusivity significantly. In tokamaks, a large fraction of particles will bounce between maxima in the magnetic field, tracing out particle trajectories called banana orbits. The width of the banana orbit is approximately the new step size and the step frequency is the ion-ion collision frequency. The physical picture is approximately as follows, shown in \Cref{fig:classical_neoc}(b): trapped particles execute banana orbits. In a collision, the particle will be scattered into a new banana orbit.

Estimating the diffusion coefficient for neoclassical transport is more involved than for classical transport \cite{Helander2002}. We concern ourselves with trapped particles undergoing banana orbits. The fraction of particles that are trapped is $f_t \approx \sqrt{2 \epsilon}$, where $\epsilon = r / R$ is the inverse aspect ratio. The effective collision frequency is $\nu_{ii} / \epsilon$ and the new step size is $l \approx \rho_{p,i} \sqrt{\epsilon}$. Here, $\rho_{p,i}$ is the poloidal gyroradius, which involves using the poloidal magnetic field rather than the total magnetic field in the gyrofrequency in \Cref{eq:Omega_c}. Putting these together, we find
\begin{equation}
\chi_i^{\mathrm{neoclassical}} \sim \sqrt{2 \epsilon} \rho_{p,i}^2 \nu_{ii}.
\end{equation}
Therefore, the ratio of neoclassical to classical diffusivity is
\begin{equation}
\frac{\chi_i^{\mathrm{neoclassical}}}{\chi_i^{\mathrm{classical}}} \sim \sqrt{2 \epsilon} \left( \frac{B}{B_p} \right)^2 \sim \sqrt{2} q^2 \epsilon^{-3/2},
\end{equation}
where $q \approx \epsilon B/B_p$ is the safety factor. In tokamaks, $q$ is typically much larger than one.

We now estimate $\tau_{E}^{\mathrm{neoclassical}}$ using
\begin{equation}
\tau_{E}^{\mathrm{neoclassical}} \approx \frac{a^2}{\chi_i^{\mathrm{neoclassical}}},
\end{equation}
for the JET discharge 42976. Assuming that $q^2 \approx 4$ for this discharge and taking a typical aspect ratio $\epsilon \approx 1/5$, we find
\begin{equation}
\tau_{E}^{\mathrm{neoclassical}} \approx 11 \; \mathrm{seconds}. 
\label{eq:tauE_neoclassical_estimate}
\end{equation}
Therefore, comparing with the experimental $\tau_E = 0.90$ seconds, the neoclassical estimate of confinement time is still an order of magnitude too long compared with experiment. Another transport mechanism is required to explain the experimentally observed confinement time.

\subsection{Turbulent Transport}

The remaining transport mechanism to explain the shorter-than-expected energy confinement time is \textit{turbulent} transport \cite{Liewer1985,Wootton1990,Dimits2000,Candy2003,Garbet2004,Boedo2009}. The nature of turbulent transport is qualitatively different from classical and neoclassical, which were predictable from the equilibrium profiles. Turbulent transport, in contrast, arises from plasma perturbations that deviate from the equilibrium. Because of the requirement for high core pressure and low edge pressure, there are \textit{pressure gradients} across the radial profiles. Pressure gradients provide a source of free energy, which can drive \textit{instabilities}. These instabilities consist of electromagnetic, density, and temperature perturbations. These perturbations grow until they \textit{saturate}, leaving the plasma in a turbulent state. Turbulence in MCF devices gives rise to \textit{gyroBohm} transport. Such transport has a step size at least as large -- and often much higher -- as the ion or electron gyroradius, and a frequency much faster than collisions.

\begin{figure}[bt]
    \centering
    \begin{subfigure}[t]{0.49\textwidth}
    \includegraphics[width=0.99\textwidth]{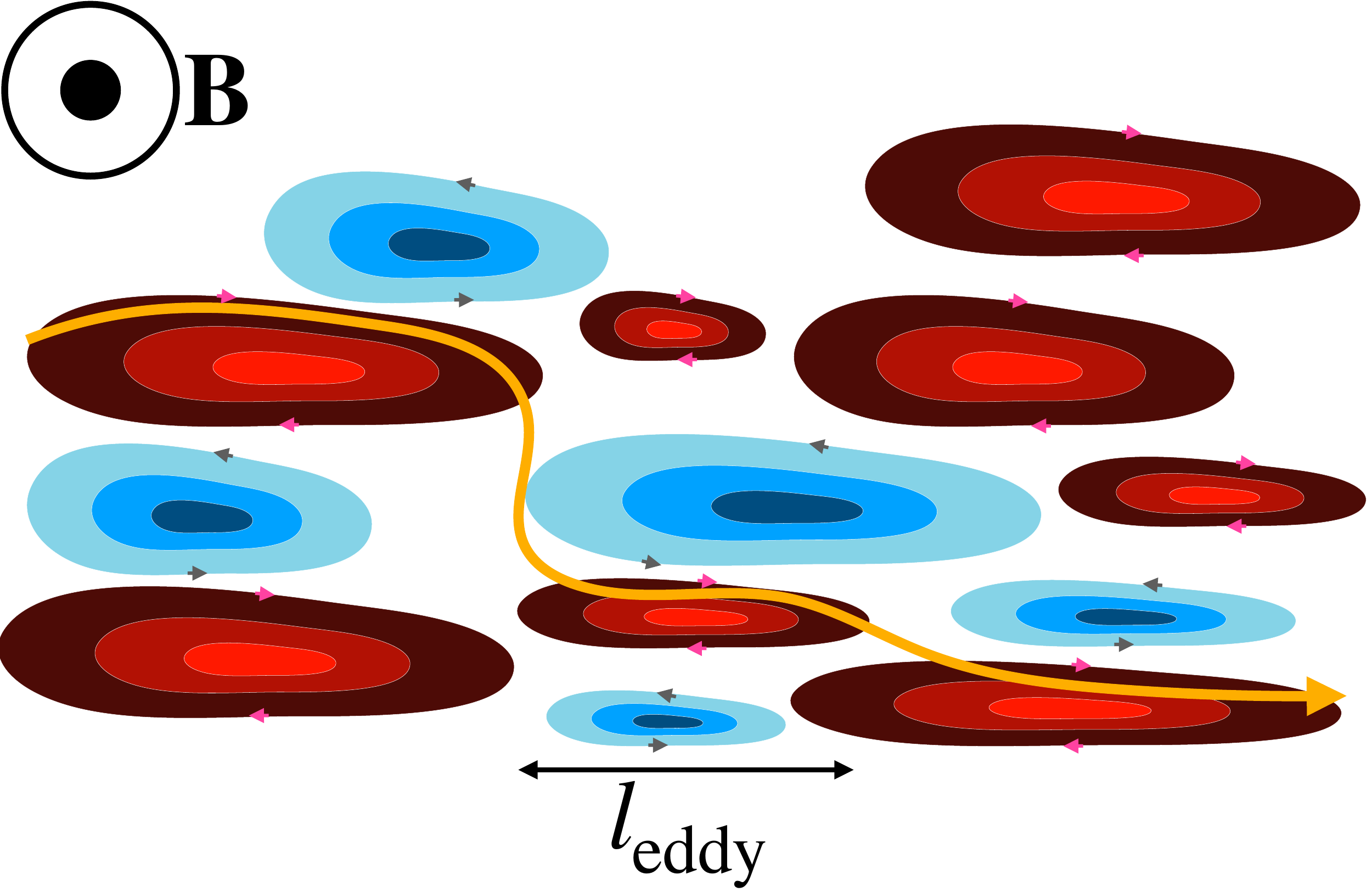}
    \caption{Turbulent random walk.}
    \end{subfigure}
    \centering
    \begin{subfigure}[t]{0.49\textwidth}
    \includegraphics[width=0.99\textwidth]{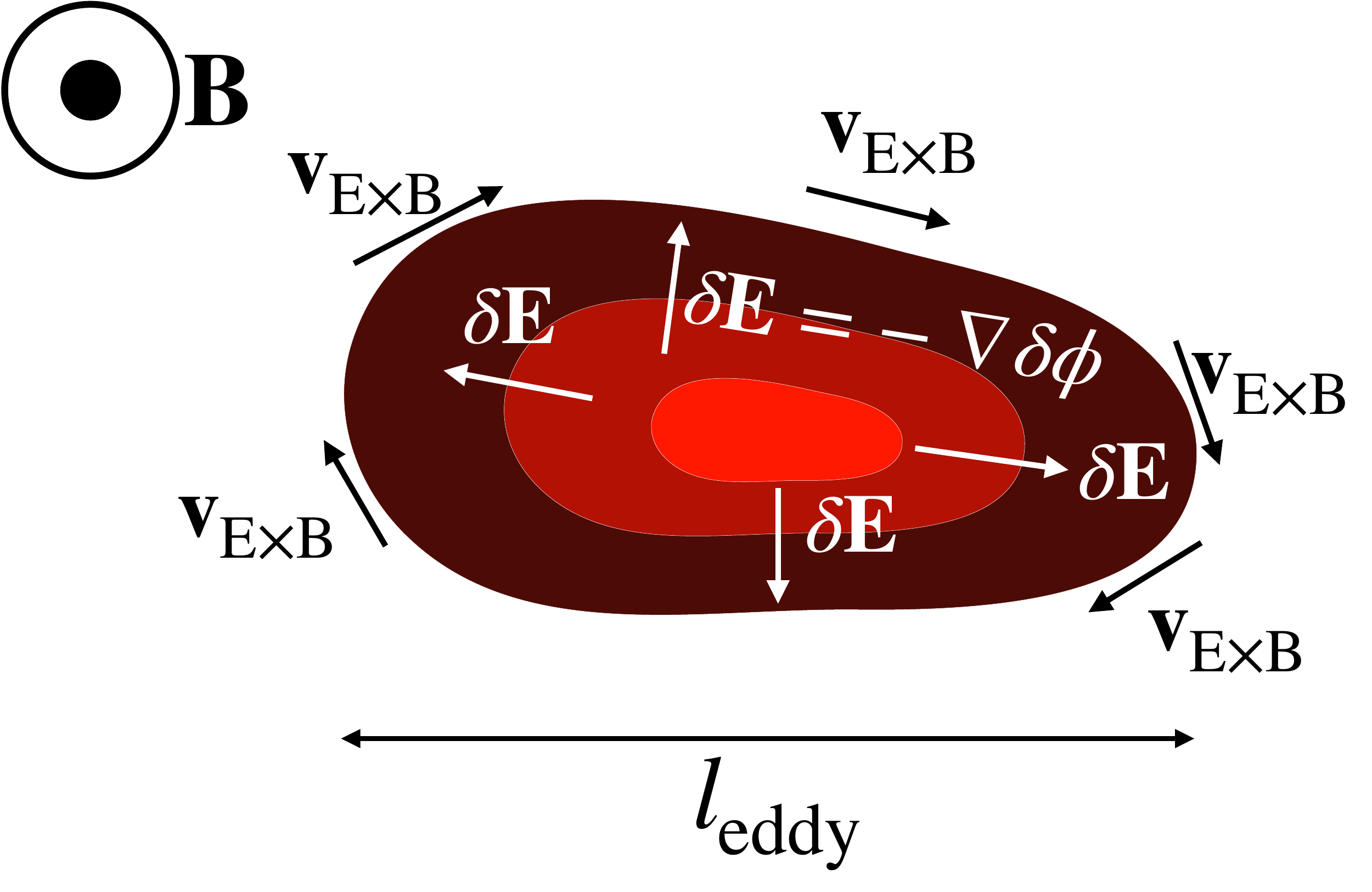}
    \caption{Turbulent eddy model.}
    \end{subfigure}
    \caption{Heuristic models for a turbulent random walk (a) and a turbulent eddy (b). Red eddies have a density slightly higher than the background mean density and blue eddies have a density slightly lower than the mean density.}
    \label{fig:transport}
\end{figure}

One mental model for turbulent transport consists of radially connected turbulent eddies shown in \Cref{fig:transport}(a). These eddies consist of overperturbations and underperturbations to the mean density, temperature, etc. For reasons that we explain shortly, these eddies to rotate in the direction of the pink and grey arrows indicated in \Cref{fig:transport}(a). Because the eddies are formed of particles themselves, they are dynamic, short-lived structures with a lifetime $\tau_{nl}$ which we call the eddy turnover time. The orange arrow in \Cref{fig:transport}(a) shows the trajectory of a particle as it moves along different eddies. Red eddies have a density slightly higher than the background mean density and blue eddies have a density slightly lower than the mean density.

We now present an argument to estimate the turbulent transport coefficient for ion heat transport. The step size is the radial scale of a turbulent eddy $l_{\mathrm{eddy}}$ and the frequency is the eddy turnover time $\tau_{nl}$,
\begin{equation}
\chi_i^{\mathrm{turb}} \sim l_{\mathrm{eddy}}^2 / \tau_{nl}.
\end{equation}
We define the nonlinear time as the time it takes particles to move across the radial eddy extent,
\begin{equation}
\tau_{nl} \equiv l_{\mathrm{eddy}}/v_{E\times B}, 
\end{equation}
where $v_{E\times B}$ is the $E \times B$ velocity,
\begin{equation}
v_{E \times B} \equiv \left| \frac{\mathbf{B} \times \nabla \delta \phi}{B^2} \right| \sim \frac{\delta \phi}{B l_{\mathrm{eddy} }}.
\label{eq:vExB_one}
\end{equation}
Physically, $v_{E\times B}$ is a drift arising from a magnetic and electric field that have a non-vanishing perpendicular component. Here, $\delta \phi$ is the fluctuating electrostatic potential. To estimate the final term in \Cref{eq:vExB_one}, we need to estimate $\delta \phi$. First, use that the normalized potential fluctuations are proportional to the eddy size divided by the device size $a$,
\begin{equation}
\frac{e \delta \phi}{T} \sim \frac{l_\mathrm{eddy}}{a}.
\end{equation}
We invoke the ambient pressure gradient argument: the perturbed potential energy $(a/l_{\mathrm{eddy}}) e \delta \phi$ is comparable to the plasma thermal energy,
\begin{equation}
\frac{a}{l_{\mathrm{eddy}}} e \delta \phi \sim T = \frac{1}{2} m v_t^2.
\label{eq:ambient_pressure}
\end{equation}
Inserting \Cref{eq:ambient_pressure} into \Cref{eq:vExB_one} gives
\begin{equation}
v_{E \times B} \sim \frac{T}{aeB }.
\label{eq:vExB_two}
\end{equation}
This gives a nonlinear time $\tau_{nl} \sim (l_{\mathrm{eddy} }/\rho) (a/v_{t})$, and therefore the turbulent diffusivity,
\begin{equation}
\chi_i^{\mathrm{turb}} \sim \frac{l_{\mathrm{eddy}}}{\rho_i} \rho_{*i} \rho_i v_{t} = \frac{l_{\mathrm{eddy}}}{\rho_i} \chi_{gB},  
\label{eq:chi_i_turb}
\end{equation}
where the gyroBohm diffusivity is
\begin{equation}
\chi_{gB} \equiv \rho_{*i} \rho_i v_{t},
\end{equation}
and
\begin{equation}
\rho_{*i} \equiv \rho_i / a.
\end{equation}
The diffusivity expression in \Cref{eq:chi_i_turb} shows that when turbulent eddies have a length equal to the ion gyroradius, transport is gyroBohm. Physically, gyroBohm diffusivity corresponds to eddies the size of the gyroradius, where the eddies themselves have an $\mathbf{ E} \times \mathbf{ B} $ drift that carries particles across the eddy structure. GyroBohm transport can be much higher than collisional transport because particles typically move across eddies much faster than they collide. A cartoon for how energy is transported around a turbulent eddy is shown in \Cref{fig:transport}(b). 

Crucially, the gyroBohm transport diffusivity in \Cref{eq:chi_i_turb} scales with $\rho_{*i}$. This observation has informed much of the thinking about how to improve confinement in future devices by decreasing $\rho_{*i}$:  (1) since $\rho_{i} \sim 1 / B$, use strong magnetic fields to increase $B$ \cite{Sorbom2015,Whyte2016,Zohm2019}, and (2) increase $a$ by making the machine larger \cite{Shimada2007}.

The ratio of turbulent to neoclassical diffusivity is
\begin{equation}
\frac{\chi_i^{\mathrm{turb}}}{\chi_i^{\mathrm{neoclassical}}} \sim \frac{1}{\sqrt{2\epsilon} } \frac{l_{\mathrm{eddy}}}{\rho_i}  \left( \frac{B}{B_p} \right)^2 \frac{v_{ti}}{a} \frac{1}{\nu_{ii}}.
\end{equation}
For ion heat transport, typically
\begin{equation}
\frac{\chi_i^{\mathrm{turb}}}{\chi_i^{\mathrm{neoclassical}}} \gtrsim 1.
\end{equation}
where $v_{ti}/a \gtrsim \nu_{ii}$. However, for electron heat transport, the turbulent eddy size is typically on the order of the ion, not the electron gyroradius \cite{Dorland2000}, $l_\mathrm{eddy} \sim \rho_i \sim (m_i / m_e) \rho_e$, giving a surprisingly high electron heat transport
\begin{equation}
\frac{\chi_e^{\mathrm{turb}}}{\chi_e^{\mathrm{neoclassical}}} \gg 1,
\end{equation}
despite initial estimates that turbulent electron heat transport is small due to the small electron gyroradius. The turbulent energy confinement time is thus approximately
\begin{equation}
\tau_{E}^{\mathrm{turb}} \approx \frac{a^2}{\chi_i^{\mathrm{turb}}}. 
\end{equation}
Comparing again with JET discharge 42976 \cite{Keilhacker1999} with the experimental $\tau_E = 0.90$ seconds, using a thermal velocity $v_\mathrm{ti} \approx 10^6$ meters / second, the turbulent diffusivity estimate is
\begin{equation}
\chi_i^{\mathrm{turb}} \approx 2 \; \mathrm{meters / second}, 
\end{equation}
and the corresponding confinement time is
\begin{equation}
\tau_{E}^{\mathrm{turb}} \approx 0.4 \; \mathrm{seconds}.
\label{eq:tau_E_turb_est}
\end{equation}
We summarize the comparison of experimental, classical, neoclassical, and turbulent confinement times in \Cref{table:confinement_times}. The classical, neoclassical, and turbulent times in \Cref{table:confinement_times} are only intended as order-of-magnitude estimates to demonstrate that energy confinement in JET is very likely determined by turbulent transport mechanisms. We emphasize that this is a very rough estimate, but it captures enough detail to be useful.

\begin{table}[bt!]
\centering
\begin{tabular}{|c|c|c|c|}
\hline
$\tau_E^\mathrm{experiment} /s $ & $\tau_{E}^{\mathrm{classical}}$ /s & $\tau_{E}^{\mathrm{neoclassical}}$ /s & $\tau_{E}^{\mathrm{turb}}$ /s \\
\hline
0.90 & 670 & 11 & 0.4 \\
\hline
\end{tabular}
\caption{Comparison of experimental confinement time for JET discharge 42976 at t = 13.13 seconds versus heuristic classical (\Cref{eq:tauE_classical_estimate}), neoclassical (\Cref{eq:tauE_neoclassical_estimate}), and turbulent (\Cref{eq:tau_E_turb_est}) confinement times. The experimental confinement time $\tau_E^\mathrm{experiment}$ is reported in \cite{Keilhacker1999}.}
\label{table:confinement_times}
\end{table}

\subsection{Power Plant Size Considerations}

Taking as given that ion thermal transport is driven primarily by turbulence, we can estimate the required size of a burning plasma in a fusion power plant. The total plasma thermal energy $W$ is roughly
\begin{equation}
W \sim n T L^3,
\end{equation}
where $L$ is the device spatial length scale and we used the previous form of the radial heat flux $q_i \sim \chi_i^{\mathrm{turb} } n_i T_i / L$. The total power loss due to turbulent transport is $P_{\mathrm{loss} } \sim q_i L^2 \sim \chi_i n_i T_i L$. The global energy confinement time $\tau_E$ is
\begin{equation}
\tau_E \equiv \frac{W}{P_{\mathrm{loss} }} \sim \frac{L^2}{\chi_i^{\mathrm{turb}}}. 
\end{equation}
One way to measure the performance of a FPP is with the Lawson criterion, requiring $n_i T_i \tau_E$ to exceed a certain bound for the plasma to ignite \cite{Lawson1957,Wurzel2022}, which we estimated in \Cref{eq:nTe},
\begin{equation}
n_i T_i \tau_E > C_{\mathrm{Lawson} }.
\label{eq:tripleproductsimple}
\end{equation}
Ignoring any scalings of $n_i$ or $T_i$ with $L$ or $B$, we see that $\chi_i^{\mathrm{turb} } \sim 1 / B^2$. Also writing a confinement enhancement/degradation $H$ in $\tau_E$ relative to a fixed nominal $\tau_{E,\mathrm{nominal} }$ as \cite{Goldston1984,Kaye1985,Yushmanov1990}
\begin{equation}
\tau_E = H \tau_{E,\mathrm{nominal} }, 
\label{eq:Hconfinementquality}
\end{equation}
we obtain
\begin{equation}
\tau_E \sim H L^2 B^2.
\end{equation}
Therefore \Cref{eq:tripleproductsimple} becomes
\begin{equation}
n_i T_i \tau_E \sim H L^2 B^2 > C_{\mathrm{Lawson} },
\label{eq:Lawsonapproximate}
\end{equation}
giving a size scaling of
\begin{equation}
L > \left( \frac{C_{\mathrm{Lawson} }}{H B^2} \right)^{1/2}.
\label{eq:size_Lawson_criterion}
\end{equation}
Therefore, for a fixed $C_{\mathrm{Lawson} }$ increasing $H$ and $B$ can both reduce device size. Note that these are very heuristic arguments, and integrated systems codes are required to predict how changes in parameters such as $H$, $L$, and $B$ feed through to plasma performance and ultimately the predicted cost as a fusion power plant. For example, in \cite{Wade2021} it was found that increasing the parameter $H$ from 0.9 to 1.7 halved the estimated cost of a fusion power plant from \$8 billion USD to \$4 billion USD, which was a much bigger effect that any other parameter variation in the system. These are tantalizing hints that ideas for improving confinement could have a significant impact on the cost of FPP. These are physics ideas, falling under the purview of fusion scientists. While the scaling exponents of $H$, $L$, and $B$ in \Cref{eq:Lawsonapproximate} are highly approximate, they capture the basic ideas. See \cite{Sheffield2001,Sorbom2015,Costley2015,Menard2016,Zohm2019,Wade2021,Schwartz2023} for a variety of more sophisticated approaches to these questions.

\subsection{Further Considerations}

Despite the optimistic message of \Cref{eq:size_Lawson_criterion} and studies such as \cite{Wade2021} showing that improving energy confinement through higher $H$ would lead to smaller fusion power plants, it is important to emphasize that increasing confinement indefinitely is not useful, and might actually be harmful. This is because there are other constraints that come into play.

One constraint is that power must be exhausted in a way that does not melt the vacuum vessel. In a burning power there will be much more power relative to the surface area available for dissipation. In D-T fusion, 1/5 of the total fusion power is carried by alpha particles that deposit power into the plasma. The total enclosed fusion power $P_{\mathrm{fus} }$ scales as $P_{\mathrm{fus} } \sim L^3$ but the surface area through which heat leaves the plasma only scales as $L^2$. However, the actual area where the plasma touches the divertor likely scales even more pessimistically with $L$ \cite{Umansky2009,Eich2013,Brunner2018}. Other techniques are being developed to radiate heat within the plasma and to protect the divertor.

In addition to heat transport, there are particle transport constraints. Further to those already mentioned in \Cref{subsec:particleconf} (helium ash and impurity removal), a pertinent constraint all tokamaks are subject to is a plasma density limit \cite{Greenwald1988} where the volume-averaged density normalized to the total plasma current cannot exceed a certain value. 

\section{Transport Framework} \label{sec:KT_GK_Transp}

In this section we introduce the theoretical background required for finding the time evolution of the magnetic equilibrium and plasma profiles. Despite the evolution of the equilibrium and profiles being slow, on transport timescales $\tau_E$, it is necessary to simulate much faster dynamics included in the gyrokinetic and neoclassical equations, which we introduce in this section.

We first introduce kinetic theory through the Fokker-Planck equation. We then describe the evolution of smaller turbulent and equilibrium dynamics in the gyrokinetic and neoclassical equations. Slower transport timescale dynamics are described in the transport and magnetic equilibrium equations. We will use a simplified treatment, neglecting certain physical effects such as rotation that are often important. 



We will use the `$\delta f$' (pronounced `delta f') formalism \cite{Beer1995}, which assumes small fluctuations compared with the equilibrium,
\begin{equation}
\frac{\delta f_s}{f_s} \sim \frac{\delta B}{B} \sim \frac{\delta E}{E} \sim \rho_{*s} = \frac{\rho_s}{a} \ll 1.
\label{eq:ordering_general}
\end{equation}
Here, $f_s$ is a particle distribution function for a species $s$ and $\delta f_s$ is its fluctuating component. This approximation is usually good in the plasma core where fluctuations are indeed small. However, this approximation usually breaks down in the edge where fluctuations can be comparable in size to the equilibrium -- one approach to solve this is `full f' gyrokinetics \cite{Grandgirard2007,Heikkinen2008,Idomura2009,Parra2010,Hakim2020}, which we don't study here.

The approach for finding the transport equations with the assumptions in \Cref{eq:ordering_general} is to exploit scale separation using an order-by-order asymptotic expansion in the parameter $\rho_{*s}$. Starting from the Fokker-Planck kinetic equation, one typically investigates at three temporal scales $\mathcal{O} \left( \rho_{*s} \Omega_s f_s \right)$, $\mathcal{O} \left( \rho_{*s}^2 \Omega_s f_s \right)$, and $\mathcal{O} \left( \rho_{*s}^3 \Omega_s f_s \right)$. Terms of the size $\mathcal{O} \left( \rho_{*s} \Omega_s f_s \right)$ change fastest. Most notably, one learns that the lowest order distribution function is a Maxwellian. Terms of the size $\mathcal{O} \left( \rho_{*s}^2 \Omega_s f_s \right)$ change slower, and at these scales one finds the gyrokinetic and neoclassical equations. Terms of the size $\mathcal{O} \left( \rho_{*s}^3 \Omega_s f_s \right)$ change slowest. These are the transport equations. The transport equations cannot be solved without information from terms found in the $\mathcal{O} \left( \rho_{*s} \Omega_s f_s \right)$ and $\mathcal{O} \left( \rho_{*s}^2 \Omega_s f_s \right)$ equations. The pedagogy of this presentation is similar to that in \cite{Abel2013}, but we only produce the main results.

There are also other approaches for deriving gyrokinetics, most notably the Hamiltonian approach; \cite{Brizard_2007} is an excellent starting point.

\subsection{Fokker-Planck Equation} \label{subsec:kinetic}

In this subsection, we introduce the Fokker-Planck equation, which describes the evolution of a system subject to electromagnetic forces. A very useful quantity is the particle distribution function $f_s(\mathbf{r}, \mathbf{v}, t)$ for a species $s$, which describes the number of particles in position space $\mathbf{r}$ and velocity space $\mathbf{v})$ at a given time $t$. For example, the total particle density is a velocity space integral of $f_s(\mathbf{r}, \mathbf{v}, t)$,
\begin{equation}
n_s (\mathbf{r}, t) = \int f_s(\mathbf{r}, \mathbf{v}, t) d \mathbf{v}, 
\end{equation}
and the momentum $m_s \mathbf{u}_s$ and temperature $T_s$ are higher order `moments,'
\begin{equation}
 m_s \mathbf{u}_s (\mathbf{r}, t) = \int \mathbf{v} f_s(\mathbf{r}, \mathbf{v}, t) d \mathbf{v}, \;\; T_s (\mathbf{r}, t) = \int \frac{m_s v^2}{2} f_s(\mathbf{r}, \mathbf{v}, t) d \mathbf{v}.
\end{equation}
Other useful quantities such as fluxes can also be obtained from moments of $f_s(\mathbf{r}, \mathbf{v}, t)$. We will discuss these later.

We are interested in the time evolution of $f_s(\mathbf{r}, \mathbf{v}, t)$, which is given by the Boltzmann equation
\begin{equation}
\frac{d f_s}{d t} = \frac{\partial f_s}{\partial t} + \frac{\partial}{\partial \mathbf{r}} \cdot \left(\frac{d \mathbf{r} }{dt} f_s \right) + \frac{\partial}{\partial \mathbf{v}} \cdot \left(\frac{d \mathbf{v} }{dt} f_s\right) = C_s + G_s,
\label{eq:conservative_kinetic}
\end{equation}
where $C_s$ is a \textit{collision operator} and $G_s$ are sources. In a plasma, we concern ourselves with forces on charged particles due to an electric field $\mathbf{E}$ and a magnetic field $\mathbf{B}$,
\begin{equation}
m_s \frac{d \mathbf{v} }{dt} = Z_s e \left( \mathbf{E} + \frac{\mathbf{v} \times \mathbf{B} }{c} \right),
\end{equation}
where $Z_s$ is the particle charge number and $e$ is the proton charge. This gives the Fokker-Planck equation,
\begin{equation}
\frac{\partial f_s}{\partial t} + \mathbf{v} \cdot \nabla f_s + \frac{Z_s e}{m} \left( \mathbf{E} + \frac{\mathbf{v} \times \mathbf{B} }{c} \right) \cdot \frac{\partial f}{\partial \mathbf{v}} = C_s + G_s.
\label{eq:fokkerplanck}
\end{equation}
Because \Cref{eq:fokkerplanck} has three unknowns -- $f_s (\mathbf{r}, \mathbf{v}, t)$, $\mathbf{E} (\mathbf{r}, t)$, and $\mathbf{B} (\mathbf{r}, t)$, we close it with Maxwell's equations,
\begin{equation}
\begin{aligned}
& \nabla \cdot \mathbf{E} = 4 \pi e \sum_s Z_s \int f_s d \mathbf{v}, \\
& \nabla \cdot \mathbf{B} = 0, \\
& \nabla \times \mathbf{E} = - \frac{1}{c} \frac{\partial \mathbf{B}}{\partial t}, \\
& \nabla \times \mathbf{B} = \frac{1}{c} \frac{\partial \mathbf{E} }{\partial t} + \frac{4 \pi e}{c} \sum_s Z_s \int \mathbf{v} f_s d \mathbf{v}.  
\end{aligned}
\label{eq:Maxwell}
\end{equation}
The system of equations in \Cref{eq:fokkerplanck,eq:Maxwell} form the basis for solving the transport problem: finding $f_s$, $\mathbf{E}$, and $\mathbf{B}$ over transport timescales on the order of $\tau_E$.

\subsection{Maxwell Equilibrium}

Taking only terms of order $\mathcal{O} \left( \rho_{*s} \Omega_s f_s \right)$ in \Cref{eq:fokkerplanck} and performing a gyroaverage shows that parallel particle streaming balances particle collisions,
\begin{equation}
v_{\parallel} \hat{\mathbf{ b} } \cdot \frac{\partial f_s}{\partial \mathbf{R_s} } = \langle C \rangle_{\mathbf{R}_s },
\end{equation}
where the magnetic field unit vector is $\hat{\mathbf{ b} } = \mathbf{B} / B$ and the parallel velocity is $v_{\parallel}$. After some algebraic manipulations \cite{Abel2013}, one arrives at
\begin{equation}
\left\langle \int d^3 v \ln f_s C \right\rangle_{\psi} = 0,
\label{eq:zeroth_order}
\end{equation}
which means that to leading order the distribution function is Maxwellian,
\begin{equation}
f_s = F_{Ms} + \ldots.
\end{equation}
The Maxwellian equilibrium distribution function (reproduced from \Cref{eq:particledistribution}) is
\begin{equation}
F_{Ms} (v) = n_s \Big{(} \frac{m_s}{2 \pi T_s } \Big{)}^{3/2} \exp \Big{(} - \frac{m_s v^2}{ 2T_s } \Big{)}.
\label{eq:particledistribution_v2}
\end{equation}
In \Cref{eq:zeroth_order} a useful operation called a flux-surface average is performed,
\begin{equation}
\langle g \rangle_\psi (\psi) \equiv \oint \frac{g}{|\nabla V|}  d S,
\end{equation}
where the integral is performed over a closed flux surface with area element $dS$ and $V$ is the volume enclosed by the flux surface. This operation is used a lot in deriving the transport equations because it simplifies many expressions. After performing the average, the only spatial dependence for a quantity $g$ is flux-surface coordinate $\psi$.

\subsection{Gyrokinetics} \label{sec:gkequations}

Gyrokinetics is a system of equations that is often a good approximation of turbulence in MCF devices \cite{Catto1978,Frieman1982,Sugama2000,Parra2008,Abel2013}. It is a very useful formalism because it can reduce computational complexity by orders of magnitude without significant loss of accuracy. This speedup is obtained by reducing velocity space from 3D to 3D, which considerably speeds up computation \cite{Beer1995}. Physically the reduction from 3D to 2D corresponds to modeling `rings' of charge rather than following the exact location of a particle throughout its gyromotion. This is achieved with the `guiding center` coordinate transformation \cite{Catto1978},
\begin{equation}
\mathbf{R}_s = \mathbf{r} - \bm{\rho}_s,
\label{eq:gktransform}
\end{equation}
where $\mathbf{r}$ is the particle position, and 
\begin{equation}
\bm{\rho}_s = \frac{\hat{\mathbf{ b} } \times \mathbf{v}}{\Omega_s},
\end{equation}
is the cyclotron motion. By subtracting $\bm{\rho}_s $ from the particle position, $\mathbf{R}_s$ describes the guiding center position of the particle.
Modeling the guiding center is generally an excellent approximation because turbulence does not usually change appreciably on the gyrofrequency timescale. Mathematically, the turbulent frequency $\omega$ is ordered to be much slower than the cyclotron frequency $\Omega_s$,
\begin{equation}
\omega \ll \Omega_s.
\end{equation}
However, while we model particle guiding center trajectories with $\mathbf{R}_s$, we still include the spatial physics of a nonzero particle gyroradius, known as Finite Larmor Radius (FLR) effects. Spatially, the perpendicular wavenumber of turbulence $k_{\perp}$ is expected to be comparable to the gyroradius,
\begin{equation}
k_{\perp} \rho_s \sim 1.
\end{equation}
This is a particularly important `kinetic' effect that vanishes when $k_{\perp} \rho_s \to 0$. For $k_{\perp} \rho_s \gg 1$, turbulence is damped because particles average over many small structures during their fast gyromotion. Mathematically, FLR effects arise from gyroaveraging a quantity $g$ over its gyrophase $\varphi$,
\begin{equation}
\langle g \rangle_{\mathbf{R}_s} \equiv  \int_0^{2\pi} g d \varphi / 2\pi.
\label{eq:gyroaverage}
\end{equation}

\begin{figure}[bt]
    \centering
    \includegraphics[width=0.65\textwidth]{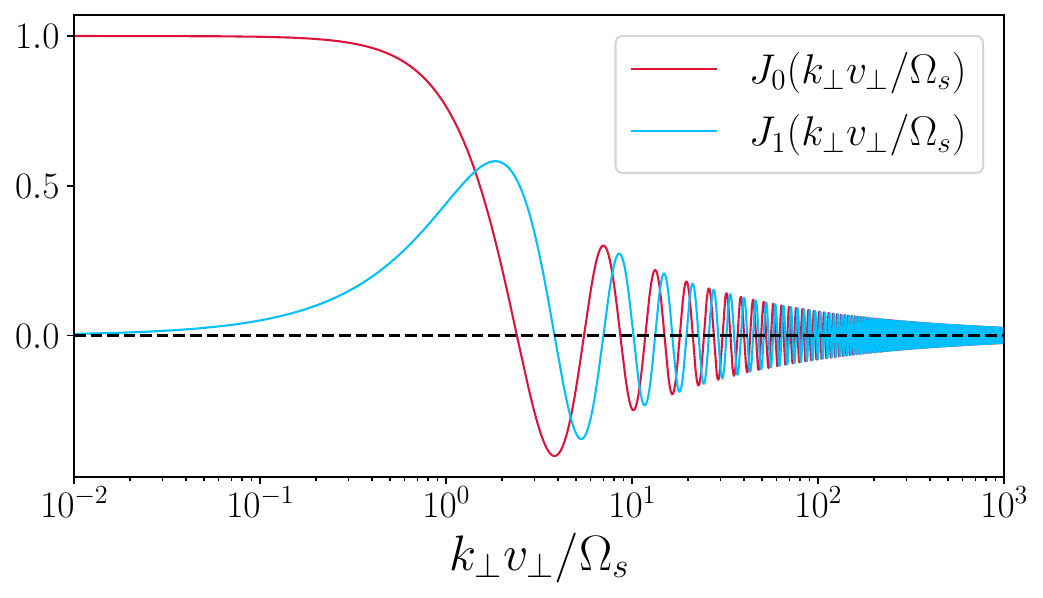}
    \caption{Bessel functions $J_0(k_{\perp} v_{\perp} / \Omega_s)$ and $J_1(k_{\perp} v_{\perp} / \Omega_s)$ that arise from gyroaveraging.}
    \label{fig:bessel}
\end{figure}

\begin{figure}[tb]
\centering
    \includegraphics[width=\textwidth]{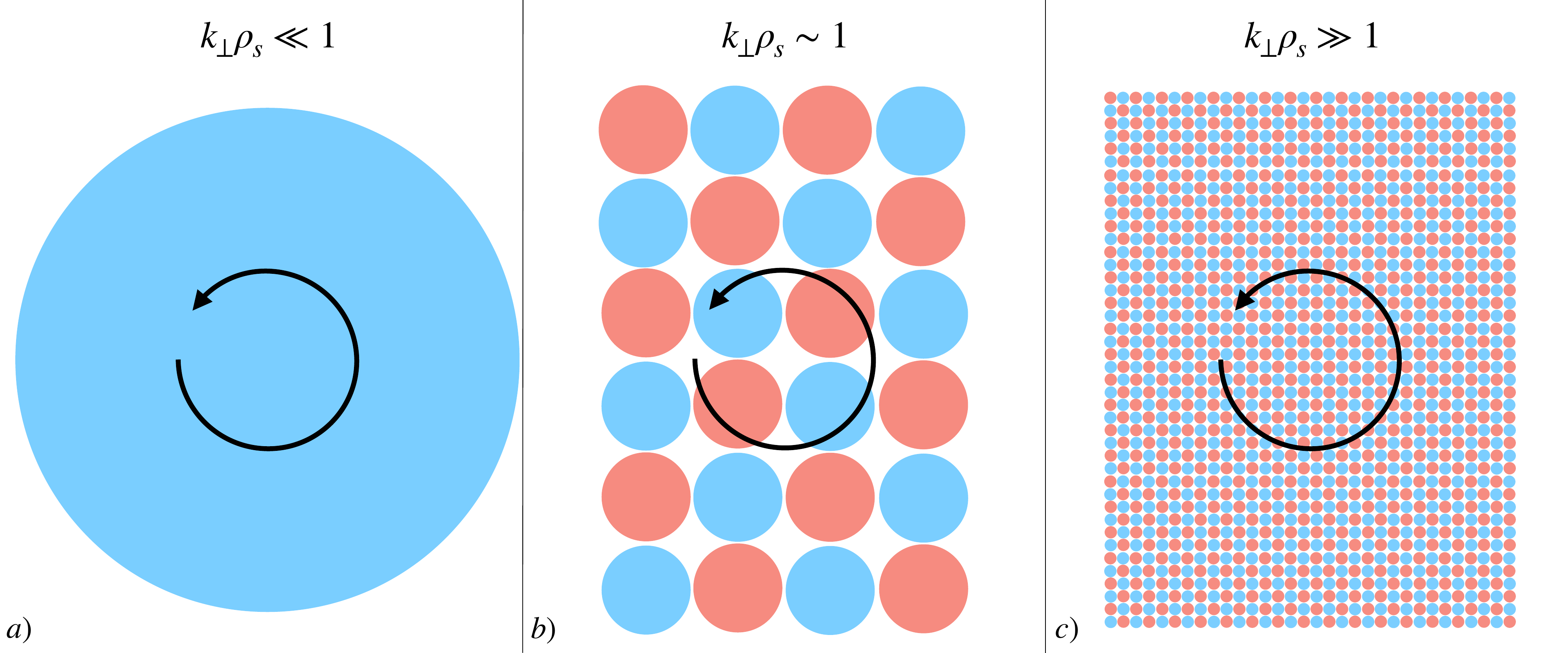}
 \caption{FLR effects at different perpendicular wavenumbers. Red and blue blobs represent turbulent overdensities and underdensities, respectively. Black arrows represent the particle's gyromotion. (a): The gyroradius is small compared with the turbulent wavelength and hence the particle sees a fairly constant density during a single gyroperiod. (b): The gyroradius is comparable to the turbulent wavelength and therefore the particle averages over several perturbations per gyroperiod, affecting its growth rate. (c): The gyroradius is large compared with the turbulent wavelength and hence the particle sees a rapidly changing turbulent density. Adapted from \cite{Parisi2020b}.}
\label{fig:FLReffects}
\end{figure}

\noindent The $\mathbf{R}_s$ subscript in $\langle g \rangle_{\mathbf{R}_s}$ indicates that the gyroaverage is performed at constant $\mathbf{R}_s$. To see how FLR effects enter physically, consider a Fourier mode $k_{\perp}$ of $g$, where $g$ represents some fluctuating quantity such as the electrostatic potential,
\begin{equation}
\hat{g}(k_{\perp}) = \int g (\mathbf{r} ) \exp (i \mathbf{k}_{\perp} \cdot \mathbf{r}) d \mathbf{r}  = \int g (\mathbf{r} ) \exp (i \mathbf{k}_{\perp} \cdot \mathbf{R}_s) \exp (i \mathbf{k}_{\perp} \cdot \bm{\rho}_s) d \mathbf{r},
\end{equation}
where we used \Cref{eq:gktransform} for the final term. Performing the gyroaverage (\Cref{eq:gyroaverage}) of the Fourier mode $\hat{g}(k_{\perp})$ converts the $\exp (i \mathbf{k}_{\perp} \cdot \bm{\rho}_s)$ term into $J_0$, a Bessel function of the first kind,
\begin{equation}
\langle \hat{g}(k_{\perp}) \rangle_{\mathbf{R}_s} = \int g (\mathbf{r} ) \exp (i \mathbf{k}_{\perp} \cdot \mathbf{R}_s) J_0 \left( \frac{k_{\perp} v_{\perp}}{\Omega_s} \right) d \mathbf{r}.
\end{equation}
The Bessel function is obtained by
\begin{equation}
J_0 \left( \frac{k_{\perp} v_{\perp}}{\Omega_s} \right) = \int_0^{2\pi} \exp (i \mathbf{k}_{\perp} \cdot \bm{\rho}_s) d \varphi / 2\pi.
\end{equation}
The higher order Bessel function $J_1$ also often appears in the gyrokinetic system of equations, occurring with higher $v_{\perp}$ moments of the plasma distribution function. We plot $J_0$ and $J_1$ in \Cref{fig:bessel}. As $x$ increases for $J_0(x)$, the turbulence wavenumber $k_{\perp}$ becomes increasingly smaller, causing the mode to average over more perturbations, decreasing the `drive' for linear instability. For $k_{\perp} \rho_s \ll 1$, a particle samples a roughly constant fluctuation amplitude during its gyromotion (see \Cref{fig:FLReffects} (a)), meaning that FLR effects are unimportant. As the turbulent wavenumber increases to be comparable to the gyroradius, $k_{\perp} \rho_s \sim 1$, the particle samples multiple fluctuations during its gyromotion (see \Cref{fig:FLReffects} (b)), which can start to damp the turbulence. When the particle gyroradius is large compared with the turbulent wavelength, $k_\perp \rho_s \gg 1$, the particle averages over a large number of fluctuations during its gyromotion (see \Cref{fig:FLReffects} (c)), causing turbulence to be significantly weaker.

Dynamics parallel to the magnetic field line are also important. In toroidal devices such as tokamaks and stellarators, particles move much faster along field lines than across them. This results in turbulence typically being more extended along magnetic field lines. Therefore, the parallel wavenumber associated with turbulence $k_{\parallel}$ is typically much smaller than the perpendicular wavenumber $k_{\perp}$, giving rise to turbulence that is spatially anisotropic,
\begin{equation}
k_{\parallel} \ll k_{\perp}.
\end{equation}

The next step is to find the gyrokinetic equation. For brevity, we don't perform a full derivation but highlight the main steps. A formal ordering of quantities is performed with the small parameter
\begin{equation}
\rho_{*s} \equiv \frac{\rho_s}{L} \ll 1,
\end{equation}
where $L$ is the device size. The standard gyrokinetic ordering is
\begin{equation}
\frac{\delta f_s}{f_s} \sim \frac{Z_s e \delta \phi}{T_s} \sim \frac{\omega}{\Omega_s} \sim \frac{\nu_s}{\Omega_s} \sim \frac{k_{\parallel }}{k_{\perp}} \sim \frac{\delta B}{B} \sim \frac{\delta E}{E} \sim \rho_{*s},
\end{equation}
where $\nu_s$ is the Coulomb collision frequency. Quantities are split into mean and fluctuating components. For example the total electric field $\tilde{\mathbf{E}}$ has a mean component $\mathbf{E}$ and fluctuating component $\delta \mathbf{E}$
\begin{equation}
\tilde{\mathbf{E}} = \mathbf{E} + \delta \mathbf{E}.
\end{equation}

\noindent We use gyrokinetic variables: the guiding center spatial coordinate $\mathbf{R}_s$, the energy $\mathcal{E}_s = m_sv^2/2$, the magnetic moment $\mu_s = m_s v_{\perp}^2 / 2 B$, the gyrophase $\varphi$, and the sign of the parallel velocity $\sigma = v_{\parallel}/ |v_{\parallel}|$. With these new coordinates, we write the kinetic equation for the fluctuating distribution function in a similar manner as \Cref{eq:conservative_kinetic},
\begin{equation}
\frac{d \delta f_s}{d t} = \frac{\partial \delta f_s}{\partial t} + \mathbf{R}_s \cdot \frac{\partial \delta f_s}{\partial \mathbf{R}_s} + \dot{\mu}_s \frac{\partial \delta f_s}{\partial \mu_s} + \dot{\mathcal{E}}_s \frac{\partial \delta f_s}{\partial \mathcal{E}_s} + \dot{\varphi} \frac{\partial \delta f_s}{\partial \varphi} = C_s.
\label{eq:conservative_kinetic_gke}
\end{equation}
It is useful to define a `non-adiabatic' distribution function
\begin{equation}
h_s = \delta f_s + \frac{Z_s e \delta \phi}{T_s} F_{Ms},
\end{equation}
which is the non-Boltzmann part of the perturbed distribution function. In order to eliminate the $\partial \delta f_s / \partial \varphi$ term, we gyroaverage \Cref{eq:conservative_kinetic_gke}, giving
\begin{equation}
\left\langle \frac{\partial \delta f_s}{\partial t} \right\rangle_{\mathbf{R}_s} + \mathbf{R}_s \cdot \frac{\partial \delta f_s}{\partial \mathbf{R}_s} + \left\langle \dot{\mathcal{E}}_s \right\rangle_{\mathbf{R}_s} \frac{\partial \delta f_s}{\partial \mathcal{E}_s} = \left\langle C_s \right\rangle_{\mathbf{R}_s},
\label{eq:conservative_kinetic_gke_gyroav}
\end{equation}
where we also used that $\left\langle \dot{\mu}_s \right\rangle_{\mathbf{R}_s}$ is small. \Cref{eq:conservative_kinetic_gke_gyroav} has an oscillating and equilibrium component.  The oscillating component of \Cref{eq:conservative_kinetic_gke_gyroav} is called the gyrokinetic equation
\begin{equation}
\frac{\partial {h}_s}{\partial t}  + \left( v_{\parallel } \hat{\mathbf{ b} } + \mathbf{v}_{Ms} + \langle \mathbf{v}_{\chi} \rangle_{\mathbf{R}_s} \right) \cdot \nabla {h}_s = \frac{Z_s e F_{Ms}}{T_s} \frac{\partial \langle {\delta \chi} \rangle }{\partial t} - \langle \mathbf{v}_{\chi} \rangle_{\mathbf{R}_s} \cdot \nabla F_{Ms} + C_s.
\label{eq:GKE}
\end{equation}
The gyrokinetic potential is
\begin{equation}
\delta \chi = \delta \phi - \frac{\mathbf{v} \cdot \mathbf{\delta A} }{c},
\end{equation}
the magnetic particle drifts are
\begin{equation}
\mathbf{v}_{Ms} =  \frac{\hat{\mathbf{ b} } }{\Omega_s} \times \left( \left[ v_{\parallel}^2 + \frac{v_{\perp}^2}{2} \right] \nabla \ln B + v_{\parallel}^2 \frac{4 \pi}{B^2} \frac{dp}{dr} \nabla r \right),
\end{equation}
and the fluctuating velocity field is
\begin{equation}
\mathbf{v}_{\chi} = \frac{c}{B} \hat{\mathbf{ b} } \times \nabla \chi.
\label{eq:vchi}
\end{equation}
The quantity $\mathbf{v}_{\chi}$ can be thought of as a generalized $\mathbf{ E} \times \mathbf{ B}$ drift, the difference being that the vector potential is also included in $\mathbf{v}_{\chi}$. Note that in \Cref{eq:GKE} we have omitted plasma rotation, which is often important in MCF plasmas. See \cite{Sugama_1998} for the gyrokinetic equation with rotation.

The gyrokinetic equation must be coupled to its equivalent order in Maxwell's equations, found by performing a similar expansion in \Cref{eq:Maxwell}.

\subsection{Neoclassical}

The equilibrium component of \Cref{eq:conservative_kinetic_gke_gyroav} is called the neoclassical drift-kinetic equation
\begin{equation}
\left(  \mathbf{v}_M \cdot \nabla + \frac{Z_s e}{c T}  v_{\parallel } \hat{\mathbf{ b} } \cdot \frac{\partial \mathbf{A}_0 }{\partial t} \right) f_0 = C[ \langle f_1 \rangle_{\mathbf{R}_s } ] - v_{\parallel } \hat{\mathbf{ b} } \cdot \nabla \langle f_1 \rangle_{\mathbf{R}_s },
\label{eq:neoclassical_eq}
\end{equation}
found by averaging \Cref{eq:conservative_kinetic_gke_gyroav} over time and space scales much longer than turbulent timescales $\omega$. The neoclassical equation describes slow perturbations at large scales to the equilibrium, in contrast to the gyrokinetic equation, which describes fast perturbations at gyroradius scales to the equilibrium. The neoclassical equation must be coupled to its equivalent order in Maxwell's equations, found by performing a similar expansion in \Cref{eq:Maxwell}.

\subsection{Transport}

The gyrokinetic system of equations describe the time evolution of fluctuating quantities such as $h_s$, $\delta f_s$, $\delta \phi$, and $\delta \mathbf{A}$. While this describes a plasma over fast timescales, the turbulent fluxes driven by turbulent fluctuations can modify the plasma profiles and the magnetic equilibrium. In order to describe this behavior self-consistently, we study the transport equations, which describe the evolution of the equilibrium distribution function $f_0$. In this tutorial, the regime with zero plasma flows. It is customary to study quantities in the transport equations averaged over both gyrophase and flux surface.

The evolution of $f_0$ is given by the particle transport equation and the energy transport equation (and also the momentum equation if plasma rotation is included). The particle equation is
\begin{equation}
\frac{\partial n_s}{\partial t} + \frac{1}{V'} \frac{\partial V' \langle \Gamma_s \rangle_{\psi}}{\partial \psi} = \langle s_s \rangle_{\psi},
\label{eq:particle_transp_flux_av}
\end{equation}
where
\begin{equation}
V' \equiv \frac{d V}{d \psi}, 
\end{equation}
and $S_s$ is the sum of particle sources and sinks for species $s$.  The energy equation is
\begin{equation}
\frac{3}{2} \frac{\partial n_s T_s}{\partial t} + \frac{1}{V'}  \frac{\partial V' \langle q_s \rangle_{\psi}}{\partial \psi} = \langle p_s \rangle_{\psi},
\label{eq:energy_transp_flux_av}
\end{equation}
where $P_s$ is the sum of energy sources and sinks.

The radial particle and heat fluxes are now scalar quantities indicating radial fluxes, containing contributions from classical, neoclassical, and turbulent transport,
\begin{equation}
\langle \Gamma_s \rangle_{\psi} \equiv \nabla \psi \cdot \int d^3v (\mathbf{v}_\chi \delta f_1 + \mathbf{v}_{Ms} \langle f_1 \rangle_{\psi} +  \bm{\rho} C[\bm{\rho} \cdot \nabla f_0]),
\end{equation}
\begin{equation}
\langle q_s \rangle_{\psi} \equiv \nabla \psi \cdot \int d^3v \frac{m_s v^2}{2} (\mathbf{v}_\chi \delta f_1 + \mathbf{v}_{Ms} \langle f_1 \rangle_{\psi} + \bm{\rho} C[\bm{\rho} \cdot \nabla f_0]). 
\end{equation}
The $\mathbf{v}_\chi \delta f_1$ term describes radial energy transport arising from turbulent perturbations contained within $\mathbf{v}_\chi$ and $\delta f_1$; the $\mathbf{v}_M \langle f_1 \rangle_{\psi}$ describes radial energy transport arising from magnetic drifts from neoclassical effects contained within $\langle f_1 \rangle_{\psi}$; the $\bm{\rho} C[\bm{\rho} \cdot \nabla f_0]$ term describes radial energy transport arising from collisions.

\subsection{Magnetic Equilibrium}

Finally, the evolution of the magnetic equilibrium must also be calculated since the magnetic equilibrium enters the gyrokinetic, neoclassical, and transport equations. This involves finding the evolution of the toroidal and poloidal fluxes (see \Cref{eq:B_eq_tok}) by solving the Grad-Shafranov equation
\begin{equation}
R^2 \nabla \cdot \left( \frac{\nabla \psi}{R^2} \right) + I \frac{d I}{d \psi} = - 4 \pi R^2 \sum_s \frac{d p_s}{d \psi}, 
\label{eq:grad_shafranov}
\end{equation}
to update $\psi$ that enters the particle and energy transport equations, and solving the toroidal component of Faraday's law
\begin{equation}
\frac{\partial}{\partial \psi} \left\langle \frac{I}{R^2}  \right\rangle_{\psi}  = - \frac{1}{V'} \frac{\partial}{\partial \psi} \left( V' \left\langle \frac{\partial \mathbf{A}_0 }{\partial t} \cdot \frac{\mathbf{B} }{c}  \right\rangle_{\psi} \right),
\label{eq:Faraday_law}
\end{equation}
and obtaining $\partial \mathbf{A}_0 /\partial t \cdot \mathbf{B}/c$ from the neoclassical equation.

We now have all of the components for studying the evolution of the plasma profiles and magnetic equilibrium.

\subsection{Example}

In this section, we show an example output from a transport solver called TRANSP \cite{Pankin2024}. In order to calculate power balance, the electron energy equation is volume-integrated from the magnetic axis up to a given flux surface,
\begin{equation}
P_{e,\mathrm{gain} } + P_{e,\mathrm{loss} } = P_{e,\mathrm{source} }.
\label{eq:powerbalance_app}
\end{equation}
The change in electron thermal energy is
\begin{equation}
P_{e,\mathrm{gain} } = \frac{3}{2} \int \frac{\partial  n_e T_e}{\partial t} dV,
\end{equation}
the conductive and convective losses are
\begin{equation}
P_{e,\mathrm{loss} } = P_{e,\mathrm{cond} } + P_{e,\mathrm{conv} } = \int \nabla \cdot \mathbf{q}_e dV,
\end{equation}
and the source terms are
\begin{equation}
P_{e,\mathrm{source} } = P_{e,\mathrm{aux} } + P_{e,\mathrm{rad} } + P_{e,\mathrm{ie} } = \int \sum_k s_{e,k} dV.
\end{equation}
Here, $P_{e,\mathrm{heat} }$ is the auxiliary electron heating power, $P_{e,\mathrm{rad} }$ is the electron radiated power, and $P_{e,\mathrm{ie} }$ is the ion-electron coupling power from electrons to ions.

\begin{figure}[bt]
    \centering
    \begin{subfigure}[t]{0.49\textwidth}
    \includegraphics[width=0.99\textwidth]{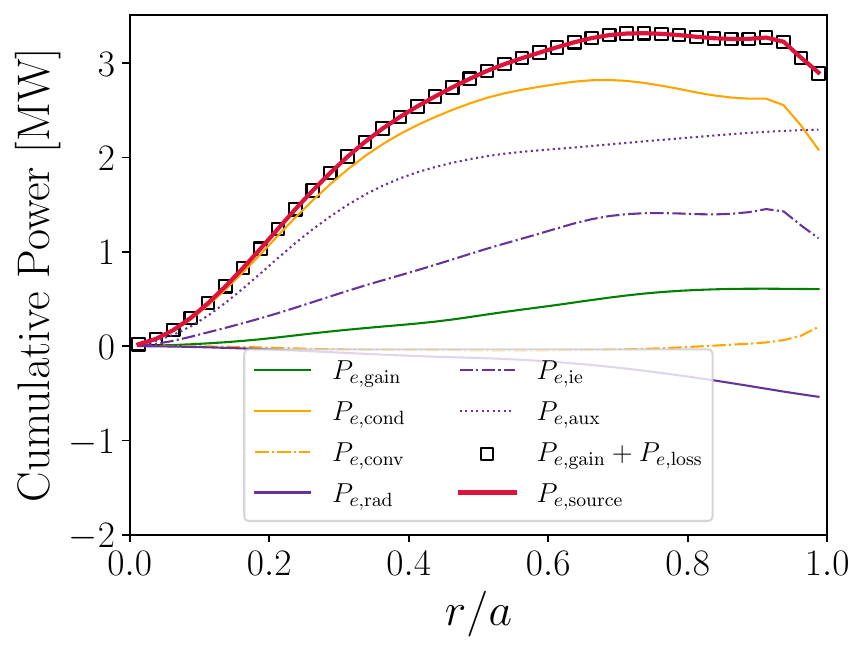}
    \caption{.}
    \end{subfigure}
    \centering
    \begin{subfigure}[t]{0.49\textwidth}
    \includegraphics[width=0.99\textwidth]{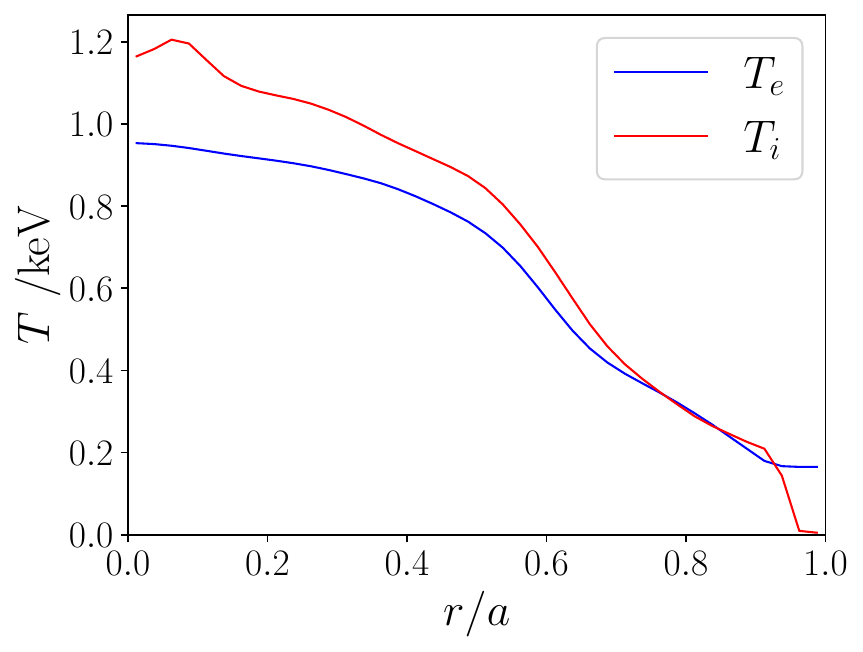}
    \caption{.}
    \end{subfigure}
    \caption{For NSTX discharge 132543 at $t = 0.614$ seconds, (a) terms in the volume-integrated electron energy transport equation versus normalized minor radius $r/a$, (b) ion and electron temperature. In (a), the time derivative term is green, the heat losses are in orange, and the sources and sinks are in purple. See \Cref{eq:powerbalance_app} for further explanation.}
    \label{fig:TRANSP_sources}
\end{figure}

In \Cref{fig:TRANSP_sources}(a), we plot these terms for NSTX discharge 132543 \cite{Berkery2024} at $t = 0.614$ seconds. In \Cref{fig:TRANSP_sources}(b), we plot the electron and ion temperature profiles.

\subsection{Further Reading}

An excellent tutorial on gyrokinetics can be found in Appendix A of \cite{Howes_2006}. For readers with more background, classic gyrokinetics references are \cite{Catto1978,Frieman1982,Sugama_1998,Parra2008,Abel2013}. For collisional and neoclassical transport \cite{Helander2002} is a good resource. For transport equations, see \cite{Sugama1996,Barnes2010trinity}.

\section{Gyrokinetic Instabilities} \label{sec:linearstab}

We have covered the main equations that describe transport in MCF devices. However, we have said very little about the physical mechanisms that give rise to turbulence. In this section, we present the most common plasma instabilities that drive heat (and some particle) transport in tokamaks and stellarators.

The physical picture for radial heat transport in tokamaks is as follows. Steep temperature and density gradients can destabilize linear instabilities. These instabilities cause plasma fluctuations to grow until they saturate in magnitude. The mechanisms that cause this saturation are an ongoing area of research.

Fusion scientists care about the transport properties of turbulence resulting from different microinstabilities. For each instability, we will provide a summary of the transport characteristics in terms of the relative heat and particle transport diffusivities, $\chi_s$ and $D_s$ for different species $s$. In \Cref{tab:modetypes} we list some features of the gyrokinetic instabilities covered in this section.

\begin{table}[htbp]
\centering
\caption{\label{tab:example} Some features of the main gyrokinetic instabilities.}
\vspace{0.5em}
\begin{tabular}{|c|c|c|c|c|c|c|c|}
\hline
Mode & $\chi_i / \chi_e$ & $D_e / \chi_e$ & $D_i / \chi_i$  & Free Energy Source & Resonance & Section \\
\hline
ITG &   $\gg 1$     &   -   &    $\ll 1$    &  $\nabla T_i$ & $v_{\parallel}$, $v_{Mi}$ & \ref{subsec:ITG} \\
\hline
ETG & $\ll 1$  & $\ll 1$  &  -  &  $\nabla T_e$ & $v_{\parallel}$, $v_{Me}$ & \ref{subsec:ETG}  \\
\hline
TEM &   $\sim1$   &   $\sim1$   &  $\sim1$    &  $\nabla T_e$ &  $v_{Me}$  & \ref{subsec:TEM}    \\
\hline
MTM & $\ll 1$  & $\ll 1$  &  -  &  $\nabla T_e$ &  $v_{Me}$ & \ref{subsec:MTM}  \\
\hline
KBM & $\sim1$  & $\sim1$  & $\sim1$   &  $\nabla p$ & $v_{Mi}$, $v_{Me}$ & \ref{subsec:KBM}  \\
\hline
PVG &   $\gg 1$     &   -  &   $\sim 1$    &  $\nabla u_{\parallel}$ & $v_{\parallel}$ & \ref{subsec:PVG} \\
\hline
Universal &   $\gg 1$     &  -  &    $\sim 1$   &  $\nabla n$ & $v_{\parallel},v_{Mi}$, $v_{Me}$ & \ref{subsec:Univ} \\
\hline
\end{tabular}
\label{tab:modetypes}
\end{table}

\subsection{Ion Temperature Gradient Instability} \label{subsec:ITG}

\begin{figure}[tb]
\centering
    \includegraphics[width=0.9\textwidth]{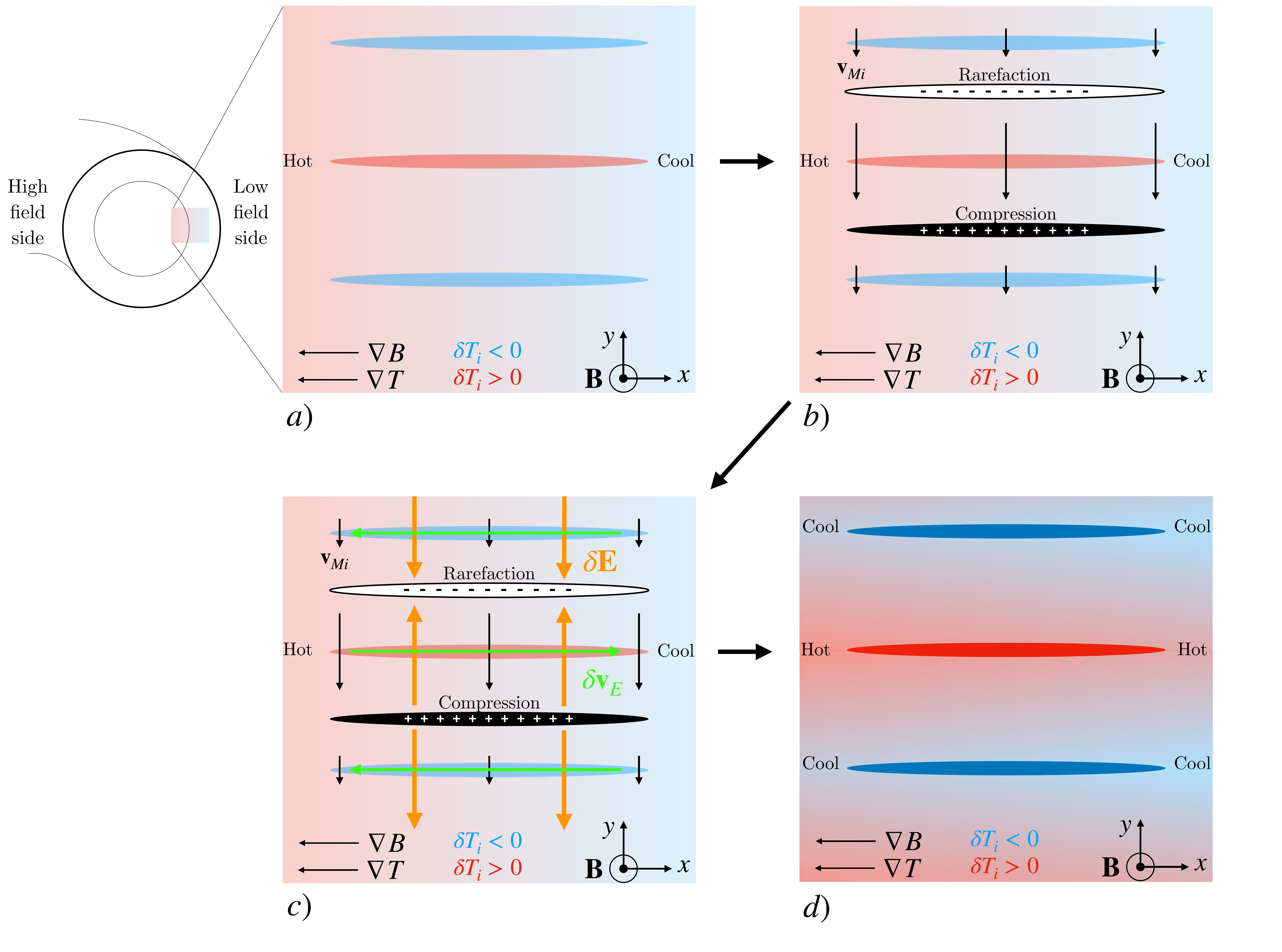}
 \caption{Mechanism for the toroidal ITG instability. (a): At the low-field side, a wave with a poloidal wavenumber creates regions of slightly hotter and colder temperature perturbations, $\delta T_i$, shown by red and blue contours. (b): The ion magnetic drift $\mathbf{v}_{Mi} $ is larger in the regions where $\delta T_i > 0$ and smaller where $\delta T_i < 0$. This makes regions of compressed and rarified plasma. In compressed regions there will be a net positive electric charge and in rarified regions a net negative electric charge. (c): The charge overdensities and underdensities create a perturbed electric field $\delta \mathbf{E}$, resulting in a perturbed $\mathbf{ E} \times \mathbf{ B}$ drift, $\delta \mathbf{v}_E$. This causes hot plasma to be sucked into regions where $\delta T_i > 0$, and cool plasma to be sucked into regions where $\delta T_i < 0$, creating a positive feedback loop. This creates the unstable state in (d), which can grow to become turbulent.}
\label{fig:ITGinstability}
\end{figure}

In this section, we describe one of the most common and virulent instabilities in magnetized toroidal devices: the ion temperature gradient (ITG) instability \cite{Rudakov1961,Nordman1990,Hammett1990,Cowley1991,Nunami2011}. ITG instability is concerning for fusion power plants because it can lower the core ion temperature, which reduces the fusion power. Analytic calculations, even for the simplest ITG instability, are quite involved. We will not reproduce the calculation in the main text; those who wish to read the technical details can read \Cref{app:ITGstab}.

There are generally two `branches' of ITG instability. The toroidal ITG branch is destabilized by the effect of ion magnetic drifts arising from toroidicity. The slab ITG branch is destabilized by the motion of ions moving along magnetic field lines. These modes have different features beyond the scope of this tutorial. See \cite{Parisi2020} for an in-depth discussion of toroidal and slab branches. In this section, we briefly describe the features of the toroidal ITG mode.

The toroidal ITG moode that is destabilized by ion magnetic drifts and steep ion temperature gradients. In \Cref{fig:ITGinstability} we present a cartoon picture of the ITG instability. In \Cref{fig:ITGinstability} (a), we show a wave at the low-field side with a poloidal wavenumber in the presence of a background temperature gradient. We represent peaks and troughs in the wave by red and blue horizontal ovals, depicting positive and negative ion temperature fluctuations, respectively. The temperature fluctuations due to the wave cause  variations in the size of the ion magnetic particle drifts in the vertical direction, shown in (b). Due to particle conservation, this causes ion density compressions and rarefactions, causing positive charge buildup and depletion due to the ion having a positive charge. Shown in (c), this creates a perturbed electric field $\delta \mathbf{E}$ that also creates a perturbed $\mathbf{ E} \times \mathbf{ B}$ drift. Due to the direction of the $\mathbf{ E} \times \mathbf{ B} $ drift, hot plasma is sucked radially outwards from the core to where there are temperature overperturbations, and cold plasma is sucked radially inwards from the edge to where there are temperature underperturbations. This exacerbates the instability, whose amplitude can eventually grow sufficiently large to cause turbulence. Note that this mechanism does not drive instability at the low-field side since the temperature gradient is in the opposite direction. This has a stabilizing effect when combined with the ion magnetic drifts. The toroidal ITG instability is usually most virulent when the magnetic curvature and temperature gradient are aligned, which occurs at the low-field side.

An important result from linear gyrokinetic analysis is the growth rate $\gamma$ of a perturbation such as the non-adiabatic ion distribution function $h_i$ (see \Cref{eq:GKE}). Given $h_i$ at an initial time $t_0$, if there is an instability $h_i$ can grow with linear dependence at $t> t_0$
\begin{equation}
h_i (t) = h_i (t_0) e^{i\gamma t},
\end{equation}
until nonlinear effects become sufficiently strong for the plasma to enter a turbulent state (see \Cref{sec:nonlinear_sec}). Generally speaking, the higher the linear growth rate, the more strongly driven the turbulence is. In \Cref{app:ITGstab} we derived an expression for the toroidal ITG growth rate (\Cref{eq:gamma_toroidal_ITG}),
\begin{equation}
\gamma \approx \sqrt{ \omega_{\kappa i}  \omega_{*e}^T} \simeq k_y \rho_i \sqrt{-\frac{v_{ti}^2}{L_{\mathrm{\kappa} } L_{T_i}} },
\label{eq:gamma_toroidal_ITG_main}
\end{equation}
where the magnetic drift frequency $\omega_{\kappa i}$ and the drift wave frequency $\omega_{*e}^T$ are
\begin{equation}
\omega_{\kappa i} \sim - k_y \rho_i \frac{v_{ti}}{L_{\mathrm{\kappa} }}, \;\;\;  \omega_{*e}^T \sim k_y \rho_i \frac{v_{ti}}{L_{Ti}},
\end{equation}
$L_{\mathrm{\kappa} }$ is the magnetic curvature length scale typically comparable in size to the major radius $R_0$, and $k_y$ is a wavenumber in direction binormal to the magnetic field. A necessary condition for instability is
\begin{equation}
\omega_{\kappa i} \omega_{*e}^T > 0,
\end{equation}
which require the magnetic curvature and ion temperature gradient have the opposite sign, $L_{\mathrm{\kappa} } L_{T_i} < 0$. The region where $L_{\mathrm{\kappa} } L_{T_i} < 0$ is known as `bad curvature' and $L_{\mathrm{\kappa} } L_{T_i} > 0$ is `good curvature.' The bad curvature region is generally on the low-field side in \Cref{fig:ITGinstability} and the good curvature region is on the high-field side.

The ITG mode typically produces significant ion heat transport, but little particle transport. Therefore, 
\begin{equation}
\frac{D_i}{\chi_i} \ll 1.  
\end{equation}
Finally, because electrons do not typically respond kinetically, ITG produces very little electron heat transport. This results in 
\begin{equation}
\frac{\chi_i}{\chi_e} \gg 1. 
\end{equation}

\subsection{Electron Temperature Gradient Instability} \label{subsec:ETG}

Linearly, electron temperature gradient (ETG) instability \cite{Jenko2000,Dorland2000,Told2008,Zocco2015,Hatch2019,Parisi2020,Adkins2022} is identical (`isomorphic') to ITG instability under the substitution,
\begin{equation}
\rho_i \to \rho_e, \;\; m_i \to m_e, \;\; h_i \to h_e, \;\; Z_i \to Z_e.
\end{equation}
In the presence of ETG driven turbulence, the plasma core electron temperature can decrease. However, there are certain features that break the isomorphism between ITG and ETG, so that the nonlinear state is quite different. A 1991 paper on ITG instability \cite{Cowley1991} and turbulence predicted that under certain conditions, the transport due to ITG turbulence would be radially elongated and cause extremely high heat transport. Thankfully, due to some technicalities ITG turbulence is not radially elongated. However, simulations of ETG turbulence found that the transport arising from ETG turbulence is radially elongated \cite{Jenko2000,Dorland2000}. This means that the ETG transport can be comparable in size to ITG turbulence. This is predicted to be the case for present-day tokamaks.

Nonlinear gyrokinetic simulations of drift‑wave turbulence show a crucial distinction between ITG-driven and ETG-driven turbulence: the way in which the zonal (flux‑surface-averaged) component of the electrostatic potential feeds back on density fluctuations. In ITG turbulence, electrons are typically treated as adiabatic on ion‑gyroradius scales. Their density perturbation is given by a modified Boltzmann response that enforces zero net electron charge on each flux surface:
\begin{equation}
\frac{\delta n_e}{n_{e}}
\;=\;
\frac{e}{T_e}\,\bigl[\delta\phi\;-\;\langle\delta\phi\rangle_{\psi}\bigr],
\label{eq:deltane_ITG_refined}
\end{equation}
where $\delta\phi$ is the fluctuating potential and $\langle\delta\phi\rangle_{\psi}$ is its flux‑surface average. Subtracting $\langle\delta\phi\rangle_{\psi}$ guarantees that the integrated electron density perturbation over the surface vanishes, preventing any unbalanced zonal charge from appearing in the adiabatic closure. In contrast, ETG turbulence occurs at electron-gyroradius scales, where ions are effectively adiabatic and respond according to a simple Boltzmann law without any zonal subtraction:
\begin{equation}
\frac{\delta n_i}{n_{i}}
\;=\;
-\,\frac{Z_i e}{T_i}\,\delta\phi,
\label{eq:deltane_ETG_refined}
\end{equation}
The subtraction in \eqref{eq:deltane_ITG_refined} has no effect until a nonzero zonal potential is present, which can only occur in a nonlinear turbulent state. In a linear analysis, $k_y=0$ is ignored because the growth rate is zero, which forces $\langle\delta\phi\rangle_{\psi}=0$ and reduces both \eqref{eq:deltane_ITG_refined} and \eqref{eq:deltane_ETG_refined} to a simple proportionality between $\delta n_s/n_{s0}$ and $\delta\phi$. Only when nonlinear mode‑mode coupling is retained does energy transfer into the $k_y=0$ `zonal' mode, generating a non-zero $\langle\delta\phi\rangle_{\psi}$. Physically, the resulting zonal shear $\partial_x\langle\delta\phi\rangle_{\psi}$ plays a dominant role in regulating ITG turbulence. We discuss zonal flows more in \Cref{sec:nonlinear_sec}

ETG-driven turbulnece typically produces significant electron heat transport, but little particle transport. Therefore, 
\begin{equation}
\frac{D_e}{\chi_e} \ll 1.  
\end{equation}
Finally, because ions do not typically respond kinetically, ITG produces very little ion heat transport. This results in 
\begin{equation}
\frac{\chi_i}{\chi_e} \ll 1. 
\end{equation}

\subsection{Trapped Electron Mode} \label{subsec:TEM}

\begin{figure}[tb]
\centering
    \includegraphics[width=\textwidth]{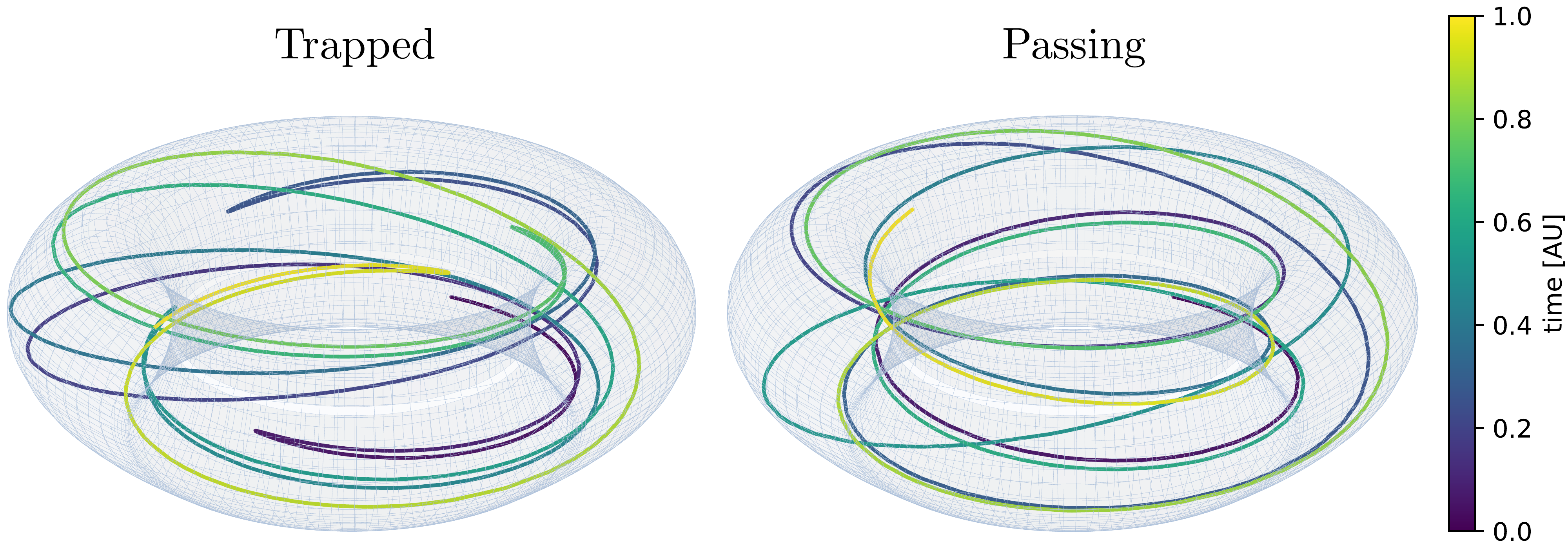}
 \caption{Trapped and passing guiding-center orbits in the ASDEX Upgrade tokamak (shot 26884 at 4300 ms) \cite{Heyn2014}. Both trapped and passing particles begin at the same location with the same energy but have different pitch. Trajectories are calculated using GORILLA \cite{Eder2020,Eder2023}.}
\label{fig:trapped_passing}
\end{figure}

\begin{figure}[tb]
\centering
    \includegraphics[width=0.9\textwidth]{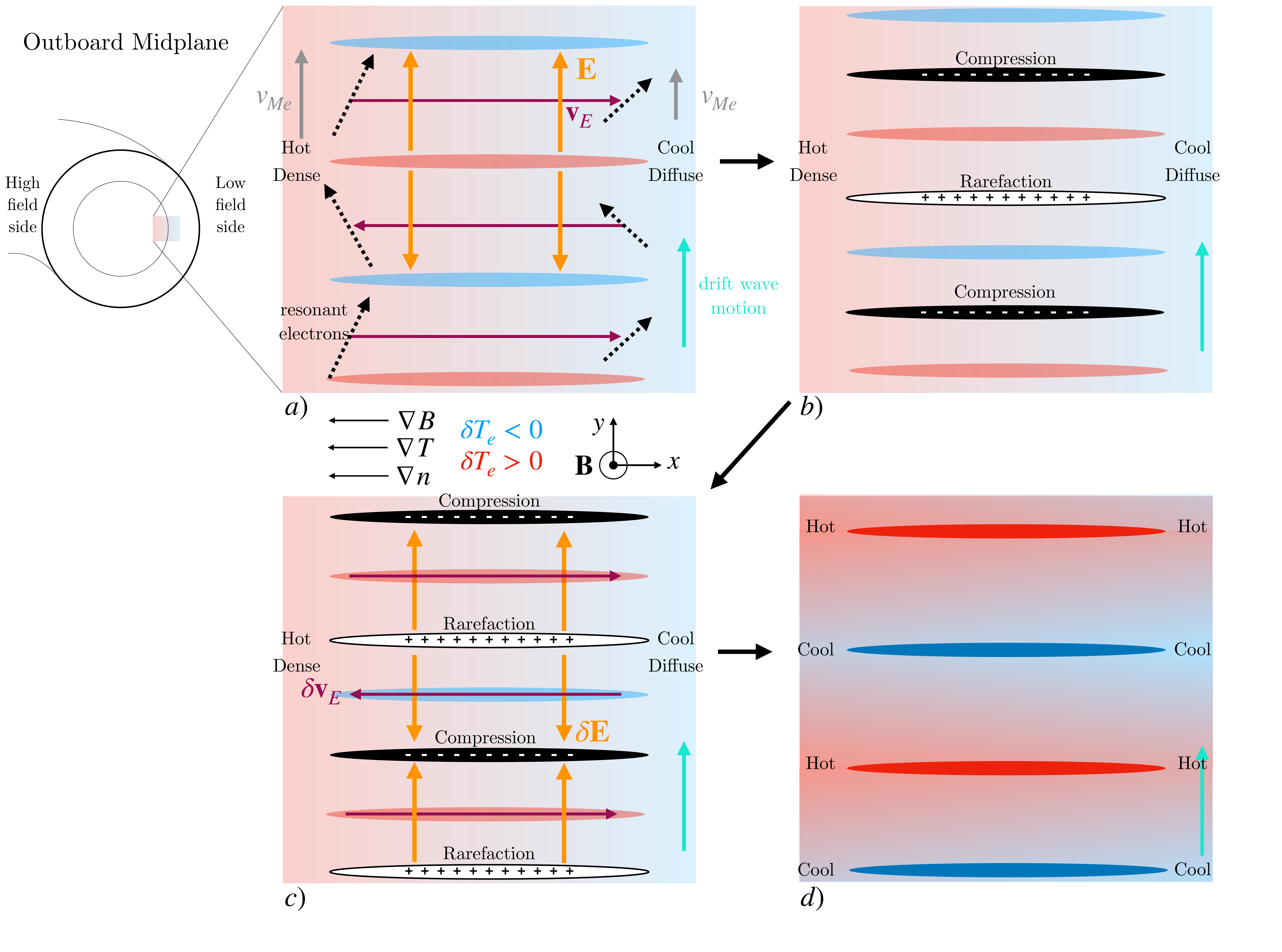}
 \caption{Mechanism for the TEM instability. See discussion above \Cref{eq:TEM_phase} for explanation of the physical mechanism.}
\label{fig:TEMinstability}
\end{figure}

In this section, we describe another common instability called the trapped electron mode (TEM) \cite{Ernst2004,Hatch2015}. It is challenging to obtain analytic results for the TEM. The TEM is destabilized by both temperature and density gradients, and is driven primarily by `trapped' electrons. In tokamaks, some particles are trapped due to variation in the magnetic field strength as they follow field lines. Trapping occurs due to conservation of two quantities: (1) the magnetic moment, $\mu = v_\perp^2 / 2 B$, and (2) the particle energy $E = m \mu B + m v_\parallel^2 / 2$ where the pitch is $\lambda = m \mu B_0 / E$ and the total speed is $v = \sqrt{2 E / m}$. The parallel velocity for a particle is
\begin{equation}
v_\parallel = \pm v \sqrt{1 - \frac{B}{B_0} \lambda}. 
\end{equation}
Any particle with $v_\parallel = 0$ during its motion along field lines will be trapped, and if its speed $v$ is sufficiently large, will bounce back. In \Cref{fig:trapped_passing} we plot trapped and passing orbits for two particles with $T = 10$ keV but different pitch. The orbit trajectories are calculated in a realistic geometry for the ASDEX upgrade tokamak using the GORILLA guiding center code \cite{Eder2020,Eder2023}.

The intuition for the TEM is that a subpopulation of trapped electrons can resonate with drift waves. Drift waves are stable waves that are common in plasmas with density gradients, and have a wave frequency $\omega = \omega_{*e}$ called the drift frequency. In \Cref{fig:TEMinstability} we consider a cartoon for the TEM instability. In (a), a drift wave is propagating in the $y$ direction on the low-field side of a tokamak. A fraction of trapped electrons also move in the $y$ direction at the same speed as the drift wave -- these resonant electrons always see the same $\mathbf{ E} \times \mathbf{ B}$ drift from the drift wave. Because of a radial density gradient, the $\mathbf{ E} \times \mathbf{ B}$ drift will create regions of more trapped electrons (labeled as `Compression') and regions of fewer trapped electrons (labeled as `Rarefaction'). Compression regions will have a net negative charge and Rarefaction regions will have a net positive charge, shown in (b). A perturbed electric field $\delta \mathbf{E}$ will form between the Compression and Rarefaction regions, giving rise to a perturbed $\mathbf{ E} \times \mathbf{ B}$ drift $\delta \mathbf{v}_E$. The perturbed drift $\delta \mathbf{v}_E$ causes hot, dense plasma to be sucked into regions where $\delta T_i > 0$, and cool, diffuse plasma to be sucked into regions where $\delta T_i < 0$, creating a positive feedback loop. This creates the unstable state in (d), which can grow to become turbulent. In this physical picture, we have assumed that density and temperature perturbations for TEM instability are in phase, that is
\begin{equation}
\frac{\delta n_e}{\delta T_e} > 0.
\label{eq:TEM_phase}
\end{equation}
Because the kinetic response of both ions and electrons matters for TEM, there is no reason to expect that the electrons or ions dominate heat transport, \begin{equation}
\frac{\chi_i}{\chi_e} \sim 1. 
\end{equation}
Similarly, because both the electron and ion kinetic response are sizable, both species can produce comparable particle and heat transport,
\begin{equation}
\frac{D_i}{\chi_i} \sim 1, \;\;\;\;\;\; \frac{D_e}{\chi_e} \sim 1.
\end{equation}

\subsection{Microtearing Mode} \label{subsec:MTM}

In this section, we describe microtearing modes (MTMs) \cite{Drake1980,Hardman2022}, which are destabilized by electron temperature gradients. Microtearing modes \textit{reconnect} magnetic field lines \cite{Yamada2010} and require collisionality to be sufficiently high to be unstable. The instability drive relies on a resonant interaction with electron parallel motion. Magnetic reconnection forms magnetic islands that enhance radial electron transport. This reconnection process allows magnetic field topology to change, breaking the frozen-in condition of ideal MHD. 

The MTM produces significant electron heat transport, primarily through the electromagnetic channel (the heat flux generated by $\delta \mathbf{A}$ fluctuations is much larger than the flux generated by $\delta \phi$ fluctuations), but little particle transport. Therefore, 
\begin{equation}
\frac{D_e}{\chi_e} \ll 1.  
\end{equation}
Finally, because ions do not typically respond kinetically, MTM produces very little ion heat transport. This results in 
\begin{equation}
\frac{\chi_i}{\chi_e} \ll 1. 
\end{equation}
MTMs have been observed on many devices, and are thought to play a particularly important role in heat transport at the plasma edge \cite{Applegate2007micro,Doerk2012,Dickinson2013,Hatch2016,Rafiq2016,Jian2019,Hassan2021_b,Hatch2021,Hardman2022,Dominski2024,Patel2025}.

\subsection{Kinetic Ballooning Mode} \label{subsec:KBM}

In this section, we describe the kinetic-ballooning-mode (KBM) \cite{Tang1980,Snyder2009,Aleynikova2018,Kennedy2023p}, which is destabilized by pressure gradients. The KBM is the kinetic analogue of the ideal ballooning mode \cite{Connor1979}. The KBM is typically characterized by a very strong stiffness \cite{Snyder2001} and so it is thought that any plasma profile subject to KBM instability will have the gradients strongly clamped to those values. This is the basis for the widely used EPED model \cite{Snyder2009}, which predicts the edge profiles in H-mode plasmas. Recently, non-stiff KBM-driven turbulence has been reported in stellarators \cite{Mulholland2024}.

Because the kinetic response of both ions and electrons matters for KBM, there is no reason to expect that the electrons or ions dominate heat transport, \begin{equation}
\frac{\chi_i}{\chi_e} \sim 1. 
\end{equation}
Similarly, because both the electron and ion kinetic response are sizable, both species can produce comparable particle and heat transport,
\begin{equation}
\frac{D_i}{\chi_i} \sim 1, \;\;\;\;\;\; \frac{D_e}{\chi_e} \sim 1.
\end{equation}
Recently, experimental measurement of KBM was reported \cite{Jian2023} on the DIII-D tokamak \cite{Luxon2002}.

\subsection{Parallel Velocity Gradient Instability} \label{subsec:PVG}

In this section, we describe the parallel velocity gradient (PVG) instability, which is destabilized by a radial gradient in the particle flow \cite{Catto1973,Kinsey2005,Peeters2005,Casson2009,Roach2009,Newton2010,Barnes2011b,Schekochihin2012}. As mentioned briefly in the introduction, tokamak plasmas can rotate toroidally with external angular momentum input using neutral beams, or in the absence of external input by redistributing momentum across flux surfaces \cite{Parra2015}. Because magnetic field lines are not fully in the toroidal direction, the toroidal flow associated with rotation has a component parallel to the magnetic field as well as a perpendicular component. A radial gradient in the perpendicular flow can stabilize unstable modes and reduce turbulent transport  \cite{Burrell1997,Barnes2011b} whereas a radial gradient in the parallel flow drives PVG.

Physically, the PVG instability is driven by adjacent plasma layers moving past each other along with different parallel flow velocities. This velocity shear is similar to the velocity jump across a shear layer in the Kelvin–Helmholtz (KH) instability in fluids. In the KH instability, perturbations at the boundary of two fluid streams can grow if there's a velocity difference. Eddies form and grow by extracting energy from the shear flow.

The PVG instability typically produces significant ion heat transport, but less particle transport,
\begin{equation}
\frac{D_i}{\chi_i} \ll 1.  
\end{equation}
Because electrons do not typically respond kinetically for PVG, it produces very little electron heat transport,
\begin{equation}
\frac{\chi_i}{\chi_e} \gg 1. 
\end{equation}

\subsection{Universal Instability} \label{subsec:Univ}

In this section, we describe the universal instability. Because many microinstabilities are stabilized by steep pressure gradients, it is tempting to use steep density gradients to stabilize turbulence. Furthermore, because the fusion power scales with the square of the fuel ion density (see \Cref{eq:powerdensity}), higher density in the plasma core resulting from steep density gradients gives much higher power. This is particularly true for stellarators that appear to not be subject to the Greenwald density limit \cite{Sudo1990,Giannone2000}.

However, there is an instability destabilized by steep density gradients \cite{Krall1965,Ross1978,Landreman2015}, the universal instability. \cite{Landreman2015} demonstrated the universal instability can be unstable a wide range perpendicular scales, from wavelengths longer than the ion gyroradius to those shorter than the electron gyroradius. This instability could limit the achievable core density values if the density gradients are too steep.

The kinetic response of both ions and electrons is important for universal modes, and therefore
\begin{equation}
\frac{\chi_i}{\chi_e} \sim 1.
\end{equation}
Because both the electron and ion kinetic responses are sizable, both species can produce comparable particle and heat transport,
\begin{equation}
\frac{D_i}{\chi_i} \sim 1, \;\;\;\;\;\; \frac{D_e}{\chi_e} \sim 1.
\end{equation}

\subsection{Nonlinear Phenomena} \label{sec:nonlinear_sec}

In this section, we briefly describe some nonlinear phenomena arising in turbulent magnetized fusion plasmas. Nonlinear physics is important because it is responsible for the turbulence that causes transport. All of the instabilities described in earlier subsections are predictable using linear equations. Only if the perturbations resulting from these instabilities grow sufficiently large in amplitude for nonlinear physics to matter, do these instabilities cause significant transport.

First, what is nonlinearity? In the gyrokinetic equation, nonlinearity arises from terms that are not linear in perturbed quantities. From the gyrokinetic equation (\Cref{eq:GKE}), there is only a single term, $\langle \mathbf{v}_{\chi} \rangle_{\mathbf{R}_s} \cdot \nabla h_{s}$. Because both $\mathbf{v}_{\chi}$ and $h_s$ are proportional to perturbed quantities, this term is quadratic in perturbations. Taking the electrostatic form of $\mathbf{v}_{\chi}$, the nonlinearity is
\begin{equation}
  \langle\mathbf v_E\rangle_{\mathbf{R}_s} \cdot \nabla h_{s}
  = \frac{c}{B} \left( \hat{\mathbf b}\times\nabla_\perp  \langle\mathbf \delta \phi \rangle_{\mathbf{R}_s} \right) \cdot \nabla h_{s},
  \label{eq:electrostatic_NL}
\end{equation}
which is the $\mathbf E\times\mathbf B$ advection of the perturbed distribution $h_s$.  Because advection in real space corresponds to a convolution in Fourier space, this nonlinearity couples different wavenumbers in the plasma.

To see why the nonlinear term couples different modes, Fourier decompose $\delta \phi$ and the distribution function
\begin{equation}
  \langle\mathbf \delta \phi \rangle_{\mathbf{R}_s} = \sum_{\mathbf k'} \Phi_{\mathbf k'}\,e^{i\mathbf k'\cdot\mathbf {\mathbf{R}_s}},
  \quad
  h_s = \sum_{\mathbf k''} H_{\mathbf k''}\,e^{i\mathbf k''\cdot\mathbf {\mathbf{R}_s}}.
\end{equation}
Substituting into \Cref{eq:electrostatic_NL} gives
\begin{equation}
  \langle\mathbf v_E\rangle_{\mathbf{R}_s} \cdot \nabla h_{s}
  = \frac{c}{B}\sum_{\mathbf k',\mathbf k''}
    \bigl(\mathbf k'\times\mathbf k''\bigr)\!\cdot\!\hat{\mathbf b}\,
    \Phi_{\mathbf k'}\,H_{\mathbf k''}
    \,e^{\,i(\mathbf k'+\mathbf k'') \cdot \mathbf{R}_s}.
\end{equation}
Fourier transforming, $\mathcal F$, gives the nonlinear term for a single wavenumber $\mathbf{k}$
\begin{equation}
  \mathcal F\{\langle\mathbf v_E\rangle_{\mathbf{R}_s}\cdot\nabla h_s\}(\mathbf k)
  = \frac{c}{B}
    \sum_{\mathbf k'+\mathbf k''=\mathbf k}
    (\mathbf k'\times\mathbf k'')\!\cdot\!\hat{\mathbf b}\,
    \Phi_{\mathbf k'}\,H_{\mathbf k''}.
    \label{eq:NLsum}
\end{equation}
Therefore, the interpretation is that two ``pump'' modes \(\mathbf k'\) (from \(\phi\)) and \(\mathbf k''\) (from \(h_s\)) nonlinearly couple to the mode at \(\mathbf k=\mathbf k'+\mathbf k''\).

An important question is why the nonlinear term doesn't keep growing bigger and bigger over time in the presence of linear instabilities such as ITG that continuously pump energy into the perturbed fields from the equilibrium. Part of the answer is that the equilibrium gradients might eventually become sufficiently weak such that there is no further free energy available to drive instability. However, another interesting phenomenon is that the turbulence can self-regulate transport via zonal flows.

A zonal flow in gyrokinetics is a shear flow, i.e. a mode with purely radial wavenumber. The wavenumber $\mathbf{k}$ can be decomposed into a radial component $x$ and binormal component $y$,
\begin{equation}
\mathbf{k} = k_x \nabla x + k_y \nabla y.
\end{equation}
For a zonal flow, $\mathbf{k} = k_x \nabla x$, and therefore the sum in \Cref{eq:NLsum} must satisfy
\begin{equation}
k'_y = - k''_y.
\end{equation}
Consider coupling two drift‐wave modes that satisfy $k'_y = - k''_y$: \(\mathbf k'=(k'_x,k'_y)\) and \(\mathbf k''=(-k'_x,-k'_y)\). This injects energy into the zonal mode at \(\mathbf k=(2k'_x,0)\):
\[
\partial_t \Phi_{\rm ZF}(2k'_x)
\;\propto\;
\frac{c}{B}\sum_{k'_y\neq0}
(\mathbf k'\times(-\mathbf k'))\!\cdot\!\hat{\mathbf b}\;
\Phi_{\mathbf k'}\,H_{-\mathbf k'}
\;\sim\;
2\,k'_x k'_y\,\frac{c}{B}\,\Phi_{\mathbf k'}\,H_{-\mathbf k'}.
\]
Thus, nonlinear drift‐wave interactions can generate $k_y=0$ shear flows, which in turn can regulate turbulence by shearing apart eddies. There is an extensive literature on zonal flows \cite{Del2000,Mckee2003,Diamond2005,Itoh2006,Fujisawa2007,Zonca2015,Hillesheim2016b,Peterson2017,Zhu2021}.

A potentially very useful phenomenon has been observed in simulations of driven driven by certain instabilities such as ITG: the critical temperature gradient that the plasma can support (with a very weak energy source) is higher than the linear critical gradient. This is known as the Dimits shift, referring to the upshift in the nonlinear critical gradient \cite{Dimits2000}. Because critical temperature gradients are logarithmic, not linear, a small increase in the critical ion temperature gradient can lead to a significant increase in ion temperature in the plasma core, which 1) increases the core fusion reactivity and therefore fusion power, and 2) increases $T_i/T_e$ (which stabilizes ITG turbulence). There is extensive work on the Dimits shift \cite{Ross2002,Mikkelsen2008,Mantica_2009,Mantica2011,Howard2014b,Citrin2014b,Zhu2020,Zarazin2021,Hallenbert2021,Ivanov2022,Nakayama2022,Li2023,Nies2024}.

\subsection{Experimental Measurements} \label{sec:experimental}

Experimental validation of theoretical and computational transport and turbulence predictions is crucial for building next-step devices. 
For fusion machines we are interested in the transport resulting from turbulent fluctuations. In order for turbulent fluctuations to produce significant transport, they must broadly do three things: (1) have large fluctuation amplitudes, (2) have appropriate fluctuation cross-phases, and (3) be coherent (maintain an appropriate cross-phase) for a sufficiently long time. The cross-phase angle between fluctuating quantities $A$ and $B$ is
\begin{equation}
\alpha_{A,B} = \Theta_A - \Theta_B,
\end{equation}
where $\Theta_A = \arg(A)$ and $A = |A| \exp (i \Theta_A)$. The electrostatic heat flux has two components that arise from the relative fluctuations of the potential and temperature, and the of the potential and density \cite{Powers1974,Tynan2009},
\begin{equation}
Q \sim \sum_{k_y} k_y \left( n | \delta T_{k_y} | | \delta \phi_{k_y} | \gamma_{T,\phi} \sin \alpha_{T,\phi} + T | \delta n_{k_y} | | \delta \phi_{k_y} | \gamma_{n,\phi} \sin \alpha_{n,\phi} \right),
\end{equation}
where $ \delta T_{k_y}$, $\delta \phi_{k_y}$, and $\delta n_{k_y}$ are the Fourier coefficients of fluctuating temperature, potential, and density for a binormal wavenumber $k_y$. The quantities $\gamma_{T,\phi}$ and $\gamma_{n,\phi}$ are the coherencies for temperature-potential and density-potential fluctuations. Heat transport vanishes when the cross phase angle satisfies $\alpha_{A,B} = n \pi$ for an integer $n$. Simultaneously measuring cross-phases, coherencies, and their corresponding amplitudes in experiment is challenging \cite{Evensen1998,Hase1999,White2008}. The first comparison of gyrokinetic turbulence simulations with measurements of the cross phase was reported in \cite{White2010}. 

We briefly describe some turbulence imaging techniques routinely used in modern machines.
Beam‑emission spectroscopy (BES) \cite{Durst1992,Mckee1999,Schmitz2008,Smith2010,Field2012} detects electron density fluctuations by measuring D‑$\alpha$ light (Balmer-series emission from excited neutral deuterium atoms) produced by an injected neutral beam. Gas‑puff imaging (GPI) injects a hydrogen or helium puff and films the resulting atomic line emission with fast cameras, mapping two‑dimensional filamentary `blob' structures and drift‑wave eddies \cite{Zweben2002,Shesterikov2013,Zweben2017,Offeddu2022}. Doppler backscattering (DBS) launches a millimeter‑wave beam that backscatters from density perturbations to measure the fluctuation power \cite{Hennequin2006,Hillesheim2015,Rhodes2016,Molina2018,Hall2022}. Microwave‑imaging reflectometry (MIR) extends this concept to 2‑D by using an antenna array, producing real‑time movies of density fluctuations \cite{Mazzucato2001,Munsat2003,Park2003,Yamaguchi2006,Lee2014b}. Correlation electron‑cyclotron emission (CECE) cross‑correlates electron‑cyclotron emission to measure electron temperature fluctuations \cite{Deng2001,White2008b,Freethy2016,Liu2018}.

\subsection{Further Reading}

\begin{itemize}
\item Ion temperature gradient instability \cite{Rudakov1961,Lee1986,Nordman1990,Cowley1991,Hammett1990,Nunami2011}, nonlinear simulations \cite{Waltz1999,Rogers2000,Watanabe2005,Chang2009,Barnes2011,Holland2021}, experiment \cite{Greaves1992,Vershkov1999,Wade2000,White2008,White2010}.
\item Electron temperature gradient instability \cite{Lee1987,Told2008,Zocco2015,Parisi2020,Adkins2022,Ren2024}, nonlinear simulations \cite{Dorland2000,Colyer2017,Parisi2022,Chapman2022}, experiment \cite{Ren2012,Mantica2021}.
\item Trapped electron mode instability \cite{Hahm1991,Clauser2022}, nonlinear simulations \cite{Ernst2004,Dannert2005,Xiao2009,Hatch2015}, experiment \cite{White2008,White2010}.
\item Microtearing mode instability \cite{Gladd1980,Drake1980,Connor1990,Hatch2021,Hardman2023}, nonlinear simulations \cite{Guttenfelder2011,Doerk2011,Hatch2016,Jian2019}.
\item Kinetic ballooning mode instability \cite{Tang1980,Wang2012,Aleynikova2018,Kennedy2023p}, nonlinear simulations \cite{Pueschel2008,Maeyama2014,Ishizawa2019,Kumar2021,Mckinney2021}.
\item Parallel velocity gradient instability \cite{Catto1973,Schekochihin2012}, nonlinear simulations \cite{Barnes2011b,Mazzi2022}
\item Universal instability \cite{Lashinsky1964,Helander2015,Landreman2015,Costello2023}.
\item Multiscale turbulence \cite{Howard2014,Maeyama2015,Howard2016b,Maeyama2017,Choi2017,Hardman2019,Pueschel2020,Belli2022,Maeyama2024}, experiment \cite{Rhodes2011,Hillesheim2013}.
\item General gyrokinetic mode considerations \cite{Kotschenreuther2019}.
\end{itemize}

\section{Codes and Workflows} \label{sec:codesworkflows}

In this section, we describe a non-exhaustive selection of commonly used codes for turbulence, neoclassical, transport, and equilibrium calculations in tokamaks and stellarators. As discussed in \Cref{sec:KT_GK_Transp}, self-consistently solving the transport problem requires coupling equations that describe physical phenomena across a wide range of spatial and temporal scales.

\subsection{Gyrokinetics}

On fast timescales, gyrokinetic codes solve the system of equations described in \Cref{sec:gkequations}, either linearly or nonlinearly. These codes differ in two key ways: local versus global domains, and $\delta f$ versus full-$f$ formulations. Local codes compute turbulence within a narrow radial region (a flux-tube), while global codes simulate a larger radial extent. $\delta f$ codes assume fluctuations are small compared to the equilibrium ($\sim \rho_*$), whereas full-$f$ codes make no such assumption.

\textbf{Local $\delta f$ gyrokinetic codes:} CGYRO \cite{Candy2003b,Candy2016}, GENE \cite{Goerler2011}, GKX \cite{Peeters2009}, GS2 \cite{Dorland2000}, GX \cite{Mandell2024}, STELLA \cite{Barnes2019}.

\textbf{Global $\delta f$ gyrokinetic codes:} GENE \cite{Goerler2011}, ORB5 \cite{Lanti2020}, XGC \cite{Ku2018}.

\textbf{Global full-$f$ gyrokinetic codes:} GKEYLL \cite{Hakim2020}, GTC \cite{Lin1998}, XGC \cite{Ku2018}.

\subsection{Neoclassical Transport}

Several codes specialize in computing neoclassical transport in both tokamaks and stellarators. These typically solve the drift-kinetic equation with appropriate collision operators.

\textbf{Neoclassical codes:} FORTEC-3D \cite{Matsuoka2011}, NEO \cite{Belli2008b}, PERFECT \cite{Landreman2014}, SFINCS \cite{Landreman2014b}.

\subsection{Integrated Transport Solvers}

These codes evolve profiles over confinement timescales by coupling turbulence, neoclassical, and source models.

\textbf{Transport codes:} T3D \cite{sachdev2024trinity3d}, TANGO \cite{Di2022global}, TGYRO \cite{Candy2009}, TRANSP \cite{Pankin2024}, Trinity \cite{Barnes2009}.

\subsection{Magnetic Equilibrium Solvers}

Magnetic equilibrium codes provide the background geometry for both gyrokinetic and transport simulations. Different tools are used for tokamaks and stellarators.

\textbf{Tokamak equilibrium codes:} CHEASE \cite{Lutjens1996}, EFIT \cite{Lao1985}, Tokamaker \cite{Hansen2024}.

\textbf{Stellarator equilibrium codes:} DESC \cite{Panici2023}, VMEC \cite{Hirshman1983}.

\section{Tokamak Confinement Regimes} \label{sec:toka_confinement}

In this section, we describe the three main confinement regimes in tokamaks, H-mode, L-mode, and I-mode. 

\subsection{H-mode}

Until the early 1980s, tokamak experiments operated in a regime now known as L-mode, or low-confinement mode. In 1982, a landmark paper \cite{Wagner1982} reported ASDEX discharges that had roughly double the energy confinement time of similar L-mode discharges and much higher core plasma pressure. These high-performance discharges, later named `H-mode', bifurcated from the L-mode discharges once the plasma heating power exceeded a threshold value, subject to other parameters values such as magnetic field strength and plasma density.

\begin{figure}[tb]
\centering
    \includegraphics[width=\textwidth]{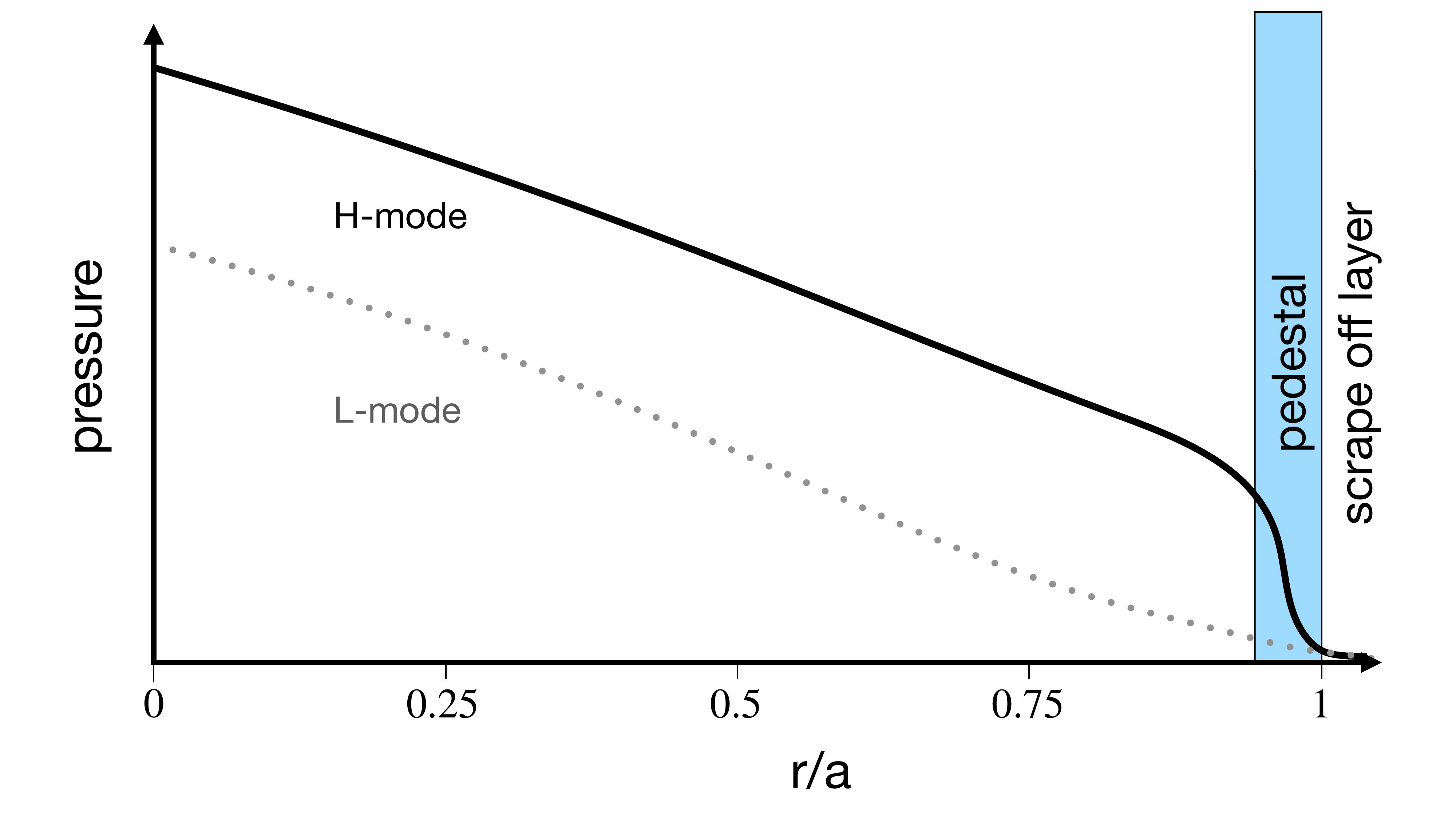}
 \caption{Cartoon of the different pressure profiles for H-mode and L-mode.}
\label{fig:Lmode_Hmode}
\end{figure}

H-mode has some very desirable characteristics, most of which can be attributed to the transport barrier that forms at the plasma edge called the pedestal, illustrated in \Cref{fig:Lmode_Hmode}. In the pedestal, the particle and density gradients are typically an order of magnitude or more higher than in L-mode discharges. This is particularly helpful because even if the critical gradients remain unchanged across the core profiles between L-mode and H-mode, the higher pressure at the pedestal top in H-mode can significantly increase the core pressure, which increases the fusion power. Second, the global energy confinement time roughly doubles in H-mode \cite{Wagner1982}, meaning that the energy diffusion coefficients decrease significantly across the plasma profiles. Third, in the highest-performing H-modes, plasma impurities (that radiate power and dilute the fuel) are flushed from the core \cite{Putterich2011} due to instabilities called edge-localized modes (ELMs) \cite{Zohm1996,Connor1998,Snyder2002,Kirk2004,Leonard2014}.

Unfortunately, ELMs are also a serious barrier to H-mode being a viable operating regime for fusion power plants. ELMs are periodic bursts of energy that can deposit tens of percent of the stored thermal plasma energy onto the divertor. While ELMs are tolerable in today's smaller machines, they are not tolerable in fusion power plants due to the much greater stored energy in the plasma \cite{Federici2019b,Creely2020,Muldrew2024,Maingi_2014,Hughes2020,Kuang2020,Viezzer2023}. There are signs of hope, however -- ELM-free or small-ELM regimes are a subject of intense investigation \cite{Viezzer2023}, which if successful could allow power plants to have higher core power density without ELMs. ELMs do have a useful feature, which is they flush out impurities that accumulate in the plasma core. Finding mechanisms for impurity flushing in the absence of ELMs is an essential research area.

\subsection{L-mode}

\begin{figure*}[bt!]
    \centering
    \begin{subfigure}[t]{0.49\textwidth}
    \includegraphics[width=0.99\textwidth]{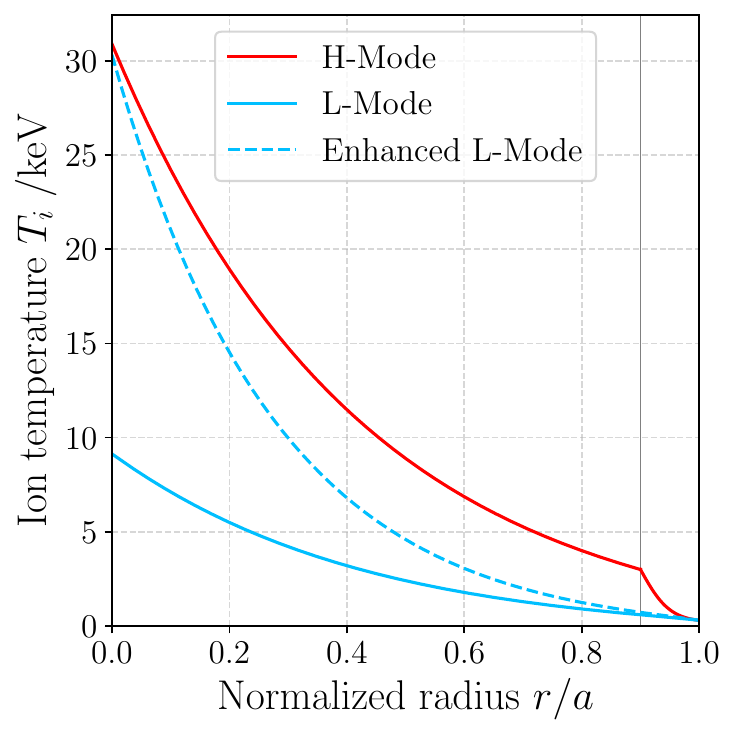}
    \caption{}
    \end{subfigure}
    \centering
    \begin{subfigure}[t]{0.49\textwidth}
    \includegraphics[width=0.99\textwidth]{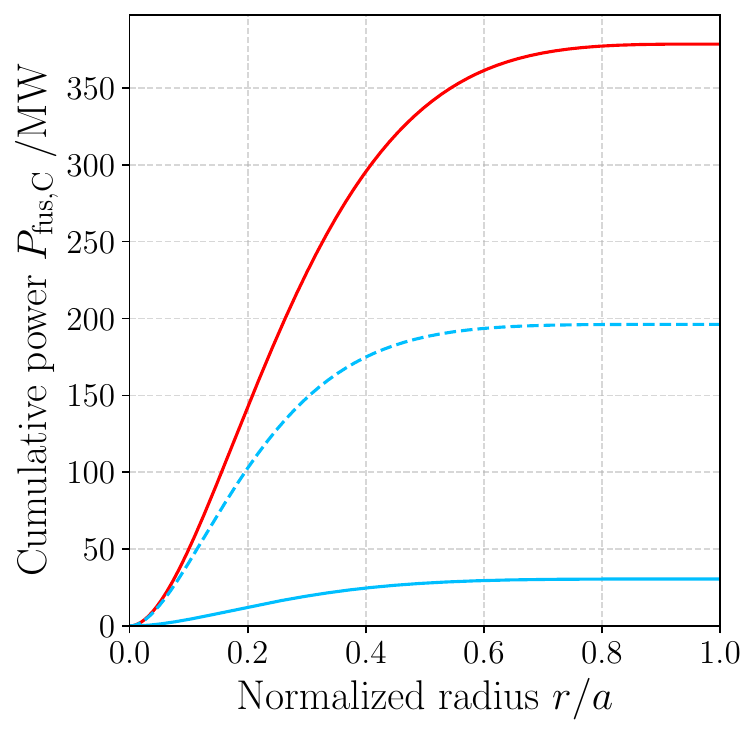}
    \caption{}
    \end{subfigure}
    \caption{(a) Calculated ion temperature profiles and (b) corresponding cumulative fusion power profiles for L-Mode, Enhanced L-mode, and H-mode (see \Cref{eq:PfusC}). For L-Mode and H-mode, we use $\alpha_\mathrm{stiff} = 3.0$, $\hat{\chi}_i = 1.0$, and $a/L_{T,i}^\mathrm{crit} = 2.4$. For Enhanced L-Mode, we use $\hat{\chi}_i = 0.5$ and $a/L_{T,i}^\mathrm{crit} = 3.6$. Density and geometry profiles are identical for all cases.}
    \label{fig:simple_example_L_H_mode}
\end{figure*}

Despite the advantages of H-mode, in the past decade there has been a renaissance for L-mode operation, the main driver being an operating mode called negative triangularity (NT) \cite{Austin2019}. The principal idea behind negative triangularity is that it forces the plasma to stay in H-mode. The leading hypothesis is that instabilities at the plasma edge prevent steep gradients from forming, preventing the emergence of a pedestal \cite{Nelson2022,Nelson2023,Nelson2024b,Nelson2024c}, although there are other explanations \cite{Burrell1992,Chankin1993,Schmitz2017,Bourdelle2014,Kramer2024}. While L-mode discharges cannot benefit from a high pedestal pressure, NT L-modes are reported to have certain transport advantages such as higher critical gradients, lower turbulent diffusivity, and lower plasma stiffness.

Recall that the primary motivation for H-mode is that it can significantly increase the total fusion power within the plasma,
\begin{equation}
P_\mathrm{fus} = \int p_\mathrm{fus} dV.  
\end{equation}
It is argued that H-mode is an efficient way to pack high power density regions into the plasma core -- by increasing pressure as quickly as possible in the plasma edge, there is a larger volume in the plasma core where $p_\mathrm{fus}$ is large. L-mode takes a different approach, which is to use higher critical gradients and lower turbulent transport to obtain a higher $p_\mathrm{fus}$ in the core.

Because the magnetic geometry of positive and negative triangularity also differs, the volume profile $V(r)$ can vary significantly \cite{parisi2025a}, especially at low aspect ratio -- negative triangularity configurations may enclose more volume near the magnetic axis, which is where $p_\mathrm{fus}$ is typically highest. This could have beneficial consequences for the total fusion power in L-mode discharges, even with the disadvantage of not having a pedestal.

We now perform a simple demonstration of the differences between L-mode and H-mode using the transport model in \Cref{sec:problem_description}. For H-mode, we set the ion temperature $T_i = 3.0$ keV at $r/a = 0.9$, whereas in L-mode we have no pedestal and use an edge temperature $T_i = 0.3$ keV. H-mode achieves nearly four times the core ion temperature despite the stiffness, turbulent diffusivity, and critical temperature gradient being identical to L-mode. We also show an Enhanced L-mode with half the turbulent diffusivity and a fifty percent higher critical gradient. Enhanced L-mode achieves a similar core temperature to H-mode but the total fusion power is roughly half that of H-mode. This demonstrates the importance of volume-integrated fusion power rather than the peak power density. We have assumed the same stiffness value, density profiles and geometry profiles for all three cases.

Finally, we mention a third promising and relatively under-explored regime, I-mode. I-mode has a temperature pedestal but not a density pedestal, and is generally ELM-free \cite{Greenwald1998,Ryter1998,Whyte2010,Hubbard_2017}. I-mode therefore has relatively long energy confinement time but short particle confinement time. A short particle confinement time can be useful for quickly flushing helium ash from the plasma core where it dilutes the fusion power.

\subsection{Prospects}

Where does this leave us for tokamak confinement regimes going forward? Machines such as ITER, designed in the 1980s and 1990s, required H-mode to obtain sufficient power density for them to obtain high plasma gain. However, given the risk of ELMs, even if the modeling demonstrates otherwise, we will not know whether we can produce FPP-relevant ELM-free H-modes until the experiment is operated, which are multi-billion dollar machines with long construction times. Given that we have new tools at our disposal to increase fusion power density such as high magnetic field \cite{Sorbom2015} and plasma shaping \cite{Nelson2023}, L-mode and I-mode merit continued investigation as serious alternatives to H-mode.

\subsection{Further Reading}

\begin{itemize}
\item Quasi-continuous exhaust regime \cite{Faitsch2021,Harrer_2022,Radovanovic_2022,Faitsch2023,Dunne2024}. 
\item Enhanced-pedestal H-mode \cite{Maingi2009,Maingi2010,Canik2013,Gerhardt_2014}.
\item Enhanced D-alpha H-mode \cite{Hubbard2001,LaBombard_2014,Gil_2020,Macwan2024}.
\item I-mode \cite{Greenwald1998,Ryter1998,Whyte2010,Hubbard_2017}. 
\item Quiescent H-mode \cite{Burrell2001,Sakamoto_2004, Suttrop_2005, Garofalo_2011,Ernst2024,Burrell2016,Chen2017b,Wilks2021b,Houshmandyar2022}.
\item Active ELM suppression strategies: resonant magnetic perturbations \cite{Kirk_2010}, vertical kicks \cite{Degeling_2003}, ELM pacing \cite{Evans_2004},  supersonic molecular beam injection \cite{LianghuaYao2001}.
\item L-mode \cite{Jardin2000,Kikuchi2019,Austin2019,Nelson2023,Paz-Soldan_2024,Wilson2024}.
\item Impurity flushing \cite{Putterich2011,Angioni2021}.
\end{itemize}

\section{Burning Plasmas} \label{sec:challenges}

In this section, we provide a survey on some of the most important outstanding confinement questions for burning plasmas \cite{Furth1990,Green2003,Gorelenkov2014,Hawryluk2019,National2021,Fasoli2023,Meyer2024}. A burning plasma is one where the fusion products constitute a significant fraction of the total plasma heating. For D-T fusion, alpha heating constitutes 50\% of all heating when the plasma gain $Q^\mathrm{sci}  = 5$ (recall $Q^\mathrm{sci} = P_\mathrm{fus} / P_\mathrm{ext}$ where $P_\mathrm{fus}$ is the fusion power and $P_\mathrm{ext}$ is the external heating power) and constitutes 100\% of all heating when the plasma has ignited ($Q^\mathrm{sci}  = \infty$). The burning regime represents a new regime not yet seen in MCF plasmas (as of 2025) -- every MCF machine so far has operated with $Q^\mathrm{sci} < 1$.

Why is this significant? The first reason is that there are new couplings between transport and sources in burning plasmas due to alpha particles produced from D-T fusion reactions. The higher the core temperature and density, the stronger the alpha heating source. And a stronger alpha heating source leads to more alpha heating and a range of fast particle instabilities. Second, the alphas are a particle source that dilute the main ion species, which reduces the fusion power. Coupling these effects is an unsolved problem because of its immense computational and theoretical complexity and uncertainty. One limiting factor is that the effect of fast particles on transport is complicated and no experiments exist for burning plasmas where we can benchmark, although there is much work from TFTR \cite{Strachan1994} and JET \cite{Keilhacker1999}, as of 2025, the only two magnetic confinement plasmas to produce significant fusion power. A recent study \cite{Wade2021} showed the capital cost of a fusion power plant is most sensitive to the energy confinement quality factor $H$ (see \Cref{eq:Hconfinementquality}). This is reason both for both caution and optimism: if burning plasmas have improved energy confinement, they could be cheaper and faster to build. But if confinement is degraded, they could be more expensive and slower. This is an area where advances in transport physics and energy confinement could be a large lever for accelerating the deployment of fusion power.

Another interesting question for burning plasmas is particle transport. A burning plasma burns deuterium and tritium fuel to produce fusion power. A critical engineering and regulatory question for fusion power is the quantity of tritium onsite, called the tritium inventory. Reducing the tritium inventory size can significantly ease engineering and regulatory requirements. It is predicted that the most effective way to decrease tritium inventory is to increase the tritium burn efficiency \cite{Abdou2021,Meschini2023,Whyte2023}. Tritium burn efficiency can be increased by increase tritium particle confinement time \cite{Abdou2021,Boozer2021,Boozer2021b,Whyte2023,Parisi2025b}, more rapidly removing helium ash from the core \cite{Reiter1990,Whyte2023}, increasing the reactivity of fusion fuel \cite{Kulsrud1986,Parisi_2024d}, and increasing fueling efficiency \cite{Tanabe2017,Abdou2021}. Similar to the discussion of energy confinement above, these are largely areas where plasma physics can drive progress.

While we largely avoided discussion of momentum transport and plasma rotation, this is an area of very high uncertainty for MCF power plant design because energy confinement is sensitive to rotation and generally benefits from higher rotation \cite{Burrell1997,Barnes2011b}. The largest driver of plasma rotation in modern machines is with neutral beam injection (NBI). However, it is unclear whether NBI will be used in MCF power plants, and even if it were, it is unlikely it will drive as much rotation as it currently does in current devices. This is important, because in the absence of NBI and other momentum inputs, it is predicted that the size of self-generated plasma flows scales with $\rho_{*} = \rho / a$ \cite{Parra2015}. This is pessimistic. Because future machines will be larger and have higher magnetic field strength (smaller $\rho_{*}$) than current machines, the available rotation and rotation shear will be much lower than is currently available.

\section{Other Important Topics} \label{sec:otherimportant}

In this section, I briefly summarize some topics that are important, but for which I did not have time to include.

\textbf{Reduced Models.} Using gyrokinetics to calculate the turbulent fluxes in the transport equations can be prohibitively expensive computationally. In order to reduce computational cost, reduced models for the turbulent fluxes have been developed \cite{Connor1995,Waltz1998,Staebler2007,Podesta2014,Bourdelle2015,Muraca2023,Hatch2024,Giacomin2025} in order to allow larger parameter scans and lower computational cost.

\textbf{Plasma Geometry.} Tokamaks and stellarators come in a wide range of shapes and sizes. These different geometric configurations have a substantial effect on confinement \cite{Goldston1984, Kaye1985, Turnbull1999, Belli2008, Marinoni2009, Ball2015, Merle2017, Austin2019, Nelson2023, Balestri_2024, Wilson2024, Berkery2024, Kim2024, Landreman2025, Parisi2025predictionelmfreeoperationspherical}.

\textbf{Scrape-Off-Layer (SOL) Turbulence.} Beyond the last-closed-flux-surface, magnetic field lines are \textit{open}: following a magnetic field line from just beyond the last-closed-flux-surface will intersect with a plasma wall. This magnetic field configuration is distinct from within the  last-closed-flux-surface, where field lines are closed and trace out flux surfaces. The SOL transport can have a significant impact on plasma exhaust and core plasma fueling \cite{Xu1998,Ono_2000,Terry2003,Fundamenski_2007,Ernst2024,Zhang2024b,Shanahan_Bold_Dudson_2024,Peret_2025}.

\textbf{Experimental Confinement Scalings.} A useful approach for predicting future tokamak energy confinement time is based on database analysis of past experiments. There are many scalings \cite{Goldston1984,Kaye1985,Yushmanov1990,Stroth1996,Kaye1997,Cordey_2005,Verdoolaege_2021}. A big question is the accuracy of the current scaling laws as future experiments enter new parameter regimes.

\textbf{Fast Particles.} Fast particles (with energy $E \gtrsim $ 1 MeV) have been reported to both drive and suppress turbulence and other instabilities \cite{Porcelli1991,Heidbrink2008,Romanelli2010,Di2018fast,Mazzi2022enhanced,Garcia2022electromagnetic}. In burning plasmas, fast particles will constitute a significant fraction of the plasma energy and density \cite{Heidbrink2002,Gorelenkov2014}. Self-consistently including the effect of fast particles in the system of transport equations discussed in \Cref{sec:KT_GK_Transp} is challenging.

\textbf{Machine Learning.} Machine learning approaches are used commonly in confinement research, for example in accelerating turbulence simulations \cite{Ma2020,Van2020}, magnetic equilibrium reconstruction \cite{Felici2011,Lao2022,Candido2023}, plasma heating modeling \cite{Wallace2022,Sanchez2024}, plasma stability prediction \cite{Piccione2020,Piccione2022} and profile prediction \cite{Boyer2021,Abbate2021,Dubbioso2023}, among other topics \cite{Smith2013,Landreman2025}.

\section{Summary} \label{sec:summary}

In this tutorial, I have attempted to provide an accessible entry point for students and researchers interested in plasma stability and transport in magnetic confinement plasmas. 

\Cref{sec:problem_description} motivated the transport problem with a simple example. We demonstrated how apparently small changes in energy transport properties could lead to big changes in the total fusion power. \Cref{sec:introduction} showed how the physics of nuclear fusion reactions and Coulomb collisions generally require a hot plasma and a confinement scheme. The deuterium-tritium nuclear fusion cross section plays a central role in determining the temperature and confinement schemes required for a successful fusion power plant. \Cref{sec:Lawson} discussed the Lawson Criterion for energy confinement and an analogous argument for particle confinement. We found the minimum density and confinement time required for fusion ignition and the minimum particle sources for a plasma with a given fusion power. \Cref{sec:confimentschemes} discussed the basics of magnetic confinement and introduced the coordinate system for tokamaks. \Cref{sec:transport_overview} estimated the confinement times of a tokamak discharge based on classical, neoclassical, and turbulent transport. We then found an approximate scaling for the minimum power plant size on the energy confinement quality. \Cref{sec:KT_GK_Transp} introduced the main equations required for finding the magnetic equilibrium and plasma profiles on long timescales. We described some features of the gyrokinetic, neoclassical, transport, and magnetic equilibrium equations. \Cref{sec:linearstab} covered some of gyrokinetic instabilities giving rise to turbulent transport in tokamaks and stellarators. We briefly discussed nonlinear physics and turbulence measurements. A detailed derivation of the ion-temperature-gradient instability is provided in \Cref{app:ITGstab}. \Cref{sec:codesworkflows} introduced some commonly used numerical tools used to solve the transport problem. \Cref{sec:toka_confinement} described the most common tokamak confinement regimes. We compared H-Mode and L-Mode fusion power profiles.  \Cref{sec:challenges} discussed some of the challenges and opportunities in burning plasmas. \Cref{sec:otherimportant} listed some important topics that I didn't have time to describe in detail.

\section{Acknowledgements}

I am indebted to Felix Parra, Michael Barnes, Steve Cowley, Paul Dellar, Walter Guttenfelder, Stan Kaye, Colin Roach, and Alex Schekochihin for their instruction in turbulence and transport. This work was supported by the U.S. Department of Energy under contract numbers DE-AC02-09CH11466, DE-SC0022270, DE-SC0022272. The United States Government retains a non-exclusive, paid-up, irrevocable, world-wide license to publish or reproduce the published form of this manuscript, or allow others to do so, for United States Government purposes.

\section{Data Availability}

Data sharing is not applicable to this article as no new data were created or analyzed in this study.

\appendix

\section{Appendix: ITG Instability} \label{app:ITGstab}

In this section we derive a dispersion relation for linear ion-temperature-gradient (ITG) instability in toroidal geometry. The goal is to find an expression for the complex frequency and the critical gradient.

We keep the effects of both magnetic drifts and parallel streaming, whose effects correspond to `toroidal' and `slab' ITG respectively. The slab ITG of this derivation closely follows \cite{Parra_2019_driftwaves,HammettParra2025} and is based on lecture courses given at Massachusetts Institute of Technology and Princeton University by F.I. Parra.

We work in the electrostatic limit where $\delta \mathbf{A}$ fluctuations are neglected, we neglect collisions, and we neglect the nonlinear term. We assume a single ion species. In this limit, the non-adiabatic distribution function from \Cref{eq:GKE} is
\begin{equation}
h_i = \frac{\omega - { \omega}_{*i} \left[ 1 + \eta_i \left( \hat{v}^2_{\parallel } +  \hat{v}^2_{\perp} - \frac{3}{2} \right) \right]  }{\omega - { k_{\parallel }} v_{\parallel } - \omega_{\kappa i}  \hat{v}_{\parallel}^2 - {\omega}_{Bi} \hat{v}_{\perp}^2/2 } J_0 \left( \frac{k_{\perp} v_{\perp}}{\Omega_i} \right) \frac{Z_i e \delta \phi}{T_i} F_{Mi}.
\label{eq:hs_general}
\end{equation}
To make analytic progress, we order the parallel streaming as much faster than the magnetic drifts (see \cite{Ivanov2023} where this was not assumed),
\begin{equation}
|\omega - k_{\parallel } v_{\parallel} | \gg | \omega_{\kappa i} \hat{v}_{\parallel}^2 + {\omega}_{Bs} \hat{v}_{\perp}^2/2|.
\label{eq:ordering_ITG}
\end{equation}
Using \Cref{eq:ordering_ITG}, the ion distribution function assuming $Z_i = 1$ is
\begin{equation}
h_i = \frac{\omega \tau + { \omega}_{*e} \left[ 1 + \eta_i \left( \hat{v}^2_{\parallel } +  \hat{v}^2_{\perp} - \frac{3}{2} \right) \right]  }{\omega - k_{\parallel } v_{\parallel} } J_0 \left( \frac{k_{\perp} v_{\perp}}{\Omega_i} \right) \frac{e \delta \phi}{T_e} F_{Mi} \left( 1 + \frac{A}{\omega - k_{\parallel } v_{\parallel} } + \ldots \right),
\label{eq:hi_ITG}
\end{equation}
where
\begin{equation}
A \equiv \omega_{\kappa s} \hat{v}_{\parallel}^2 + {\omega}_{Bs} \hat{v}_{\perp}^2/2, \;\;\;\;\;\; \tau \equiv \frac{ T_e }{ T_i}.
\end{equation}
For this analysis of ITG instability, we limit ourselves to the perpendicular scales $k_{\perp} \rho_e \ll 1$. At these scales, we expect $|k_{\parallel } v_{\parallel}| \gg  \omega_{*e}, \omega$. Using this assumptions, the electrons are `adiabatic': $h_e = 0$. Physically, this corresponds to electrons rapidly streaming along field lines with a modified Boltzmann response,
\begin{equation}
\delta f_e = - \frac{e \delta \phi}{T_e} F_{Me}.
\label{eq:deltafe_ITG}
\end{equation}
The perturbed quasineutrality equation is
\begin{equation}
\sum Z_s \delta n_s = 0.
\end{equation}
Using \Cref{eq:hi_ITG,eq:deltafe_ITG}, perturbed quasineutrality is
\begin{equation}
\sum Z_s \delta n_s = \sum Z_s \int \delta f_s d^3 v = - \frac{e \delta \phi}{T_i} \int F_{Mi} d^3v + \int h_{i} d^3v - \frac{e \delta \phi}{T_e} \int F_{Me} d^3v,
\label{eq:perturbed_QN_ITG}
\end{equation}
We also consider quasineutrality for the equilibrium density,
\begin{equation}
\sum Z_s n_s = 0,
\end{equation}
which gives $n_i = n_e$. Therefore, perturbed quasineutrality in \Cref{eq:perturbed_QN_ITG} is
\begin{equation}
\int h_{i} d^3v - \frac{e \delta \phi}{T_e} n_e \left( 1 + \tau \right) = 0.
\label{eq:QN_ITG}
\end{equation}
The next step is to split the integral for $h_i$ into three parts corresponding to the term independent of magnetic drifts, the curvature drift term, and the grad-B drift term,
\begin{equation}
\int h_{i} d^3v = \int T_1 d^3v + \int T_1 \frac{\omega_{\kappa i} }{\omega - k_{\parallel } v_{\parallel}} \hat{v}_{\parallel}^2 d^3v + \int T_1 \frac{\omega_{\nabla B i} }{\omega - k_{\parallel } v_{\parallel}} \frac{ \hat{v}_{\perp}^2}{2} d^3v
\label{eq:ITG_QN_three_integrals}
\end{equation}
where
\begin{equation}
T_1 = \frac{\omega \tau + { \omega}_{*e} \left[ 1 + \eta_s \left( \hat{v}^2_{\parallel } +  \hat{v}^2_{\perp} - \frac{3}{2} \right) \right]  }{\omega - k_{\parallel } v_{\parallel}} J_0 \left( \frac{k_{\perp} v_{\perp}}{\Omega_s} \right) \frac{e \delta \phi}{T_s} F_{Ms}.
\end{equation}

\subsubsection{Useful Integrals} \label{subapp:usefulintegrals}

Before evaluating the three integrals in \Cref{eq:ITG_QN_three_integrals}, we introduce a few useful relations. The first is the well-known plasma dispersion function \cite{Faddeeva_1961}
\begin{equation}
Z(\zeta) = \frac{1}{\sqrt{\pi}} \int^{\infty}_{-\infty} \frac{dx e^{-x^2}}{x- \zeta},
\end{equation}
for $\mathrm{Im}(\zeta) > 0$ where
\begin{equation}
\zeta \equiv \frac{\omega}{k_{\parallel } v_{ts}} \;\;\;\;\;\;\; x \equiv \frac{v_{\parallel}}{v_{ts}}.
\end{equation}
Because we will taking integrals with moments of $v_{\parallel}$, it is also helpful to use
\begin{equation}
\frac{1}{\sqrt{\pi}} \int^{\infty}_{-\infty} \frac{dx x^2 e^{-x^2}}{x- \zeta} = 2\zeta + \zeta^2 Z(\zeta), \;\;\;\; \frac{1}{\sqrt{\pi}} \int^{\infty}_{-\infty} \frac{dx x^4 e^{-x^2}}{x- \zeta} = 6\zeta^2 + \zeta^4 Z(\zeta).
\end{equation}
There will also be integrals with two powers of $\omega - k_{\parallel} v_{\parallel}$ in the denominator, giving integrals of the form
\begin{equation}
\frac{1}{\sqrt{\pi}} \int_{-\infty}^\infty \frac{x^a e^{-x^2}}{(x - \zeta)^{2}} \, dx,
\end{equation}
where $a \geq 0$ is an even integer. To handle these integrals, we take derivatives of the dispersion function. The n-th derivative of $Z(\zeta)$ is
\begin{equation}
Z^{(n)}(\zeta) = (-1)^n n! \frac{1}{\sqrt{\pi}} \int_{-\infty}^\infty \frac{e^{-x^2}}{(x - \zeta)^{n+1}} \, dx,
\end{equation}
which give the handy recurrence relation valid for $n \geq 1$,
\begin{equation}
Z^{(n+1)}(\zeta) = -2 \left( Z^{(n)}(\zeta) + \zeta Z^{(n-1)}(\zeta)  \right),
\end{equation}
where
\begin{equation}
Z^{(0)}(\zeta) = Z(\zeta), \;\;\; Z^{(1)}(\zeta) = - 2 \left( 1 + \zeta Z(\zeta) \right).
\end{equation}
Therefore, the zeroth moment is
\begin{equation}
\frac{1}{\sqrt{\pi}} \int_{-\infty}^\infty \frac{e^{-x^2}}{(x - \zeta)^{2}} \, dx = 2 \left( 1 + \zeta Z(\zeta) \right).
\end{equation}
After some more algebra, we find the second moment
\begin{equation}
\frac{1}{\sqrt{\pi}} \int_{-\infty}^\infty \frac{x^2 e^{-x^2}}{(x - \zeta)^{2}} \, dx  = 1 + 2 \zeta Z - 2 \zeta^2- 2 \zeta^3 Z,
\end{equation}
and the fourth moment
\begin{equation}
\frac{1}{\sqrt{\pi}} \int_{-\infty}^\infty \frac{x^4 e^{-x^2}}{(x - \zeta)^{2}} \, dx  = \frac{1}{2}  + 3 \zeta^2 + 4 \zeta^3 Z - 2 \zeta^4 - 2 \zeta^5 Z.
\end{equation}

Another useful integral involves the Bessel functions,
\begin{equation}
\int_0^{\infty} y J_0^2 \left( a y \right) \exp \left( - b y^2 \right) dy = \frac{1}{2b} \Gamma_0 \left( \frac{a^2}{2b} \right),
\end{equation}
its cubic moment,
\begin{equation}
\int_0^{\infty} y^3 J_0^2 \left( a y \right) \exp \left( - b y^2 \right) dy = \frac{ \left( 2b - a^2 \right)  \Gamma_0 \left( \frac{a^2}{2b} \right) + a^2 \Gamma_1 \left( \frac{a^2}{2b} \right)}{4b^3},
\end{equation}
and its quintic moment,
\begin{equation}
\int_0^{\infty} y^5 J_0^2 \left( a y \right) \exp \left( - b y^2 \right) dy = \frac{ \left( 2b - a^2 \right)^2  \Gamma_0 \left( \frac{a^2}{2b} \right) + a^2 \left( 3b - a^2 \right) \Gamma_1 \left( \frac{a^2}{2b} \right)}{4b^5}.
\end{equation}


\subsubsection{Dispersion Relation}

Using the integral expressions in \Cref{subapp:usefulintegrals}, the first integral term in \Cref{eq:ITG_QN_three_integrals} is 
\begin{equation}
\begin{aligned}
& \frac{1}{n_e} \frac{T_e}{e \delta \phi}  \int T_1 d^3v = \\
& - Z (\zeta_i) \Gamma_0 \left( \tau \zeta_i + \frac{\omega_{*e} }{k_{\parallel } v_{ti}} \left( 1 - \frac{3}{2} \eta_i \right)  \right) \\ 
& - \Gamma_0 \frac{\omega_{*e} }{k_{\parallel } v_{ti}} \eta_i \left( \zeta_i + \zeta_i^2 Z(\zeta_i) \right) \\
& - \frac{\omega_{*e} }{k_{\parallel } v_{ti}} \eta_i Z (\zeta_i) \left( \Gamma_0 + b_i \left( \Gamma_1 - \Gamma_0 \right)  \right).
\end{aligned}
\label{eq:first_ITG_QN}
\end{equation}
The second integral is $Z(\zeta_i) \to 1 + 2 \zeta_i Z + 2 \zeta^2_i+ 2 \zeta_i^3 Z(\zeta_i)$ or $Z(\zeta_i) \to 1 + 6 \zeta_i^2 + 4 \zeta_i^3 Z(\zeta_i) + 2 \zeta_i^4 + 2 \zeta_i^5 Z(\zeta_i)$
\begin{equation}
\begin{aligned}
& \frac{1}{n_e} \frac{T_e}{e \delta \phi}  \int T_1 \frac{\omega_{\kappa i} }{\omega - k_{\parallel} v_{\parallel }} \hat{v}_{\parallel}^2 d^3v = \\
& - \left[ 1 + 2 \zeta_i Z - 2 \zeta^2_i- 2 \zeta_i^3 Z(\zeta_i)) \right] \Gamma_0 \left( \tau \zeta_i + \frac{\omega_{*e} }{k_{\parallel } v_{ti}} \left( 1 - \frac{3}{2} \eta_i \right)  \right) \frac{\omega_{\kappa i} }{k_{\parallel } v_{ti}} \\
& - \Gamma_0 \frac{\omega_{*e} }{k_{\parallel } v_{ti}} \eta_i \left( \frac{1}{2}  + 3 \zeta^2_i+ 4 \zeta^3_i Z - 2 \zeta^4_i - 2 \zeta^5_i Z \right) \frac{\omega_{\kappa i} }{k_{\parallel } v_{ti}} \\
& - \frac{\omega_{*e} }{k_{\parallel } v_{ti}} \eta_i \left[ 1 + 2 \zeta_i Z - 2 \zeta^2_i- 2 \zeta_i^3 Z(\zeta_i) \right] \left( \Gamma_0 + b_i \left( \Gamma_1 - \Gamma_0 \right)  \right) \frac{\omega_{\kappa i} }{k_{\parallel } v_{ti}}.
\end{aligned}
\label{eq:second_ITG_QN}
\end{equation}
The third integral is $\Gamma_0 \to \Gamma_0 + b_i \left( \Gamma_1 - \Gamma_0 \right) $, $Z(\zeta_i) \to 2 \left( 1 + \zeta_i Z(\zeta_i) \right)$, $Z(\zeta_i) \to 1 + 2 \zeta_i Z + 2 \zeta^2_i+ 2 \zeta_i^3 Z(\zeta_i)$, and
\begin{equation}
\Gamma_0 + b_i \left( \Gamma_1 - \Gamma_0 \right) \to 4 \left[ \Gamma_0 + b_i \left( \frac{3}{2} \Gamma_1 - 2 \Gamma_0 \right) + b_i^2 \left( \Gamma_0 - \Gamma_1 \right)  \right],
\end{equation}
giving
\begin{equation}
\begin{aligned}
& \frac{1}{n_e} \frac{T_e}{e \delta \phi}  \int T_1 \frac{1}{2} \frac{\omega_{\nabla B i} }{\omega - k_{\parallel} v_{\parallel }} \hat{v}_{\perp}^2 d^3v = \\
& - 2 \left( 1 + \zeta_i Z(\zeta_i) \right) \left( \Gamma_0 + b_i \left( \Gamma_1 - \Gamma_0 \right) \right)  \left( \tau \zeta_i + \frac{\omega_{*e} }{k_{\parallel } v_{ti}} \left( 1 - \frac{3}{2} \eta_i \right)  \right) \frac{\omega_{\nabla B i} }{2 k_{\parallel } v_{ti}} \\
& -  4 \left[ \Gamma_0 + b_i \left( \frac{3}{2} \Gamma_1 - 2 \Gamma_0 \right) + b_i^2 \left( \Gamma_0 - \Gamma_1 \right)  \right] \frac{\omega_{*e} }{k_{\parallel } v_{ti}} \eta_i \left( 1 + 2 \zeta_i Z - 2 \zeta^2_i- 2 \zeta_i^3 Z(\zeta_i) \right) \frac{\omega_{\nabla B i} }{2 k_{\parallel } v_{ti}} \\
& - \frac{\omega_{*e} }{k_{\parallel } v_{ti}} \eta_i 2 \left( 1 + \zeta_i Z(\zeta_i) \right) \left( \Gamma_0 + b_i \left( \Gamma_1 - \Gamma_0 \right)  \right) \frac{\omega_{\nabla B i} }{2 k_{\parallel } v_{ti}}.
\end{aligned}
\label{eq:third_ITG_QN}
\end{equation}

Putting it all together, the dispersion relation from \Cref{eq:QN_ITG} is 
\begin{equation}
\begin{aligned}
& 1 + \tau  \left( 1 +  \Gamma_0 \zeta_i Z (\zeta_i) \right) \\ 
& + \frac{\omega_{*e} }{k_{\parallel } v_{ti}} \Gamma_0 Z (\zeta_i) \left( 1 - \frac{3}{2} \eta_i \right) \\
& + \frac{\omega_{*e} }{k_{\parallel } v_{ti}} \eta_i \Gamma_0 \left( \zeta_i + \zeta_i^2 Z(\zeta_i) \right) \\
& +  \frac{\omega_{*e} }{k_{\parallel } v_{ti}} \eta_i \left( \Gamma_0 + b_i \left( \Gamma_1 - \Gamma_0 \right)  \right) Z (\zeta_i)  \\
& + \frac{\omega_{\kappa i} }{k_{\parallel } v_{ti}} \Gamma_0 \left[ 1 + 2 \zeta_i Z - 2 \zeta^2_i- 2 \zeta_i^3 Z(\zeta_i)) \right]  \left( \tau \zeta_i + \frac{\omega_{*e} }{k_{\parallel } v_{ti}} \left( 1 - \frac{3}{2} \eta_i \right)  \right)  \\
& + \frac{\omega_{\kappa i} }{k_{\parallel } v_{ti}} \frac{\omega_{*e} }{k_{\parallel } v_{ti}} \eta_i \Gamma_0 \left( \frac{1}{2}  + 3 \zeta^2_i+ 4 \zeta^3_i Z - 2 \zeta^4_i - 2 \zeta^5_i Z \right) \\
& + \frac{\omega_{\kappa i} }{k_{\parallel } v_{ti}} \frac{\omega_{*e} }{k_{\parallel } v_{ti}} \eta_i \left( \Gamma_0 + b_i \left( \Gamma_1 - \Gamma_0 \right)  \right) \left[ 1 + 2 \zeta_i Z - 2 \zeta_i^2 - 2 \zeta_i^3 Z(\zeta_i) \right] \\
& + \frac{\omega_{\nabla B i} }{ k_{\parallel } v_{ti}} \left( \Gamma_0 + b_i \left( \Gamma_1 - \Gamma_0 \right) \right) \left( 1 + \zeta_i Z(\zeta_i) \right)  \left( \tau \zeta_i + \frac{\omega_{*e} }{k_{\parallel } v_{ti}} \left( 1 - \frac{3}{2} \eta_i \right)  \right) \\
& + 2 \frac{\omega_{\nabla B i} }{ k_{\parallel } v_{ti}} \frac{\omega_{*e} }{k_{\parallel } v_{ti}} \eta_i \left[ \Gamma_0 + b_i \left( \frac{3}{2} \Gamma_1 - 2 \Gamma_0 \right) + b_i^2 \left( \Gamma_0 - \Gamma_1 \right)  \right] \left( 1 + 2 \zeta_i Z - 2 \zeta_i^2 - 2 \zeta_i^3 Z(\zeta_i) \right)  \\
& + \frac{\omega_{\nabla B i} }{ k_{\parallel } v_{ti}} \frac{\omega_{*e} }{k_{\parallel } v_{ti}} \eta_i \left( \Gamma_0 + b_i \left( \Gamma_1 - \Gamma_0 \right)  \right) \left( 1 + \zeta_i Z(\zeta_i) \right) = 0.
\end{aligned}
\label{eq:ITG_disp_rln_full}
\end{equation}

\subsection{Toroidal ITG Instability}

We now calculate the toroidal ITG instability by taking the limit
\begin{equation}
|\zeta_i| \gg 1,
\end{equation}
and by ordering
\begin{equation}
\omega_{*e} \sim \zeta_i.
\end{equation}
In the $|\zeta_i| \gg 1$ limit, the plasma dispersion function can be expanded as
\begin{equation}
Z (\zeta_i) \sim - \frac{1}{\zeta_i} - \frac{1}{2\zeta_i^3} - \frac{3}{4\zeta_i^5} - \frac{15}{8 \zeta_i^7}  + \ldots.
\end{equation}
In this limit, we will expand the frequently occurring expressions with $Z(\zeta_i)$.
\begin{equation}
1 + \zeta_i Z(\zeta_i) \sim - \frac{1}{2\zeta_i^2} - \frac{3}{4 \zeta_i^3} +  \ldots,
\end{equation}
\begin{equation}
\zeta_i + \zeta_i^2 Z(\zeta_i) \sim - \frac{1}{2\zeta_i} - \frac{3}{4 \zeta_i^2} +  \ldots,
\end{equation}
\begin{equation}
1 + 2 \zeta_i Z (\zeta_i) - 2 \zeta^2_i- 2 \zeta_i^3 Z(\zeta_i)) \sim \frac{1}{2 \zeta_i^2} + \ldots, 
\end{equation}
\begin{equation}
\frac{1}{2}  + 3 \zeta^2_i+ 4 \zeta^3_i Z - 2 \zeta^4_i - 2 \zeta^5_i Z \sim \frac{3}{4 \zeta_i^2} + \ldots .
\end{equation}
Substituting these expressions into quasineutrality (\Cref{eq:ITG_disp_rln_full}), and for further simplification taking the long-wavelength limit $b_i \to 0$, $\Gamma_0 \to 1$, $\Gamma_1 \to 0$, gives
\begin{equation}
\begin{aligned}
& \omega^2 - \omega  \omega_{*e}  + \frac{ \omega_{\kappa i}  \omega_{*e}}{2} \left( \eta_i + 1 \right) + \frac{ \omega_{\nabla B i} \omega_{*e}}{2} \left( \eta_i - 1 \right) = 0.
\end{aligned}
\label{eq:ITG_disp_rln_toroidal_limit_driftkinetic_simp}
\end{equation}
Finally, it is often assumed that the grad-B and curvature drift frequencies are equal, $\omega_{\nabla B i} = \omega_{\kappa i}$. In this limit, \Cref{eq:ITG_disp_rln_toroidal_limit_driftkinetic_simp} becomes
\begin{equation}
\begin{aligned}
& \omega^2 - \omega \omega_{*e} + \omega_{\kappa i}  \omega_{*e} \eta_i = 0.
\end{aligned}
\label{eq:ITG_disp_rln_toroidal_limit_driftkinetic_simp_equaldrifts}
\end{equation}
The frequency in \Cref{eq:ITG_disp_rln_toroidal_limit_driftkinetic_simp_equaldrifts} has the solution
\begin{equation}
 \omega = \frac{ \omega_{*e}\pm \sqrt{ \left( \omega_{*e} \right)^2 - 4 \omega_{\kappa i}  \omega_{*e} \eta_i }}{2},
\end{equation}
with instability arising when
\begin{equation}
4 \omega_{\kappa i} \eta_i > \omega_{*e},
\end{equation}
with a growth rate
\begin{equation}
\gamma = \frac{ \sqrt{ 4 \omega_{\kappa i}  \omega_{*e} \eta_i - \left( \omega_{*e} \right)^2 }}{2}.
\label{eq:gamma_tor}
\end{equation}
The growth rate in \Cref{eq:gamma_tor} has some curious features. In the limit where $\eta_i \gg 1$, the ion temperature gradient is much steeper than density gradient. Assuming
\begin{equation}
\omega_{*e} \sim \omega_{\kappa i}, \;\;\;\; \eta_i \gg 1,
\end{equation}
and approximating the magnetic drift frequency as
\begin{equation}
\omega_{\kappa i} \approx k_y \rho_i \frac{v_{ti}}{R}, 
\end{equation}
the growth rate is approximately
\begin{equation}
\gamma \approx \sqrt{ \omega_{\kappa i}  \omega_{*e} \eta_i} \simeq k_y \rho_i \sqrt{\frac{v_{ti}^2}{R L_{T_i}} }.
\label{eq:gamma_toroidal_ITG}
\end{equation}
Also note that a necessary condition for instability is
\begin{equation}
\omega_{\kappa i} \omega_{*e} \eta_i > 0,
\end{equation}
which is the condition that the magnetic curvature and ion temperature gradient have the same sign. The region where $\omega_{\kappa i} \omega_{*e} > 0$ is known as `bad curvature' and $\omega_{\kappa i} \omega_{*e} < 0$ is `good curvature.' Generally, ITG is driven in bad-curvature regions, although electromagnetic effects (neglected above) can destabilize ITG in good-curvature regions \cite{Ivanov2025}.

The toroidal ITG mode is analytically complicated, and an in-depth treatment is beyond the scope of this tutorial. There are many excellent resources on the ITG instability including \cite{Romanelli1989,Biglari1989,Plunk2014,Rodriguez2025}.

\subsection{Slab ITG Instability}

We can also take the `slab' ITG limit, which is found when setting the magnetic drift frequencies in \Cref{eq:ITG_disp_rln_full} to zero. This gives the dispersion relation
\begin{equation}
\begin{aligned}
& 1 + \tau  \left( 1 +  \zeta_i Z (\zeta_i) \Gamma_0 \right) + \frac{\omega_{*e} }{k_{\parallel } v_{ti}} Z (\zeta_i) \Gamma_0 \left( 1 - \frac{3}{2} \eta_i \right) \\
& + \frac{\omega_{*e} }{k_{\parallel } v_{ti}} \eta_i \Gamma_0 \left( \zeta_i + \zeta_i^2 Z(\zeta_i) \right) +  \frac{\omega_{*e} }{k_{\parallel } v_{ti}} \eta_i Z (\zeta_i) \left( \Gamma_0 + b_i \left( \Gamma_1 - \Gamma_0 \right)  \right) = 0.
\end{aligned}
\label{eq:ITG_disp_rln_slab}
\end{equation}

\subsubsection{Drift Wave}

We recover the `drift-wave' limit, which is when $b_i \ll 1, \zeta_i \gg 1, \eta_i = 0$, and $\tau \ll 2 \zeta_i^2$ we obtain,
\begin{equation}
1 - \frac{\omega_{*e}}{k_{\parallel}  v_{ti}} \frac{1}{\zeta_i} \approx 0,
\end{equation}
and so we obtain the drift-wave relation
\begin{equation}
\omega \approx \omega_{*e}.
\end{equation}
This is a stable wave known as a drift wave.

\subsubsection{Instability}

Next, we consider a mode that has a positive growth rate. In order to obtain analytic results, we follow the treatment in \cite{Parra_2019_driftwaves} using the limits
\begin{equation}
\eta_i \gg 1, \; \zeta_i \gg 1, \; b_i \ll 1, \; \Gamma_0 \to 1, \; \Gamma_1 \to 0, \; \omega \sim \omega_{*e},
\end{equation}
and the dispersion relation becomes cubic in $\omega$
\begin{equation}
\omega^3 - \omega_{*e} \omega^2 - \omega_{*e} \eta_i \frac{k_{\parallel} ^2 T_i}{m_i} = 0,
\label{eq:ITG_cubic}
\end{equation}
where to make all of the terms of the same order for $\omega \sim \omega_{*e}$, we had to assume
\begin{equation}
\zeta_i \sim \sqrt{\eta_i} \gg 1.
\end{equation}
Next, we take the limit $\eta_i \to \infty$, and assuming that $\omega > \omega_{*e}$. This gives $\left( \omega/\omega_{*e} \right)^3 > \left( \omega/\omega_{*e} \right)^2$, so \Cref{eq:ITG_cubic} becomes
\begin{equation}
\omega^3 = \eta_i \frac{k_{\parallel} ^2 T_i}{m_i} \omega_{*e},
\end{equation}
which has three roots,
\begin{equation}
\omega = \bigg{(} \eta_i \frac{k_{\parallel} ^2 T_i}{m_i} \omega_{*e} \bigg{)}^{1/3} e^{\frac{2\pi}{3}i}, \;\;\;\; \bigg{(} \eta_i \frac{k_{\parallel} ^2 T_i}{m_i} \omega_{*e} \bigg{)}^{1/3} e^{\frac{4\pi}{3}i}, \;\;\;\; \bigg{(} \eta_i \frac{k_{\parallel} ^2 T_i}{m_i} \omega_{*e} \bigg{)}^{1/3}.
\end{equation}
The first root is unstable, since $\mathrm{Im}(\omega) > 0$,
\begin{equation}
\omega = \bigg{(} \eta_i \frac{k_{\parallel} ^2 T_i}{m_i} \omega_{*e} \bigg{)}^{1/3} e^{\frac{2\pi}{3}i} =  \bigg{(} \eta_i \frac{k_{\parallel} ^2 T_i}{m_i} \omega_{*e} \bigg{)}^{1/3} (-\frac{1}{2}+ i \frac{\sqrt{3}}{2} ).
\label{eq:freqroot}
\end{equation}
Writing the complex frequency as a sum of a real frequency $\omega_R$ and growth rate $\gamma$, 
\begin{equation}
\omega = \omega_R + i \gamma,
\end{equation}
\Cref{eq:freqroot} gives that the growth rate increases with increasing $k_{\parallel} $,
\begin{equation}
\gamma = \frac{\sqrt{3}}{2} \bigg{(} \eta_i \frac{k_{\parallel} ^2 T_i}{m_i} \omega_{*e} \bigg{)}^{1/3} \sim k_{\parallel} ^{2/3}.
\end{equation}

\subsubsection{Landau Damping}

\begin{figure}
\centering
  \includegraphics[width=0.50\linewidth]{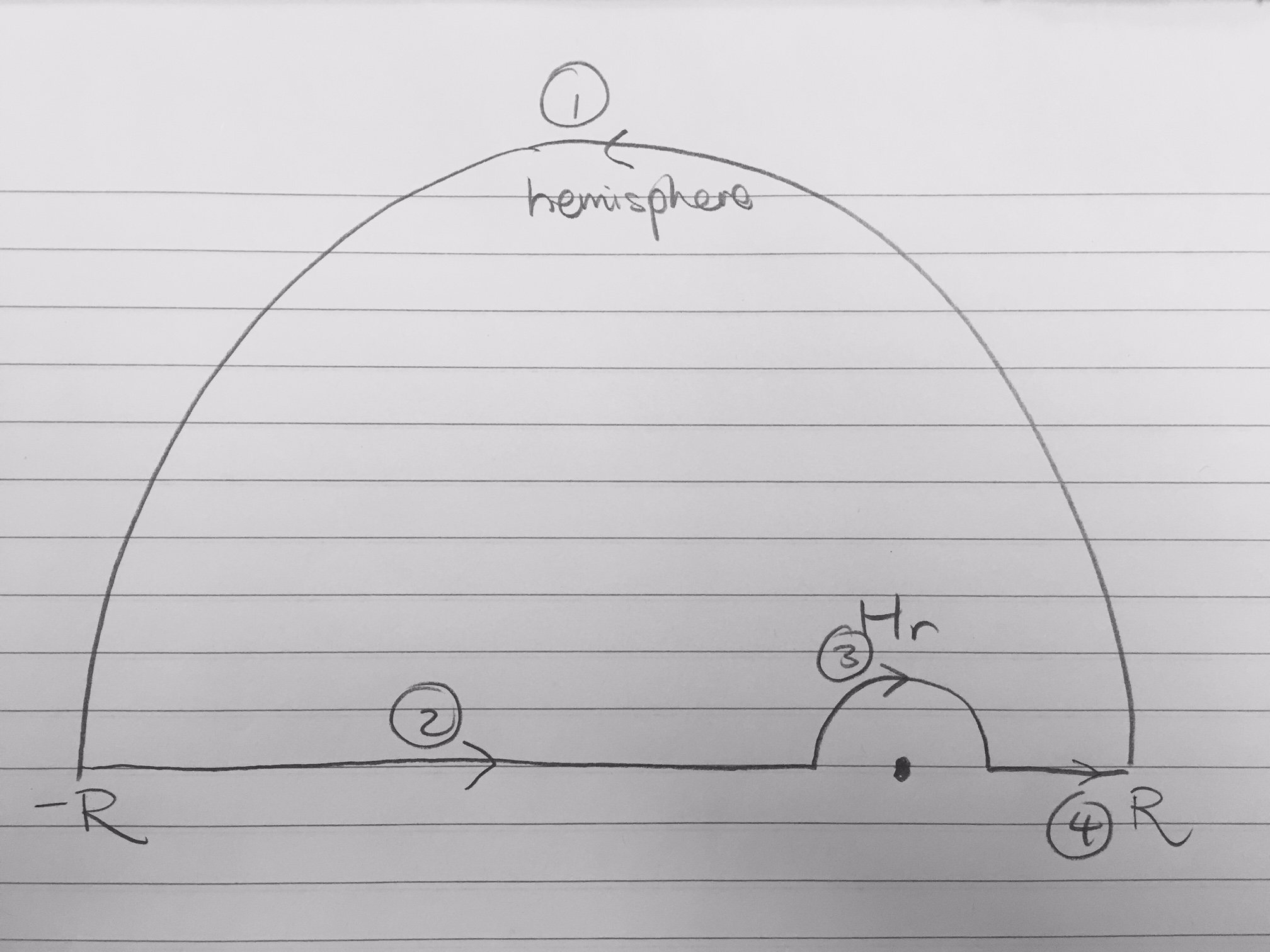}
  \caption{Contour for $Z(\zeta)$ integral, with the pole at $p = \zeta$.}
\label{fig:zzetaintegral}
\end{figure}

However, $\gamma \sim k_{\parallel} ^{2/3}$ cannot hold for all values of $k_{\parallel} $, because another kinetic effect called Landau damping \cite{Landau1946} decreasing the growth rate of modes at sufficiently high $k_{\parallel}$ values. Therefore, we want to obtain the $k_{\parallel} $ at which the growth rate goes to zero ($\gamma = 0$). First, for Landau damping to occur the frequency is set to
\begin{equation}
\omega \sim k_{\parallel}  v_{ti} \sim \eta_i \omega_{*e}.
\label{eq:kz_ordering}
\end{equation}
Using this ordering for the dispersion relation, we obtain in the limit $\eta_i \gg 1$,
\begin{equation}
1 + \tau \bigg{(} 1+ \zeta_i Z(\zeta_i) \Gamma_0 \bigg{)} + \frac{w_{*e} \eta_i}{k_{\parallel}  v_{ti}} \bigg{(} \zeta_i (1+\zeta_i Z(\zeta_i))\Gamma_0 + b_i Z(\zeta_i)(\Gamma_1 - \Gamma_0) - \frac{1}{2} Z(\zeta_i) \Gamma_0 \bigg{)} = 0.
\label{eq:dispersionetagg1}
\end{equation}
To obtain the $k_{\parallel} $ for which $\gamma = 0$, we require $\omega$ is real. When $\omega$ real, the dispersion function becomes
\begin{equation}
Z(\zeta) = \frac{1}{\sqrt{\pi}} \int_{-\infty}^{\infty} \frac{dp e^{-p^2}}{p - \zeta}, 
\end{equation}
which we can evaluate using contour integration. First, let $f(p)$ equal
\begin{equation}
f(p) = \frac{1}{\sqrt{\pi}} \frac{e^{-p^2}}{p- \zeta}.  
\end{equation}
Next, choosing the standard upper hemisphere contour, with a pole at $p = \zeta$ on the real axis (refer to Figure \ref{fig:zzetaintegral}) \footnote{There is a pole on the real axis, so
$$ Z(\zeta) = \mathcal{P} \int_{-\infty}^{\infty} f(p) $$ 
$$ \oint f(p) dp =  \int_{-\infty}^{\infty} f(p) + \int_{\text{hemisphere}} f(p) $$
but
$$ \int_{\text{hemisphere}} f(p) = 0 ,$$
so
$$ \oint f(p) dp = \mathcal{P} \int_{-\infty}^{\infty} f(p) + \int_{H_r} f(p) .$$
Now, letting $ z - \zeta = \epsilon e^{i \theta}$,
$$ \int_{H_r} f(p)= \frac{1}{\sqrt{\pi}}  \lim_{\epsilon \to 0} \int_{\zeta-\epsilon}^{\zeta+\epsilon} \frac{e^{-z^2}}{z-\zeta} dz = \frac{1}{\sqrt{\pi}} \lim_{\epsilon \to 0} \int_{-\pi}^0 \frac{e^{-z^2}}{\epsilon e^{i \theta}} \epsilon i \theta e^{i \theta} d \theta  = \frac{1}{\sqrt{\pi}} \lim_{\epsilon \to 0} \int_{-\pi}^0 e^{-(\epsilon e^{i \theta} - \zeta)^2} i \theta e^{i \theta} d \theta = i \sqrt{\pi} e^{-\zeta^2_i} .$$
Next, using the residue theorem,
$$ \oint f(p) dp = 2 \pi i \sum_k \mathrm{Res}(f(p), \zeta_k) = \pi i \lim_{p \to \zeta} \bigg{(} (p-\zeta) f(p) \bigg{)} = 2 i \sqrt{\pi} e^{-\zeta^2_i} = \int_{-\infty}^{\infty} f(p) +  i \sqrt{\pi} e^{-\zeta^2_i} .$$
Therefore
$$ \int_{-\infty}^{\infty} f(p) = i \sqrt{\pi} e^{-\zeta^2_i} .$$
},

\begin{equation}
Z(\zeta) = \mathcal{P} \oint f(p) dp = 2 \pi i \sum_k \mathrm{Res}(f(p), \zeta_k) = \pi i \lim_{p \to \zeta} \bigg{(} (p-\zeta) f(p) \bigg{)} = \sqrt{\pi} i e^{-\zeta^2_i}.
\end{equation}
\begin{equation}
Z(\zeta) = i \sqrt{\pi} e^{-\zeta^2_i}
\end{equation}
However, all remaining terms in the dispersion relation are real if $\omega$ is real. So, we just find the coefficients to $Z(\zeta)$, and set them equal to zero. So, we have
\begin{equation}
\bigg{[} \tau \zeta_i \Gamma_0 +   \frac{w_{*e} \eta_i}{k_{\parallel}  v_{ti}} \bigg{(} \zeta_i^2 \Gamma_0 + b_i (\Gamma_1 - \Gamma_0) - \frac{1}{2} \Gamma_0 \bigg{)}  \bigg{]} \sqrt{\pi} i e^{-\zeta^2_i} = 0.
\label{eq:im}
\end{equation}
Thus, we require that the real and imaginary parts of the dispersion relation are individually equal to zero. For the real part,
\begin{equation}
1 + \tau  + \frac{w_{*e} \eta_i}{k_{\parallel}  v_{ti}} \zeta_i \Gamma_0 = 0,
\end{equation}
which gives a relation for $\zeta_i$:
\begin{equation}
\zeta_i = - \left( 1 + \tau \right) \frac{1}{\Gamma_0} \frac{k_{\parallel}  v_{ti}}{\eta_i \omega_{*e}}. 
\label{eq:zetaeq1}
\end{equation}
Next, substituting \Cref{eq:zetaeq1} into the imaginary component of the dispersion relation (\Cref{eq:im}),
\begin{equation}
\bigg{(} \frac{k_{\parallel}  v_{ti}}{\eta_i \omega_{*e}} \bigg{)}^2 \bigg{(} - \tau (1 + \tau ) + (1+ \tau )^2 \frac{1}{\Gamma_0}  \bigg{)} + b_i (\Gamma_1 - \Gamma_0) - \frac{\Gamma_0}{2} = 0.
\end{equation}
This gives the $k_{\parallel} $ at which the growth rate goes to zero,
\begin{equation}
\frac{k_{\parallel}  v_{ti}}{\eta_i \omega_{*e}} = \pm \left( \Gamma_0 \frac{b_i (\Gamma_1 - \Gamma_0) + \frac{\Gamma_0}{2}}{(1+ \tau )(1+ \tau (1 - \Gamma_0) ) }  \right)^{1/2}.
\end{equation}
Letting $b_i \ll 1$,
\begin{equation}
\frac{k_{\parallel}  v_{ti}}{\eta_i \omega_{*e}} \approx \bigg{(} \frac{1}{2 (1+ \tau )}  \bigg{)}^{1/2}.
\end{equation}
Finally, plugging this into the equation for $\zeta_i$, Equation \ref{eq:zetaeq1} we obtain, when assuming that the frequency is only real (i.e. no damping/instability), we obtain the frequency at which the growth rate goes to zero:
\begin{equation}
\omega = - \frac{ b_i (\Gamma_1 - \Gamma_0) + \frac{\Gamma_0}{2} }{1 + \tau (1 - \Gamma_0) } \eta_i \omega_{*e} =_{{}_{b_i \ll 1}} - \frac{1}{2} \eta_i \omega_{*e}.
\end{equation}

\subsubsection{Low $k_{\parallel} $ Growth Rate}

Next, we calculate the mode at $k_{\parallel}  \to 0$, with $\eta \gg 1$, and we drop the restriction of $\gamma = 0$. Recall that for $b_i = 0$, we found $\gamma \sim k_{\parallel} ^{2/3}$. Here, we take $\zeta_i \gg 1$. For the dispersion relation in \Cref{eq:dispersionetagg1}, we use the limits $|\mathrm{Re}(\zeta_i)| \gg 1$, $|\mathrm{Re}(\zeta_i)| |\mathrm{Im}(\zeta_i)| \ll 1$ to expand the dispersion function,
\begin{equation}
Z(\zeta_i) = - \frac{1}{\zeta_i} - \frac{1}{2 \zeta_i^3} + ... + i \sqrt{\pi} e^{-\zeta_i^2}. 
\label{eq:plasdisperzetahigh}
\end{equation}
The dispersion relation is
\begin{equation}
\begin{aligned}
& 1 + \tau \bigg{(} 1+ \zeta_i Z(\zeta_i) \Gamma_0 \bigg{)} + \frac{w_{*e} \eta_i}{k_{\parallel}  v_{ti}} \bigg{(} \zeta_i (1+\zeta_i Z(\zeta_i))\Gamma_0 + b_i Z(\zeta_i)(\Gamma_1 - \Gamma_0) - \frac{1}{2} Z(\zeta_i) \Gamma_0 \bigg{)} = 0 \\
& 1 + \tau +  \frac{w_{*e} \eta_i}{k_{\parallel}  v_{ti}} \zeta_i \Gamma_0 + Z(\zeta_i) \bigg{[} \tau \zeta_i  \Gamma_0 + \frac{w_{*e} \eta_i}{k_{\parallel}  v_{ti}} \bigg{(} \zeta_i^2 \Gamma_0 + b_i (\Gamma_1 - \Gamma_0) - \frac{1}{2} \Gamma_0 \bigg{)} \bigg{]} = 0.
\end{aligned}
\label{eq:inter_DR}
\end{equation}
Inserting the plasma dispersion function in Equation \ref{eq:plasdisperzetahigh}, we find\footnote{Note that
\begin{equation}
\begin{aligned}
& \mathrm{Re} \bigg{(} Z(\zeta_i) \bigg{(} \zeta_i^2 \Gamma_0 + b_i (\Gamma_1 - \Gamma_0) - \frac{1}{2} \Gamma_0 \bigg{)} \bigg{)} \approx (- \frac{1}{\zeta_i} - \frac{1}{2 \zeta_i^3}) \bigg{(} \zeta_i^2 \Gamma_0 + b_i (\Gamma_1 - \Gamma_0) \bigg{)} \bigg{)} \\
& \approx - \zeta_i \Gamma_0 - \frac{1}{\zeta_i} b_i (\Gamma_1 - \Gamma_0)
\end{aligned}
\end{equation}
}
\begin{equation}
\begin{aligned}
& 0 = 1 + \tau +  \frac{w_{*e} \eta_i}{k_{\parallel}  v_{ti}} \zeta_i \Gamma_0 \\
& + ( - \frac{1}{\zeta_i} - \frac{1}{2 \zeta_i^3} + i \sqrt{\pi} e^{-\zeta_i^2} ) \bigg{[} \tau \zeta_i  \Gamma_0 + \frac{w_{*e} \eta_i}{k_{\parallel}  v_{ti}} \bigg{(} \zeta_i^2 \Gamma_0 + b_i (\Gamma_1 - \Gamma_0) - \frac{1}{2} \Gamma_0 \bigg{)} \bigg{]} = 0 \\
& \approx 1 + \tau (1- \Gamma_0) \\
& - \frac{w_{*e} \eta_i}{k_{\parallel}  v_{ti}} \frac{1}{\zeta_i} \bigg{(} b_i (\Gamma_1 - \Gamma_0) - \frac{\Gamma_0}{2} \bigg{)} + i \sqrt{\pi} e^{-\zeta_i^2} ) \bigg{[} \tau \zeta_i  \Gamma_0 + \frac{w_{*e} \eta_i}{k_{\parallel}  v_{ti}} \bigg{(} \zeta_i^2 \Gamma_0 + b_i (\Gamma_1 - \Gamma_0) - \frac{1}{2} \Gamma_0 \bigg{)} \bigg{]}.
\end{aligned}
\label{eq:dispersionhighzeta}
\end{equation}
Next, assuming that $\omega_R \gg \gamma$, the real part of \Cref{eq:dispersionhighzeta} is
\begin{equation}
1 + \tau (1- \Gamma_0) + \eta_i \frac{\omega_{*e}}{\omega_R} \bigg{(} b_i (\Gamma_1 - \Gamma_0)  \bigg{)} = 0.
\end{equation}
This gives the real frequency
\begin{equation}
\omega_R =  \frac{\eta_i}{\omega_{*e}} \frac{ b_i (\Gamma_0 - \Gamma_1) }{1 + \tau (1- \Gamma_0)}.
\end{equation}
The imaginary part of the dispersion relation i
\begin{equation}
- \gamma \eta_i \frac{\omega_{*e}}{\omega_R^2}  b_i (\Gamma_1 - \Gamma_0) + \eta_i \sqrt{\pi} e^{- \frac{\omega_R^2}{k_{\parallel} ^2 v_{ti}^2} } \frac{\omega_R^2 \omega_{*e}}{k_{\parallel} ^3 v_{ti}^3} \Gamma_0  = 0.
\end{equation}
Rearranging for the growth rate gives
\begin{equation}
\gamma =  \frac{\omega_R^4 }{k_{\parallel} ^3 v_{ti}^3} \frac{  \sqrt{\pi} e^{- \frac{\omega_R^2}{k_{\parallel} ^2 v_{ti}^2} }  }{ b_i (\Gamma_1 - \Gamma_0) } \Gamma_0,
\label{eq:gamma_lowkz}
\end{equation}
where we used $w_R \gg \gamma$ and $\zeta_i \gg 1$. Therefore \Cref{eq:gamma_lowkz} gives the ITG growth in the $k_{\parallel}  \to 0$ limit.

\subsubsection{Critical $\eta_i$}

In this section, we find an expression for the critical value of $\eta_i$, which occurs when $\gamma = 0$. We no longer assume that $\eta_i$ is large, instead ordering $\eta_i \sim 1$. 

When $\gamma = 0$, the only imaginary components in the dispersion relation (\Cref{eq:inter_DR}) come from coefficients of $Z(\zeta_i)$,
\begin{equation}
\tau \zeta_i \Gamma_0 + \frac{\omega_{*e}}{k_{\parallel}  v_{ti}} \bigg{(}  \Gamma_0 + \eta_i (\zeta_i^2 \Gamma_0 + b_i (\Gamma_1 - \Gamma_0) - \frac{1}{2}  \Gamma_0 ) \bigg{)} = 0.
\label{eq:dispersiongeneraimag}
\end{equation}
Therefore for $\gamma = 0$ the real components of the dispersion relation are
\begin{equation}
1 + \tau + \eta_i \frac{\omega_{*e}}{k_{\parallel}  v_{ti}} \zeta_i \Gamma_0 = 0,
\end{equation}
giving the $\zeta_i$ for which $\gamma = 0$
\begin{equation}
\zeta_i = - \frac{k_{\parallel}  v_{ti}}{\eta_i \omega_{*e}} \frac{1}{\Gamma_0} \bigg{(} 1 + \tau \bigg{)}.
\end{equation}
Substituting $\zeta_i$ into \Cref{eq:dispersiongeneraimag} gives the zero growth rate dispersion relation
\begin{equation}
- \tau  \frac{k_{\parallel}  v_{ti}}{\eta_i \omega_{*e}} \frac{1}{\Gamma_0} \bigg{(} 1 + \tau \bigg{)} \Gamma_0 + \frac{\omega_{*e}}{k_{\parallel}  v_{ti}} \bigg{(}  \Gamma_0 + \eta_i (\bigg{[} \frac{k_{\parallel}  v_{ti}}{\eta_i \omega_{*e}} \frac{1}{\Gamma_0} \bigg{(} 1 + \tau \bigg{)} \bigg{]}^2 \Gamma_0 + b_i (\Gamma_1 - \Gamma_0) - \frac{1}{2}  \Gamma_0 ) \bigg{)} = 0.
\end{equation}
The growth rate vanishes when the frequency satisfies
\begin{equation}
\omega = - \eta_i \omega_{*e} \Gamma_0 \frac{ b_i (1 - \Gamma_1/ \Gamma_0) - 1/\eta_i  + 1/2  }{ 1 + \tau (1 - \Gamma_0) }.
\end{equation}
and $k_{\parallel}  v_{ti}$ is
\begin{equation}
k_{\parallel}  v_{ti} = \pm \eta_i \omega_{*e} \Gamma_0 \bigg{[} \frac{ b_i (1 - \Gamma_1/\Gamma_0) - 1/\eta_i  + 1/2 }{  \left( 1 + \tau \right)  \left( 1 + \tau (1 - \Gamma_0) \right)  }  \bigg{]}^{1/2}.
\label{eq:kzvti_eq_ITG}
\end{equation}
To find the critical value of $\eta_i$ where the growth rate is zero for all $k_{\parallel} $ values, we set $k_{\parallel}  = 0$ in \Cref{eq:kzvti_eq_ITG}. This gives the critical $\eta_i$ value for which the slab ITG mode is stabilized
\begin{equation}
\eta_{i, \mathrm{crit}} = \frac{1}{b_i (1 - \Gamma_1/\Gamma_0) + 1/2 }.
\end{equation}
In the long-wavelength limit where $b_i \to 0$, this gives the well-known result that $\eta_{i, \mathrm{crit}} = 2$.

\bibliographystyle{plain} %
\bibliography{Master_EverythingPlasmaBib}

\end{document}